\documentclass[aps,prd,twocolumn,preprintnumbers,showpacs,nofootinbib,amssymb]{revtex4}
\usepackage{graphicx}
\usepackage{amsmath}
\usepackage{amssymb}
\usepackage{amsfonts}
\usepackage{bm}
\usepackage{color}

\def\be{\begin{equation}}
\def\ee{\end{equation}}
\def\bea{\begin{eqnarray}}
\def\eea{\end{eqnarray}}

\begin{document}

\title{Derivation of the core mass -- halo mass relation of fermionic and
bosonic dark matter halos
from an effective thermodynamical model}
\author{Pierre-Henri Chavanis}
\affiliation{Laboratoire de Physique Th\'eorique, Universit\'e de Toulouse,
CNRS, UPS, France}

\begin{abstract} 

We consider the possibility that dark matter halos are made of quantum
particles such as fermions or bosons in the form of Bose-Einstein condensates.
In that case, they generically have a ``core-halo'' structure
with a quantum core that depends on the type of particle considered and a halo
that is relatively independent of the dark matter particle and that is
similar to the NFW profile of cold dark matter. The quantum core is
equivalent to a polytrope of index $n=3/2$ for fermions, $n=2$ for
noninteracting bosons, and $n=1$ for bosons with a repulsive self-interaction
in the Thomas-Fermi limit.
We model the halo by an isothermal gas with an effective temperature $T$. We
then derive the core mass -- halo mass relation $M_c(M_v)$ of dark matter halos
from an
effective thermodynamical model by maximizing the entropy $S(M_c)$  with respect
to the
core mass $M_c$ at fixed total
mass and total energy. We obtain a general relation, valid for an arbitrary
polytropic
core, that is equivalent to the ``velocity dispersion tracing'' relation
according to which
the velocity dispersion in the core $v_c^2\sim GM_c/R_c$ is of the same order as
the velocity
dispersion in the halo $v_v^2\sim GM_v/r_v$. We provide therefore a
justification of this relation from thermodynamical arguments. In the case of
fermions, we obtain a relation
$M_c\propto M_v^{1/2}$  that agrees with the relation found numerically by
Ruffini
{\it et al.} [Mon. Not. R.
Astron. Soc. {\bf 451}, 622 (2015)]. In the
case of
noninteracting bosons, we
obtain a relation $M_c\propto M_v^{1/3}$ that agrees with the relation
found numerically  by Schive {\it et al.} [Phys. Rev. Lett {\bf
113}, 261302 (2014)].
In the case of  bosons with a repulsive self-interaction in the Thomas-Fermi
limit, we
predict a relation $M_c\propto M_v^{2/3}$ that
still has to be confirmed numerically. Using a Gaussian ansatz, we obtain a
general approximate
core mass -- halo mass relation $M_c(M_v)$ that is valid for bosons with
arbitrary repulsive or
attractive self-interaction. For an attractive self-interaction, we determine
the maximum halo mass $(M_v)_{\rm max}$ that can harbor a stable quantum core
(dilute axion ``star''). Above that mass, the quantum core collapses.
Finally, we argue that the fundamental mass
scale of the bosonic dark matter particle is
$m_{\Lambda}=\hbar\sqrt{\Lambda}/c^2=2.08\times 10^{-33}\,
{\rm eV/c^2}$ and that the fundamental mass scale of the fermionic dark matter
particle is $m_{\Lambda}^*=({\Lambda\hbar^3}/{Gc^3})^{1/4}=\sqrt{m_{\Lambda}M_P}
=5.04\times 10^{-3}\, {\rm eV/c^2}$, where $\Lambda$ is the cosmological
constant and $M_P$ is the Planck mass. Their ratio is
$m_{\Lambda}^*/m_{\Lambda}=(c^5/G\hbar\Lambda)^{1/4}=2.42\times 10^{30}$ which
explains the difference of mass between fermions and bosons in dark matter
models. The
actual value
of the dark
matter particle mass is equal to these mass scales multiplied by a large factor
that
we obtain from our model.

\end{abstract}


\maketitle

\section{Introduction}

The nature of dark matter (DM) is still unknown and remains one of the greatest
mysteries of modern cosmology. The standard cold dark matter (CDM) model
works remarkably well at large (cosmological) scales and is consistent with ever
improving measurements of the cosmic microwave background (CMB) from WMAP and
Planck missions \cite{planck2013,planck2016}. However, it encounters serious
problems at small (galactic) scales. In particular, it predicts that DM halos
should be cuspy \cite{nfw}, with a density diverging as $r^{-1}$ for
$r\rightarrow 0$,  while observations reveal that they have a flat 
core density \cite{observations}. On the other hand, the CDM model predicts
an over-abundance of small-scale structures (subhalos/satellites), much more
than what is observed around the Milky Way \cite{satellites}. These problems are
referred to as the ``cusp problem'' and  ``missing satellite problem''. The
expression ``small-scale crisis of CDM'' has been coined.

In order to solve these problems, some authors have
proposed to take  the quantum nature
of the DM particle into account.\footnote{See our
previous papers \cite{prd1,modeldm} for an exhaustive list of references (more
than
$200$) on the
subject. See also the reviews
\cite{srm,rds,chavanisbook,marshrevue,leerevue,braatenrevue}.} Indeed, quantum
mechanics creates an effective
pressure even at zero thermodynamic temperature ($T_{\rm th}=0$) that may
balance the gravitational
attraction at small
scales and lead to cores instead of cusps. The DM particle
could be a fermion, like a massive neutrino,  with a mass $\sim 170\, {\rm
eV/c^2}$ (see
Appendix D of \cite{suarezchavanis3}). It could also be
a boson in the form of a Bose-Einstein condensate (BEC), like an ultralight
axion, with a mass in the range
$2.19\times 10^{-22}\, {\rm eV}/c^2<m<1.10\times 10^{-3}\, {\rm
eV}/c^2$ depending
whether the bosons are  noninteracting or self-interacting (see
Appendix D of \cite{suarezchavanis3}).

In these quantum models, DM halos have a ``core-halo'' structure which results
from a
process of violent collisionless relaxation \cite{lb} and gravitational cooling
\cite{seidel94,gul0,gul}.
The core stems from the equilibrium between
quantum pressure and gravitational
attraction. For fermions, the
quantum pressure arises from the Pauli exclusion principle like in the case of
white dwarfs and neutron stars. For
bosons, the quantum pressure arises from the
Heisenberg uncertainty principle or from the 
repulsive self-interaction of the bosons like in the case of boson stars.
Quantum mechanics stabilizes the halo against
gravitational collapse,\footnote{This is true for the nonrelativistic systems
that we consider here. For general relativistic systems, there is a maximum
mass $M_{\rm max}^{\rm GR}$ \cite{ov,kaup,rb,colpi,tkachev,chavharko} above
which the system collapses towards a black hole. In our case, we will
find that $M_c\ll M_{\rm max}^{\rm GR}$ so that a Newtonian approach is
sufficient. On the other hand, if the bosons have an attractive
self-interaction, like in the case of the axion \cite{marshrevue}, there exists
a maximum mass
$M_{\rm max}$  \cite{prd1} for the quantum core even in the Newtonian regime.
Above that limit, the quantum core (dilute axion star) undergoes gravitational
collapse.} leading to a flat core instead of a
cusp. The quantum core is equivalent to a polytrope of index
$n=3/2$ for fermions, $n=2$ for noninteracting
bosons, and $n=1$ for bosons with a repulsive self-interaction in the
Thomas-Fermi (TF) limit. It is responsible for the finite density
of the DM halos at the center. The core mass-radius relation is $M_c
R_c^3=1.49\times 10^{-3}\, 
h^6/G^3m^8$ for fermions, $M_c R_c=5.25\, 
\hbar^2/Gm^2$ for noninteracting bosons, and $R_c=\pi(a_s\hbar^2/Gm^3)^{1/2}$
for self-interacting bosons in the TF limit. On the other hand, the
halo is relatively independent of quantum effects and is similar to the
Navarro-Frenk-White (NFW)
profile \cite{nfw} produced in CDM simulations or to the empirical Burkert
profile
\cite{observations} deduced from the observations. It is responsible for the
flat rotation curves of the
galaxies at large distances. We shall approximate this halo by an isothermal
atmosphere with an
effective temperature $T$.\footnote{We approximate the
atmosphere by an isothermal sphere but we stress
that the temperature is effective and does not correspond to the true
thermodynamic temperature (which is almost equal to zero). In particular, the
atmosphere does not correspond to a statistical equilibrium state resulting from
a ``collisional'' evolution of the quantum particles of mass $m$ which would
be much too long (much larger than the age of the Universe) \cite{modeldm}. It
may rather correspond to an out-of-equilibrium thermodynamical state -
or quasistationary state - resulting from a collisionless free fall evolution
(independent
of $m$) like in Lynden-Bell's
theory of violent relaxation \cite{lb}. In the case of fuzzy DM, the
approximately isothermal atmosphere (due to quantum interferences of excited
states)
may result either from a collisionless relaxation or from the
``collisional'' evolution of quasiparticles (granules)
of the size of the solitonic core and of mass $m_*\gg m$ as argued in
\cite{ch2,ch3,hui,bft}.} In that case, the density decreases at large
distance as $\rho\propto r^{-2}$ \cite{bt}, instead of $r^{-3}$ for the NFW and
Burkert  profiles,
leading
exactly to flat rotation curves for $r\rightarrow +\infty$. For
sufficiently large halos, the halo mass-radius relation is $M_h=1.76\,
\Sigma_0
r_h^2$ \cite{modeldm} where
\begin{eqnarray}
\label{p5}
\Sigma_0=\rho_0 r_h=141\, M_{\odot}/{\rm
pc}^2
\end{eqnarray} 
is the universal surface density of DM halos deduced from the observations
\cite{kormendy,spano,donato}. Ultracompact halos like dSphs ($r_h\sim
1\, {\rm kpc}$ and $M_h\sim 10^{8}\, M_\odot$) are dominated by
the quantum core and have almost no atmosphere. Large halos like the Medium
Spiral ($r_h\sim
10\, {\rm kpc}$ and $M_h\sim 10^{11}\, M_\odot$) are dominated by the isothermal
atmosphere.

In a recent paper \cite{modeldm}, we have developed of model of DM halos
made of bosons with a repulsive self-interaction in the TF limit. We have
obtained a generic phase diagram (see Fig. 49 of  \cite{modeldm})
determining the structure of the DM halos (measured by the core mass $M_c$) as a
function of their mass $M_h$. There is a minimum halo mass $(M_h)_{\rm min}$
corresponding to the ground  state of the boson gas ($T=0$) at which the DM halo
is a
purely quantum object without isothermal atmosphere ($M_c\simeq M_h$). We
will call it the ``minimum halo''. Larger halos
have a ``core-halo'' structure with a quantum core, representing
a ``nucleus'' or a ``bulge'', and an isothermal atmosphere.
We found a branch along which the core mass
$M_c$ decreases as the halo mass $M_h$ increases. Rapidly, the core mass
becomes negligible and the halos behave as purely isothermal halos without
quantum core. However, we found a critical point $(M_h)_{\rm CCP}$, that we
interpreted as a canonical critical point, at which a bifurcation
occurs. On the
new branch,
the core mass $M_c$ increases as the halo mass $M_h$ increases. On that branch,
we found another  critical point at a higher mass $(M_h)_{\rm MCP}$, that we
interpreted as
a microcanonical critical point, above which the quantum core becomes unstable
and is replaced by a supermassive black hole resulting from a gravothermal
catastrophe \cite{lbw} followed by a dynamical instability of general
relativistic origin \cite{balberg}.
Considering the bifurcated branch with the ``core-halo'' structure,
we developed an effective thermodynamical model to analytically predict the
core mass -- halo mass relation
$M_c(M_h)$.\footnote{This thermodynamical model was originally introduced in
\cite{pt} to
analytically obtain the caloric curves of self-gravitating fermions.}
 We showed that
this relation is equivalent to the ``velocity dispersion tracing'' relation
according to which
the velocity dispersion in the core $v_c^2\sim GM_c/R_c$ is of the same order as
the velocity
dispersion in the halo $v_h^2\sim GM_h/r_h$ \cite{mocz,bbbs,modeldm}. We could
provide therefore a
justification of this relation from thermodynamical arguments.

In the present paper, we extend this thermodynamical model to the
case of DM halos made of fermions and to the case of DM halos
made of noninteracting bosons. To unify the formalism, we model
the quantum core as a polytrope of arbitrary index $n$
(with $n=3/2$ for fermions, $n=2$ for noninteracting bosons, and $n=1$ for
self-interacting bosons in the TF limit) and we model the atmosphere as an
isothermal gas with a
uniform density confined within a ``box'' of radius $R$.
The radius of the box is
identified with the halo radius $r_h$ and the total mass of the system
contained within the box (core $+$ halo) is identified with the halo mass $M_h$.
They are related by $M_h=1.76\, \Sigma_0
r_h^2$ \cite{modeldm}. We analytically compute the entropy
$S(M_c)$ of the
system. By maximizing $S(M_c)$ as a function of $M_c$ for a given value of
the mass $M_h$, radius $r_h$ and energy $E_h$, we
obtain the core mass $M_c$ as a function of the halo mass $M_h$. We find this
relation
to be always (for any
value of $n$) equivalent to the velocity dispersion tracing relation, thereby
generalizing our previous result \cite{modeldm}. Using a Gaussian ansatz, we
obtain a general
approximate
relation $M_c(M_h)$ that is valid for bosons with arbitrary repulsive or
attractive self-interaction. For an attractive self-interaction, we determine
the maximum halo mass $(M_h)_{\rm max}$ that can harbor a stable quantum core
(dilute axion ``star''). Finally, we use our results to predict the
fundamental mass scale of the bosonic or fermionic DM particle in terms of the
fundamental constants of physics.

The paper is organized as follows. In Sec. \ref{sec_qm} we consider models of
DM halos with a quantum
(fermionic or bosonic) core and an isothermal atmosphere. In Sec. \ref{sec_gwe}
we show that these models can be obtained in a unified manner from a 
generalized wave equation introduced in \cite{ggp,modeldm,nottalechaos}. In Sec.
\ref{sec_ana} we obtain the core mass -- halo
mass relation of DM halos from an analytical thermodynamical model. This
relation is valid for a general polytropic core. In Sec.
\ref{sec_vdt} we show that this relation is equivalent to the velocity
dispersion tracing relation. In Secs. \ref{sec_app} and \ref{sec_ga} we
specifically apply these results to DM halos made of fermions, noninteracting
bosons and self-interacting bosons.

\section{Quantum models of DM halos}
\label{sec_qm}

In this section, we review quantum models of DM halos made of fermions
or bosons. If DM halos are quantum objects, there must be a minimum halo radius
$R$ and a minimum halo mass $M$  in the Universe corresponding to the ground
state ($T=0$) of the self-gravitating quantum gas. This
result is in agreement with the observations. Indeed, there are apparently no DM
halos with a radius smaller than $R\sim 1\, {\rm kpc}$ and a mass smaller
than $M\sim 10^8\, M_\odot$, the typical values of the radius
and mass of dSphs like Fornax. This observational
result cannot be explained by the CDM model which predicts
the existence of DM halos at all scales. 

Ultracompact dwarf DM halos just have a quantum core without
atmosphere (ground state). Larger
DM halos
have a core-halo structure with a quantum core corresponding to the ground state
($T=0$) of the quantum gas and an ``atmosphere'' resulting from
violent relaxation and gravitational cooling. The atmosphere has a
density profile that can be fitted by the empirical Burkert profile or by
NFW profile. In this paper, we shall
approximate this density profile by an isothermal profile of effective
temperature $T$. It is the atmosphere that fixes the
size of large halos and explains why their radius $r_h$ increases with their
mass
as $M_h\propto r_h^2$ (since $\Sigma_0$ is constant). By contrast, the radius
$R_c$ of the quantum
core usually decreases or remains constant as its mass $M_c$ increases (see
below).

To determine the parameters of the DM particle, we proceed as follows (see
Appendix D of \cite{suarezchavanis3}).\footnote{Our aim here is not to make an
accurate model of DM halos. Therefore, an order of magnitude of the DM
particle parameters is sufficient.} We assume that the smallest
halo that has been observed, with a typical radius and a typical mass
\begin{eqnarray}
\label{fornax}
R\sim 1\, {\rm kpc},\qquad  M\sim 10^8\, M_\odot, \qquad  ({\rm Fornax}),
\end{eqnarray}
corresponds to the ground state of
a self-gravitating quantum gas. Using the mass-radius relation $M(R)$ of the
self-gravitating quantum gas at $T=0$, we can  obtain  the parameters of the  DM
particle. We can then check that the
nonrelativistic treatment used in this paper is valid by showing that $M_c\ll
M_{\rm max}^{\rm GR}$.

\subsection{Fermionic DM}
\label{sec_fdm}

\begin{figure}[!h]
\begin{center}
\includegraphics[clip,scale=0.3]{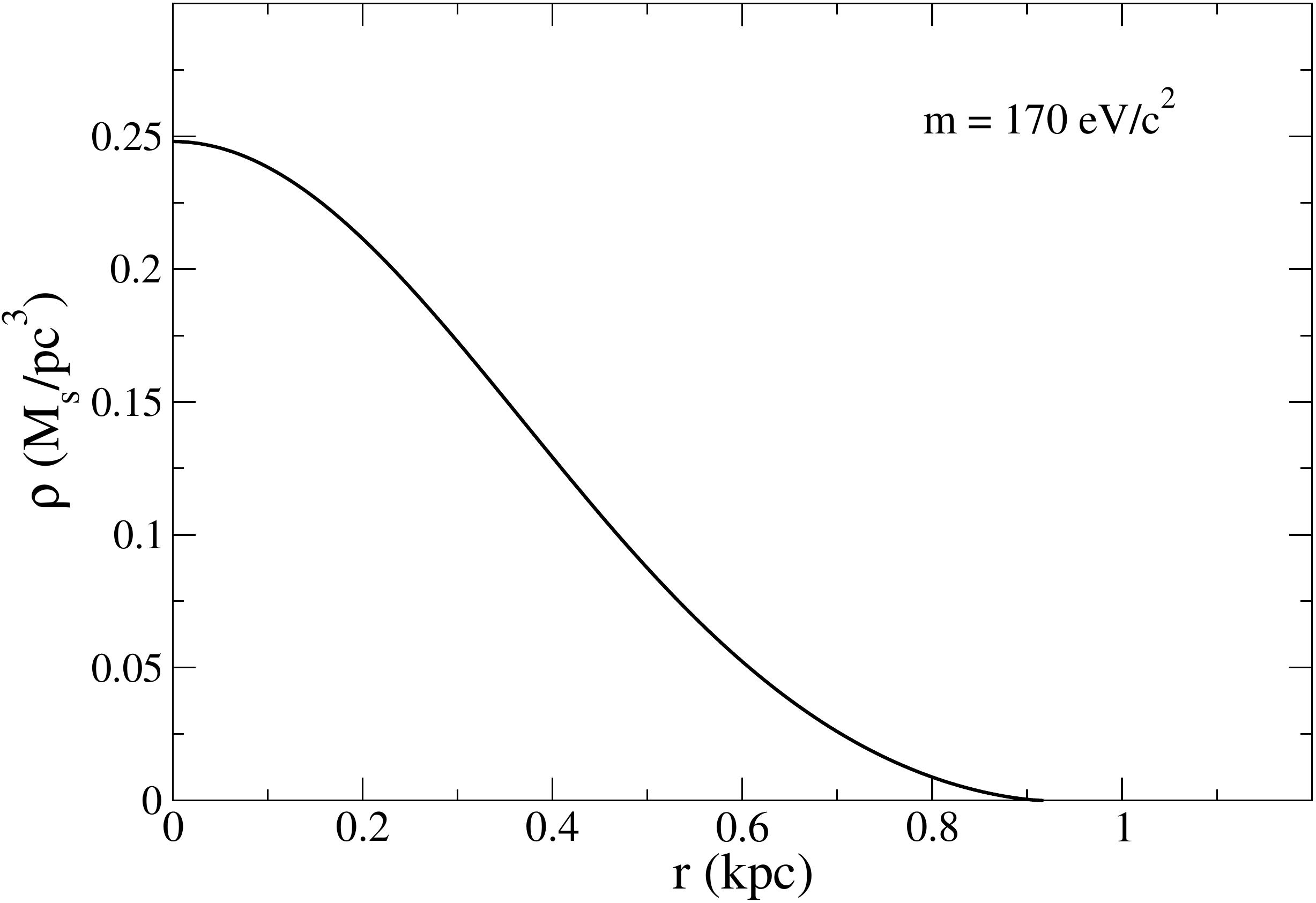}
\caption{Density profile of the ``minimum halo'' (ground state) of surface
density $\Sigma_0=\rho_0 r_h=141\, M_{\odot}/{\rm
pc}^2$ made of fermions of mass $m=170\, {\rm eV}/c^2$.}
\label{rhofermions}
\end{center}
\end{figure}

The equation of state of a nonrelativistic Fermi gas at $T=0$ is
\cite{chandrabook}
\begin{eqnarray}
\label{fdm1}
P=\frac{1}{20}\left (\frac{3}{\pi}\right )^{2/3}\frac{h^2}{m^{8/3}}\rho^{5/3}.
\end{eqnarray}
This is  a polytropic equation of state of index $\gamma=5/3$ (i.e. $n=3/2$) and
polytropic constant
\begin{eqnarray}
\label{fdm2}
K=\frac{1}{20}\left (\frac{3}{\pi}\right )^{2/3}\frac{h^2}{m^{8/3}}.
\end{eqnarray}
In the TF approximation, which amounts to neglecting the quantum potential, the
fundamental differential equation of hydrostatic
equilibrium
determining the density profile of a nonrelativistic fermion star\footnote{In
this paper, the name ``fermion star'' refers to ultracompact DM halos
made of fermions or to the fermionic core of larger DM halos (the same
comment applies to
the names ``boson stars'', ``BEC stars'' and ``axion stars'' used below).} at
$T=0$ 
with the equation of
state (\ref{fdm1}) writes (see Appendix \ref{sec_gl})
\begin{equation}
\label{fdm2b}
\frac{1}{8}\left (\frac{3}{\pi}\right
)^{2/3}\frac{h^2}{m^{8/3}}\Delta\rho^{2/3}=-4\pi G\rho.
\end{equation}
It can be reduced to the Lane-Emden equation (\ref{gl7}) of
index $n=3/2$. This profile has a compact
support (see Fig. \ref{rhofermions}) and the fermion star is stable. The 
mass-radius relation is
\begin{eqnarray}
\label{fdm3}
M_c R_c^3=\frac{9\omega_{3/2}}{8192\pi^4}\frac{h^6}{G^3m^{8}},
\end{eqnarray}
where $\omega_{3/2}=132.3843$. On the
other hand, from  Eqs. (\ref{vt7}) and (\ref{vt9}) which reduce to\footnote{In
the main text, we denote by $E_c$, $U_c$ and $W_c$ what are called $E_{\rm
tot}$, $U$ and $W$ in the Appendices.}
\begin{eqnarray}
\label{fdm4}
E_c=U_c+W_c,
\end{eqnarray}
\begin{eqnarray}
\label{fdm5}
2U_c+W_c=0,
\end{eqnarray}
and from the Betti-Ritter formula (\ref{br5}), we obtain
\begin{eqnarray}
\label{fdm6}
E_c=-U_c=\frac{1}{2}W_c=-\frac{3}{7}\frac{GM_c^2}{R_c}.
\end{eqnarray}
Combined with Eq. (\ref{fdm3}), we find that the energy of a
nonrelativistic
fermion star at
$T=0$ (ground state) is
\begin{eqnarray}
\label{fdm7}
E_c=-\frac{3}{7}\left (\frac{8192\pi^4}{9\omega_{3/2}}\right
)^{1/3}\frac{G^2m^{8/3}M_c^{7/3}}{h^2}.
\end{eqnarray}

Let us assume that the smallest DM halo that we know, with mass $M$ and
radius $R$, corresponds to the ground state of a nonrelativistic
self-gravitating Fermi gas. From the 
mass-radius relation (\ref{fdm3}), we get 
\begin{equation}
\frac{m}{{\rm eV}/c^2}=2.27\times 10^4 \left (\frac{\rm pc}{R}\right
)^{3/8}\left (\frac{M_{\odot}}{M}\right )^{1/8}.
\label{fdm8}
\end{equation}
Using the reference values of
$M$ and $R$ corresponding to Fornax [see Eq. (\ref{fornax})], we
find a fermion mass (see
Appendix D of \cite{suarezchavanis3}):
\begin{equation}
m=170 \, {\rm eV}/c^2.
\label{fdm9}
\end{equation}

Alternatively, using the results of Appendix \ref{sec_p} and taking
$m=170\, {\rm eV}/c^2$ in the numerical applications, we find that the minimum
halo radius,
the minimum halo mass and the maximum central density are
\begin{equation}
\label{fpv2}
(r_{h})_{\rm min}=1.50\, \left (\frac{\hbar^6}{G^3m^8\Sigma_0}\right
)^{1/5}=570\, {\rm pc},
\end{equation}
\begin{equation}
\label{fpv3}
(M_{h})_{\rm min}=4.47\, \left (\frac{\hbar^{12}\Sigma_0^3}{G^6m^{16}}\right
)^{1/5}=9.09\times 10^7\, M_{\odot}.
\end{equation}
\begin{equation}
\label{fpv3b}
(\rho_{0})_{\rm max}=0.667\, \left (\frac{\Sigma_0
m^{4/3}G^{1/2}}{\hbar}\right
)^{6/5}=0.248\, M_{\odot}/{\rm pc}^3,
\end{equation}
where $\Sigma_0$ is the universal density of DM halos given by Eq. (\ref{p5}).
These values can be improved if we have a more reliable expression of the
fermion mass $m$, but they are sufficient for our purposes (the same comment
applies to the bosonic models considered below).

The maximum mass of a fermion star at $T=0$ set by general
relativity is $M_{\rm
max}=0.384\, (\hbar
c/G)^{3/2}/m^2$ and its minimum radius is $R_{\rm
min}=8.73 \,
GM_{\rm max}/c^2$ \cite{ov}. They can be written as
\begin{equation}
\frac{M_{\rm max}}{M_{\odot}}=6.26\times 10^{17}\left (\frac{{\rm
eV}/c^2}{m}\right
)^2,\quad \frac{R_{\rm min}}{{\rm km}}=12.9 \frac{M_{\rm max}}{M_{\odot}}.
\label{fdm10}
\end{equation}
For a fermion of mass $m=170\, {\rm eV}/c^2$, we obtain $M_{\rm max}=2.17\times
10^{13}\,
M_{\odot}$
and $R_{\rm min}=8.85\, {\rm pc}$.  The maximum mass is much larger than the
typical core mass of a DM halo. Assuming that a fermion star at $T=0$
describes the quantum core of a DM halo, we conclude that such cores are
nonrelativistic since $M_c\ll M_{\rm max}$ in general. Since the maximum mass
is much larger than the core mass, gravity can
be treated within a Newtonian framework.

\subsection{Noninteracting bosonic DM}
\label{sec_ni}

\begin{figure}[!h]
\begin{center}
\includegraphics[clip,scale=0.3]{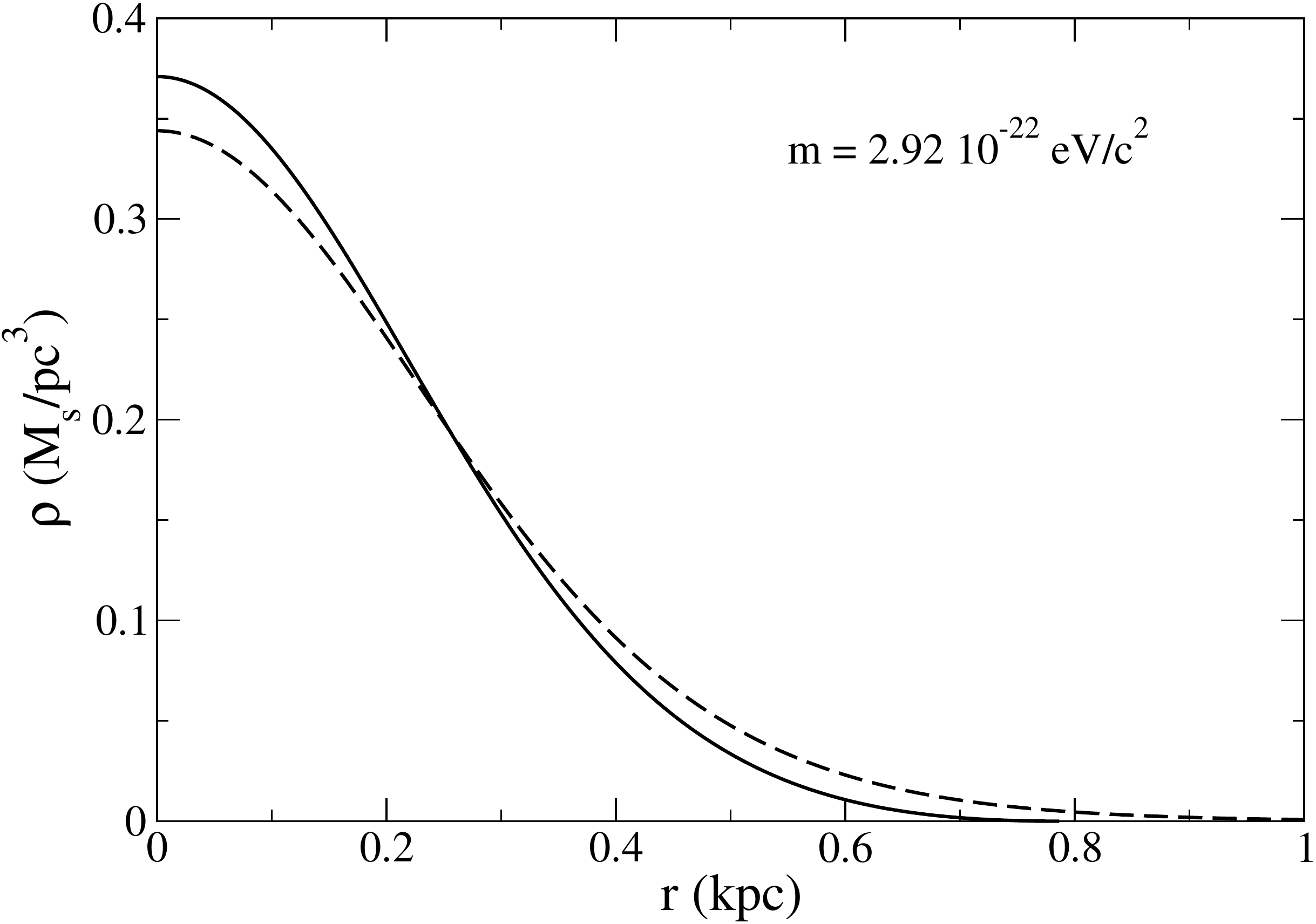}
\caption{Density profile of the ``minimum halo'' (ground state) of surface
density $\Sigma_0=\rho_0 r_h=141\, M_{\odot}/{\rm
pc}^2$ made of
noninteracting bosons of mass $m=2.92\times 10^{-22}\, {\rm eV}/c^2$. We have
plotted the polytropic profile with a compact support obtained in
Appendix \ref{sec_comp} (solid line) and the more conventional profile
computed in
\cite{rb,membrado,gul0,gul,prd2,ch2,ch3,pop,hui} for which $(r_{h})_{\rm
min}=410\, {\rm pc}$, $(M_{h})_{\rm min}=4.52\times 10^7\,
M_{\odot}$ and $(\rho_0)_{\rm max}=0.344\, M_{\odot}/{\rm
pc}^3$ (dashed line taken from \cite{prd2,modeldm}). }
\label{rhobosonNI}
\end{center}
\end{figure}

We consider a gas of noninteracting bosons at $T=0$ forming a BEC. The
wavefunction of a self-gravitating BEC without self-interaction is governed by
the Schr\"odinger-Poisson equation \cite{prd1}. Using Madelung's hydrodynamic
representation of the Schr\"odinger equation \cite{madelung}, we find that the
fundamental
differential equation of quantum hydrostatic equilibrium
determining the density profile of the BEC
is \cite{prd1}
\begin{equation}
\label{ni1}
\frac{\hbar^2}{2m^2}\Delta
\left (\frac{\Delta\sqrt{\rho}}{\sqrt{\rho}}\right )=4\pi G\rho.
\end{equation}
This  equation can be solved numerically 
\cite{rb,membrado,gul0,gul,prd2,ch2,ch3,pop,hui}. The density profile of a
noninteracting BEC star at
$T=0$ (ground state) extends to infinity (see Fig. \ref{rhobosonNI}) and the BEC
star is stable. The
mass-radius relation
is
\begin{equation}
M_c (R_c)_{99}=9.95\, \frac{\hbar^2}{G m^2},
\label{ni2}
\end{equation}
where $(R_c)_{99}$ is the radius enclosing $99\%$ of the mass. From
Eqs. (\ref{vt1}) and (\ref{vt6}) which reduce to
\begin{eqnarray}
\label{ni3}
E_c=\Theta_Q^c+W_c,
\end{eqnarray}
\begin{eqnarray}
\label{ni4}
2\Theta_Q^c+W_c=0,
\end{eqnarray}
we obtain
\begin{eqnarray}
\label{ni5}
E_c=-\Theta_Q^c=\frac{1}{2}W_c.
\end{eqnarray}
From numerical computations \cite{rb,membrado,prd2}, we find that the total
energy of a
noninteracting BEC star at
$T=0$ (ground state) is
\begin{eqnarray}
\label{ni6}
E_c=-0.0543\, \frac{G^2M_c^3m^2}{\hbar^2}.
\end{eqnarray}
Actually, we find in Appendix \ref{sec_comp} that there exists
another solution of Eq. (\ref{ni1}). It has a compact support (see Fig.
\ref{rhobosonNI}) and its profile
corresponds to a polytrope of index $\gamma=3/2$ (i.e. $n=2$). Its energy is
lower than the
energy  of the solution considered here,  suggesting
that it is more stable, even if comparing the energies of stable states may not
be decisive in view of the very long lifetime of metastable states in systems
with long-range interactions. In the following, in order to develop a unified
description of fermions and bosons based
on polytropic equations of state, we will use the solution from
Appendix  \ref{sec_comp}. However, as far as scalings and orders of magnitude
are concerned, we would get similar results by using the more conventional (but
maybe less stable) solution of this section.

Let us assume that the smallest DM halo that we know, with mass $M$ and
radius $R$,
corresponds to the ground state of a  noninteracting BEC star. From
the 
mass-radius relation (\ref{ni2}), we get
\begin{equation}
\frac{m}{{\rm eV}/c^2}=9.22\times 10^{-17} \left (\frac{\rm pc}{R}\right
)^{1/2}\left (\frac{M_{\odot}}{M}\right )^{1/2}.
\label{ni7}
\end{equation}
Using the reference values of
$M$ and $R$ corresponding to Fornax [see Eq. (\ref{fornax})], we
find a boson mass (see
Appendix D of \cite{suarezchavanis3}):
\begin{equation}
m=2.92\times 10^{-22}\, {\rm eV}/c^2. 
\label{ni8}
\end{equation}

Alternatively, using the results of Appendix \ref{sec_p} and taking
$m=2.92\times 10^{-22}\, {\rm eV}/c^2$ in the numerical applications, we find
that the minimum
halo radius,
the minimum halo mass and the maximum central density are (computed from the
polytropic solution of Appendix \ref{sec_comp})
\begin{equation}
\label{nipv2}
(r_{h})_{\rm min}=0.913\, \left (\frac{\hbar^2}{Gm^2\Sigma_0}\right
)^{1/3}=378\, {\rm pc},
\end{equation}
\begin{equation}
\label{nipv3}
(M_{h})_{\rm min}=1.61\, \left (\frac{\hbar^4\Sigma_0}{G^2m^4}\right
)^{1/3}=3.89\times 10^7\, M_{\odot}.
\end{equation}
\begin{equation}
\label{nipv3b}
(\rho_0)_{\rm max}=1.09\,
\frac{G^{1/3}m^{2/3}\Sigma_0^{4/3}}{\hbar^{2/3}}=0.371\, M_{\odot}/{\rm
pc}^3,
\end{equation}
where $\Sigma_0$ is the universal density of DM halos given by Eq. (\ref{p5}).

The maximum mass of a noninteracting boson star at $T=0$ set by
general relativity
is $M_{\rm max}=0.633\, \hbar c/Gm$ and its minimum radius is $R_{\rm min}=9.53
\, GM_{\rm max}/c^2$ \cite{kaup,rb}. They can be written as
\begin{equation}
\frac{M_{\rm max}}{M_{\odot}}=8.48\times 10^{-11}\frac{{\rm
eV}/c^2}{m},\quad \frac{R_{\rm min}}{{\rm km}}=14.1
\frac{M_{\rm max}}{M_{\odot}}.
\label{ni9}
\end{equation}
For a boson of mass $m=2.92\times
10^{-22}\, {\rm eV}/c^2$, we obtain $M_{\rm max}=2.90\times 10^{11}\, M_{\odot}$
and $R_{\rm min}=0.133\, {\rm pc}$. The maximum mass is much larger than the
typical core mass of a DM halo. Assuming that a BEC star at $T=0$
(soliton) describes the quantum core of a DM halo, we conclude that
such cores
are nonrelativistic since $M_c\ll M_{\rm max}$ in general. Since the maximum
mass
is much larger than the core mass, gravity can
be treated within a Newtonian framework.

\subsection{Bosonic DM with a repulsive self-interaction in the TF limit}
\label{sec_tf}

We consider a gas of self-interacting bosons at $T=0$ forming a BEC.
The wavefunction of a self-gravitating BEC with a quartic self-interaction is
governed
by the GPP equations \cite{prd1}. 
The equation of state of a self-interacting
BEC is \cite{prd1}
\begin{eqnarray}
\label{tf1}
P=\frac{2\pi a_s\hbar^2}{m^3}\rho^{2},
\end{eqnarray}
where $a_s$ is the scattering length of the bosons. This is a
polytropic
equation of state of index $\gamma=2$ (i.e. $n=1$) and
polytropic constant
\begin{eqnarray}
\label{tf2}
K=\frac{2\pi a_s\hbar^2}{m^3}.
\end{eqnarray}
We assume that the self-interaction is
repulsive ($a_s>0$). Using
Madelung's
hydrodynamic
representation of the GPP equations, and taking the quantum potential into
account, we find that the
fundamental
differential equation of quantum hydrostatic equilibrium
determining the density profile of the BEC 
is \cite{prd1} 
\begin{equation}
\label{att1}
-\frac{\hbar^2}{2m^2}\Delta
\left (\frac{\Delta\sqrt{\rho}}{\sqrt{\rho}}\right )+\frac{4\pi
a_s\hbar^2}{m^3}\Delta\rho=-4\pi G\rho.
\end{equation}
This  equation can be solved numerically 
\cite{prd2}. The density profile of a noninteracting BEC star at
$T=0$ (ground state) extends to infinity. The 
mass-radius relation has been obtained in \cite{prd1,prd2}.

\begin{figure}[!h]
\begin{center}
\includegraphics[clip,scale=0.3]{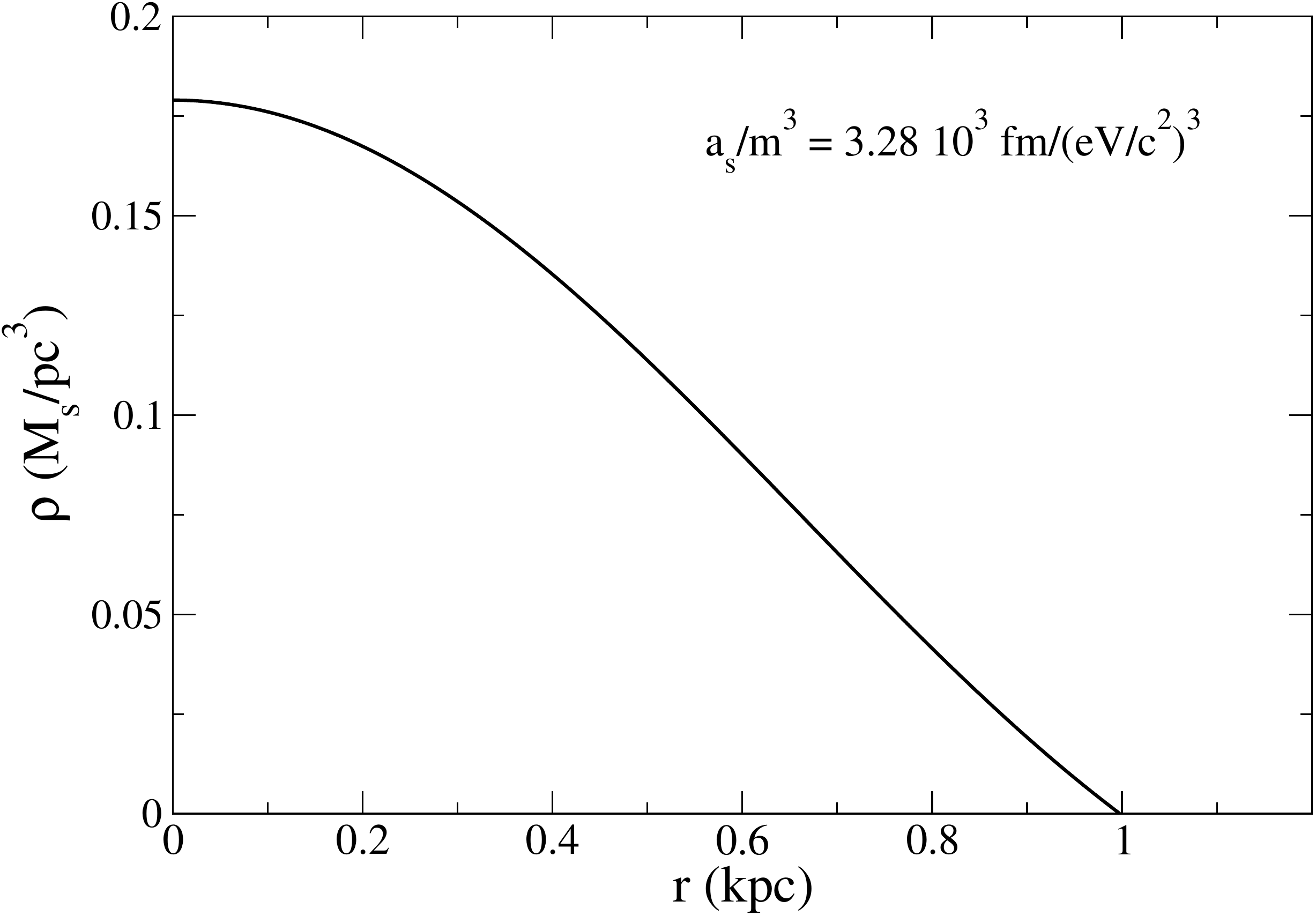}
\caption{Density profile $\rho=\rho_0 \sin(\pi r/R_c)/(\pi
r/R_c)$ of the ``minimum halo'' (ground state) of surface
density $\Sigma_0=\rho_0 r_h=141\, M_{\odot}/{\rm
pc}^2$ made of
self-interacting bosons in the TF limit with $a_s/m^3=3.28\times 10^3\, {\rm
fm\, (eV/c^2)^{-3}}$.}
\label{rhobosonTF}
\end{center}
\end{figure}

In the TF approximation, which amounts to neglecting the
quantum potential, the fundamental differential equation of hydrostatic
equilibrium
determining the density profile of a self-interacting BEC star at $T=0$ 
with the equation of
state (\ref{tf1}) writes  (see Appendix \ref{sec_gl}) 
\begin{equation}
\label{tf2b}
\frac{4\pi a_s\hbar^2}{m^3}\Delta\rho=-4\pi G\rho.
\end{equation}
It can be reduced to the Lane-Emden equation (\ref{gl7}) of
index $n=1$ which has a simple analytical solution \cite{chandrabook}. This
profile has a compact
support (see Fig. \ref{rhobosonTF}) and the  self-interacting BEC star is
stable. A self-gravitating BEC 
with a
repulsive self-interaction in the TF approximation has a unique
radius \cite{tkachev,maps,leekoh,goodman,arbey,bohmer,prd1},
\begin{eqnarray}
R_c=\pi\left (\frac{a_s\hbar^2}{Gm^3}\right )^{1/2},
\label{tf3}
\end{eqnarray}
that is independent of its mass. From Eqs. (\ref{vt7}) and
(\ref{vt9}) which reduce to
\begin{eqnarray}
\label{tf4}
E_c=U_c+W_c,
\end{eqnarray}
\begin{eqnarray}
\label{tf5}
3U_c+W_c=0,
\end{eqnarray}
and from the Betti-Ritter formula (\ref{br5}), we obtain
\begin{eqnarray}
\label{tf6}
E_c=-2U_c=\frac{2}{3}W_c=-\frac{1}{2}\frac{GM_c^2}{R_c}.
\end{eqnarray}
Combined with Eq. (\ref{tf3}), we find that the energy of a
self-interacting BEC
star at $T=0$ (ground state) in the TF limit  is 
\begin{eqnarray}
\label{tf7}
E_c=-\frac{1}{2\pi}\frac{G^{3/2}m^{3/2}M_c^{2}}{a_s^{1/2}\hbar}.
\end{eqnarray}

Let us assume that the smallest DM halo that we know, with mass $M$ and
radius $R$,
corresponds to the ground state of a self-interacting BEC star. From Eq.
(\ref{tf3}), we get 
\begin{equation}
\frac{a_s}{\rm fm}\left (\frac{{\rm eV}/c^2}{m}\right
)^3=3.28\times 10^{-3} \left (\frac{R}{\rm pc}\right )^{2}.
\label{tf8}
\end{equation}
This formula depends only on $R$. Using the reference value of $R$
corresponding to Fornax [see Eq. (\ref{fornax})], we
find that the ratio $a_s/m^3$ of the boson parameters $a_s$ and
$m$ is given by (see
Appendix D of \cite{suarezchavanis3}):
\begin{eqnarray}
\frac{a_s}{\rm fm}\left (\frac{{\rm eV}/c^2}{m}\right )^3=3.28\times 10^3.
\label{tf9}
\end{eqnarray}
In order to determine the mass of the boson, we need another
relation. In this respect, we note that $m$ and $a_s$ must
satisfy the constraint $\sigma/m<1.25\, {\rm cm}^2/{\rm g}$ set by the
Bullet Cluster \cite{bullet}, where $\sigma=4\pi a_s^2$ is the self-interaction
cross section. In  Appendix D of \cite{suarezchavanis3} we have 
considered two extreme cases corresponding
to
an upper boson mass 
\begin{equation}
m=1.10\times 10^{-3}\, {\rm eV}/c^2,\qquad  a_s=4.41\times
10^{-6}\, {\rm fm},
\label{tf10}
\end{equation}
and a lower boson mass
\begin{equation}
m=2.92\times 10^{-22}\, {\rm eV}/c^2,\quad a_s=8.13\times 10^{-62}\, {\rm fm}.
\label{tf11}
\end{equation}
The upper boson mass (\ref{tf10}) corresponds to the bound
$\sigma/m=1.25\, {\rm cm}^2/{\rm g}$. The lower boson mass (\ref{tf11})
corresponds to the transition between the TF regime and the noninteracting
regime.
We note that when a self-interaction between the bosons is allowed, a large
mass window is open. In particular, a repulsive self-interaction allows one to
have a larger boson mass than in the noninteracting case. As discussed in
Appendix D.4 of \cite{suarezchavanis3} this may be
interesting in view of the fact that the mass of a noninteracting boson ($m\sim
1-10\times 10^{-22}\, {\rm eV}/c^2$) is in
tension with observations of the Lyman-$\alpha$ forest \cite{hui}. {\it This
tension could reflect the fact that bosons have a repulsive
self-interaction.}

Alternatively, using the results of Appendix \ref{sec_p} and taking
$a_s/m^3=3.28\times 10^3\, {\rm fm\, (eV/c^2)^{-3}}$ in the numerical
applications, we find
that the minimum
halo radius,
the minimum halo mass and the maximum central density are
\begin{equation}
\label{sipv2}
(r_{h})_{\rm min}=2.47\, \, \left (\frac{a_s\hbar^2}{Gm^3}\right
)^{1/2}=786\, {\rm pc},
\end{equation}
\begin{equation}
\label{sipv3}
(M_{h})_{\rm min}=13.0\, \frac{a_s\hbar^2\Sigma_0}{G m^3}=1.86\times 10^8\,
M_{\odot}.
\end{equation}
\begin{equation}
\label{sad1}
(\rho_{0})_{\rm max}=0.404\,  \left (\frac{Gm^3\Sigma_0^2}{a_s\hbar^2}\right
)^{1/2}=0.179\, M_{\odot}/{\rm
pc}^3,
\end{equation}
where $\Sigma_0$ is the universal density of DM halos given by Eq. (\ref{p5}).

The maximum mass of a self-interacting boson star set by general relativity is
$M_{\rm max}=0.307\,
\hbar c^2\sqrt{a_s}/(Gm)^{3/2}$ and its minimum radius is
$R_{\rm min}=6.25 \, GM_{\rm max}/c^2$ \cite{colpi,tkachev,chavharko}. They can
be written as
\begin{equation}
\frac{M_{\rm max}}{M_{\odot}}=1.12\, \left (\frac{a_s}{{\rm fm}}\right
)^{1/2}\left (\frac{{\rm
GeV}/c^2}{m}\right )^{3/2},
\label{tf12}
\end{equation}
\begin{equation}
\frac{R_{\rm min}}{{\rm km}}=9.27
\frac{M_{\rm max}}{M_{\odot}}.
\label{tf13}
\end{equation}
We note that these results do not depend 
on the specific mass $m$ and scattering length $a_s$ of the bosons, but only on
the ratio $a_s/m^3$. For a ratio $(a_s/{\rm
fm})({\rm
eV}/mc^2)^3=3.28\times 10^3$, we obtain $M_{\rm max}=2.03\times
10^{15}\, M_{\odot}$ and $R_{\rm min}=609\, {\rm pc}$. The maximum mass is much
larger
than the
typical core mass of a DM halo. Assuming that a self-interacting BEC
star at $T=0$ describes the quantum core of a DM halo, we conclude that
such cores
are nonrelativistic since $M_c\ll M_{\rm max}$ in general. Since the maximum
mass
is much larger than the core mass, gravity can
be treated within a Newtonian framework.

\subsection{Bosonic DM with an attractive self-interaction}
\label{sec_att}

We consider a gas of self-interacting bosons at $T=0$ forming a
BEC.
The wavefunction of a self-gravitating BEC with a quartic self-interaction is
governed
by the GPP equations \cite{prd1}.
The equation of state of a self-interacting
BEC is given by Eq. (\ref{tf1}).
We assume that the self-interaction is
attractive ($a_s<0$). This is the case for the axion \cite{marshrevue}. Using
Madelung's
hydrodynamic
representation of the GPP equations, and taking the quantum potential into
account, we find that the
fundamental
differential equation of quantum hydrostatic equilibrium
determining the density profile of the BEC 
is given by Eq. (\ref{att1}) \cite{prd1}. 
This  equation can be solved numerically 
\cite{prd2}. The density profile of a noninteracting BEC star at
$T=0$ (ground state) extends to infinity. The 
mass-radius relation has been obtained in \cite{prd1,prd2}. There is a maximum
mass \cite{prd1}
\begin{eqnarray}
\label{intro4}
M_{\rm max}^{\rm exact}=1.012\frac{\hbar}{\sqrt{Gm|a_s|}}
\end{eqnarray}
corresponding to a minimum stable radius 
\begin{eqnarray}
\label{intro5}
(R_{99}^*)^{\rm exact}=5.5\left
(\frac{|a_s|\hbar^2}{Gm^3}\right )^{1/2}.
\end{eqnarray}
When $M_c>M_{\rm max}$ the axion star is expected to collapse and form a dense
axion star, a black hole or a bosenova as discussed
in \cite{braaten,cotner,bectcoll,ebycollapse,tkachevprl,helfer,phi6,visinelli,
moss}.

\subsection{Mass-radius relation of isothermal DM halos}
\label{sec_iso}

In the previous subsections, we have focused on the quantum core of DM halos.
Ultracompact dwarf DM halos  just have a quantum core (ground state). Larger
DM halos have a ``core-halo'' structure with a quantum core surrounded by an
atmosphere. We have seen that the structure of the quantum core strongly depends
on the nature of the DM particle. By contrast, the structure of the atmosphere
is
relatively independent of the DM particle. We assume that it has an
isothermal equation of state 
\begin{equation}
P=\rho \frac{k_B T}{m}
\label{ih1}
\end{equation}
with an effective temperature $T$.
For sufficiently
large DM halos, the isothermal atmosphere dominates the core. Indeed,
the DM halo mass $M_h$ is much larger than the core mass $M_c$ and  it
is a good
approximation to assume that the DM halo is purely
isothermal.\footnote{We have in mind, for example,  the Medium Spiral ($R\sim
10\, {\rm kpc}$ and $M\sim 10^{11}\, M_\odot$).} Therefore, from the
``outside'', large DM halos behave as classical isothermal spheres. If
$M_h$ represents the halo mass and $r_h$ the halo radius as defined in
Appendix \ref{sec_p}, then the mass-radius and temperature-radius relations of
isothermal DM
halos are \cite{modeldm}
\begin{equation}
\label{iso1}
M_h=1.76\, \Sigma_0 r_h^2,\qquad 
\frac{k_B T}{m}=0.954\, G\Sigma_0r_h,
\end{equation}
where $\Sigma_0$ is the universal surface density of DM
halos given by Eq. (\ref{p5}). On the other hand, the circular velocity at the
halo radius is
\begin{equation}
\label{iso1b}
v_h^2=\frac{GM_h}{r_h}=1.76\, \Sigma_0 G  r_h=1.33\, G \Sigma_0^{1/2}M_h^{1/2}.
\end{equation}

\section{A generalized wave equation}
\label{sec_gwe}

\subsection{Coarse-grained dynamics}

The previous results can be obtained in a unified manner from the generalized
GPP equations \cite{ggp,modeldm,nottalechaos}
\begin{eqnarray}
\label{gw1}
i\hbar \frac{\partial\psi}{\partial
t}=-\frac{\hbar^2}{2m}\Delta\psi+m\Phi\psi+\frac{K\gamma
m}{\gamma-1}|\psi|^{2(\gamma-1)}\psi\nonumber\\
+2k_B
T\ln|\psi|\psi
-i\frac{\hbar}{2}\xi\left\lbrack \ln\left (\frac{\psi}{\psi^*}\right
)-\left\langle \ln\left (\frac{\psi}{\psi^*}\right
)\right\rangle\right\rbrack\psi,
\end{eqnarray}
\begin{equation}
\label{gw2}
\Delta\Phi=4\pi G |\psi|^2.
\end{equation}
As discussed in
more detail in \cite{ggp,modeldm}, the thermal ($T$) and dissipative
($\xi$) terms present in the generalized GPP equations (\ref{gw1}) and
(\ref{gw2})
parametrize
the complicated processes of violent relaxation \cite{lb} and gravitational
cooling \cite{seidel94} experienced by a collisionless system of
self-gravitating
fermions or bosons. As a result, the generalized GPP equations (\ref{gw1}) and
(\ref{gw2}) describe the evolution of the system on a ``coarse-grained'' scale. 

\subsection{Madelung transformation }

Making the Madelung transformation 
\begin{equation}
\label{mad1}
\psi({\bf r},t)=\sqrt{{\rho({\bf r},t)}} e^{iS({\bf r},t)/\hbar},
\end{equation}
\begin{equation}
\label{mad1b}
\rho=|\psi|^2,\qquad {\bf u}=\frac{\nabla S}{m},
\end{equation}
we can show that the generalized GPP equations (\ref{gw1}) and
(\ref{gw2}) are equivalent to the fluid
equations
\begin{equation}
\label{mad2}
\frac{\partial\rho}{\partial t}+\nabla\cdot (\rho {\bf u})=0,
\end{equation}
\begin{equation}
\label{mad3}
\frac{\partial {\bf u}}{\partial t}+({\bf u}\cdot \nabla){\bf
u}=-\frac{1}{\rho}\nabla P-\nabla\Phi-\frac{1}{m}\nabla
Q-\xi{\bf u},
\end{equation}
\begin{equation}
\label{mad4}
\Delta\Phi=4\pi G\rho,
\end{equation}
where
\begin{equation}
\label{mad5}
Q=-\frac{\hbar^2}{2m}\frac{\Delta
\sqrt{\rho}}{\sqrt{\rho}}
\end{equation}
is the quantum potential and $P$ is the pressure determined by the equation of
state 
\begin{equation}
\label{mad6}
P=K\rho^{\gamma}+\rho \frac{k_B T}{m}\qquad (\gamma=1+1/n).
\end{equation}
This equation of state has a linear part and a polytropic part. The linear
(isothermal) equation of state accounts for effective thermal effects. 
The polytropic equation of state takes into
account the self-interaction of the
bosons or the quantum pressure arising from the Pauli exclusion principle for
fermions. The equation of state (\ref{mad6}) defines a composite model of DM
halos
with a core-halo structure. The polytropic equation of state
dominates in the core where the density is high and the isothermal equation of
state dominates in the halo where the density is low (we assume that
$\gamma>1$). As a result, the
corresponding DM
halos present a quantum (fermionic/bosonic) core surrounded by an
isothermal envelope. The quantum core solves the cusp problem and the
isothermal envelope leads to flat rotation curves. This model has been studied
in detail in \cite{modeldm} for self-interacting BECs. Its extension to
noninteracting BECs and fermions will be presented in a forthcoming paper.

\subsection{Condition of quantum hydrostatic
equilibrium}

The equilibrium state of the hydrodynamic equations (\ref{mad2}) and 
(\ref{mad3}) satisfies the condition
of quantum hydrostatic equilibrium \cite{ggp}
\begin{eqnarray}
\label{ch1}
\frac{\rho}{m}\nabla Q+\nabla P+\rho\nabla\Phi={\bf 0}.
\end{eqnarray}
It
describes the balance
between the quantum potential arising from the Heisenberg uncertainty
principle,  the quantum pressure (due to the Pauli exclusion principle for
fermions
or due to the self-interaction of the bosons), the pressure due to the
effective temperature, and the
gravitational attraction. Combining Eq.
(\ref{ch1}) with the Poisson equation (\ref{mad4}), we obtain the fundamental
differential equation of quantum hydrostatic equilibrium \cite{ggp}
\begin{equation}
\label{ch2}
\frac{\hbar^2}{2m^2}\Delta
\left (\frac{\Delta\sqrt{\rho}}{\sqrt{\rho}}\right )-\nabla\cdot \left
(\frac{\nabla P}{\rho}\right )=4\pi G\rho.
\end{equation}
For the
equation of state (\ref{mad6}),  it takes the form
 \begin{equation}
\label{ch3}
\frac{\hbar^2}{
2m^2}\Delta
\left (\frac{\Delta\sqrt{\rho}}{\sqrt{\rho}}\right
)-\frac{K\gamma}{\gamma-1}\Delta\rho^{\gamma-1}-\frac{k_B
T}{m}\Delta\ln\rho=4\pi
G\rho.
\end{equation}
This differential equation determines the general equilibrium density profile
$\rho({\bf r})$ of a quantum DM halo in our model \cite{ggp,modeldm}. This
profile generically has
a
core-halo
structure with a polytropic core and an isothermal halo.

In the core, the differential equation (\ref{ch3}) reduces to
\begin{equation}
\label{ch4}
\frac{\hbar^2}{
2m^2}\Delta
\left (\frac{\Delta\sqrt{\rho}}{\sqrt{\rho}}\right
)-\frac{K\gamma}{\gamma-1}\Delta\rho^{\gamma-1}=4\pi
G\rho.
\end{equation}
It determines the structure of the quantum core as described in
Secs. \ref{sec_fdm}-\ref{sec_att}.

In the halo, the differential equation reduces to
 \begin{equation}
\label{ch5}
-\frac{k_B
T}{m}\Delta\ln\rho=4\pi
G\rho.
\end{equation}
It is equivalent to the Emden equation \cite{chandrabook}. It determines the
structure of the isothermal atmosphere of large DM
halos as described in Sec. \ref{sec_iso}.

\subsection{Free energy}

The free energy associated with the generalized GPP equations
(\ref{gw1}) and
(\ref{gw2}) or equivalently with the
hydrodynamic equations (\ref{mad2}) and 
(\ref{mad3}) is
\begin{eqnarray}
\label{ch6}
F=E_*-TS=\Theta_c+\Theta_Q+U+W-TS.
\end{eqnarray}
The energy $E_*$ is the sum of the classical kinetic energy
\begin{eqnarray}
\label{ch7}
\Theta_c=\int \rho\frac{{\bf u}^2}{2}\, d{\bf r},
\end{eqnarray}
the quantum kinetic energy 
\begin{eqnarray}
\label{ch8}
\Theta_Q=\frac{\hbar^2}{8m^2}\int
\frac{(\nabla\rho)^2}{\rho}\, d{\bf
r},
\end{eqnarray}
the internal energy associated with the polytropic equation of
state
\begin{eqnarray}
\label{ch9}
U=\frac{K}{\gamma-1}\int \rho^{\gamma}\, d{\bf r},
\end{eqnarray}
and the gravitational energy
\begin{eqnarray}
\label{ch10}
W=\frac{1}{2}\int \rho \Phi\, d{\bf r}.
\end{eqnarray}
On the other hand,
\begin{eqnarray}
\label{ch11}
S=-k_B
\int\frac{\rho}{m}(\ln\rho-1)\, d{\bf r}
\end{eqnarray}
is the Boltzmann entropy associated with
the isothermal equation of state.

The generalized GPP equations
(\ref{gw1}) and
(\ref{gw2}) or equivalently the
hydrodynamic equations (\ref{mad2}) and 
(\ref{mad3}) satisfy an
$H$-theorem \cite{ggp}
\begin{eqnarray}
\label{ch12}
\dot F=-\xi\int \rho {\bf u}^2\, d{\bf r}=-2\xi\Theta_c\le 0.
\end{eqnarray}
The free energy decreases monotonically when $\xi>0$ (or is constant when
$\xi=0$). At equilibrium, we have $\dot F=0$ implying
${\bf
u}={\bf 0}$. Then, Eq. (\ref{mad3}) leads to the condition of quantum
hydrostatic equilibrium (\ref{ch1}). When $\xi>0$, using Lyapunov's direct
method, one can show that the system relaxes, for $t\rightarrow +\infty$,
towards a stable equilibrium state which is a (local) minimum of free energy at
fixed mass. 

The extremization of the free energy at fixed mass, corresponding
to the variational principle $\delta F-\frac{\mu}{m}\delta M=0$ where $\mu$ is a
Lagrange
multiplier, returns the condition of quantum hydrostatic equilibrium
(\ref{ch1}). Furthermore, (local) minima of free energy are stable while maxima
or saddle points are unstable.

The generalized GPP equations (\ref{gw1})  and (\ref{gw2}) are associated with a
canonical description in which the temperature $T$ is fixed. It is possible to
modify these equations so that the temperature $T(t)$ evolves in time in order 
to conserve the energy $E$ (see Appendix I of \cite{ggp}). This corresponds to
a 
microcanonical description. As is well-known \cite{ijmpb}, the equilibrium
states are the
same in the microcanonical and canonical ensembles. However, their stability
may be different in case of ensembles inequivalence. In particular,
equilibrium states that are unstable in the canonical ensemble may be stable
in the microcanonical (this is because the microcanonical ensemble is more
constrained than the canonical ensemble). For example, equilibrium states with a
negative specific heat are always unstable in the canonical ensemble while they
may be stable in the microcanonical ensemble. This property will play a
fundamental role in the following analysis.

\section{Analytical model of DM
halos with a
polytropic core and an isothermal atmosphere}
\label{sec_ana}

In this section, we develop an approximate analytical model of DM
halos with a quantum core surrounded by an isothermal atmosphere. For
simplicity, we assume that the density of the isothermal atmosphere is
uniform. In all the DM models discussed in Sec. \ref{sec_qm} the quantum core
can be described by a
polytropic equation of state. Therefore, by considering a  polytropic core with
an arbitrary index $n$, we can account for a wide diversity of situations and,
in particular, unify the treatment of fermionic and bosonic DM halos.
We shall enclose the system within a box of radius $R$. The box is necessary to
have a finite mass $M$. In order to connect this model with real DM halos, we
shall identify the box radius $R$ with the halo radius $r_h$ and the mass $M$
with the halo mass $M_h$. Therefore, we set 
\begin{equation}
\label{def}
M=M_h,\qquad R=r_h.
\end{equation}
The mass and the radius of sufficiently large DM halos are related to each
other by the first relation of Eq. (\ref{iso1}).

\subsection{Polytropic core}
\label{sec_pc}

We modelize the core of a DM halo by a pure polytrope of index $n$. Its mass
$M_c$ and its
radius $R_c$ satisfy the mass-radius relation (see Appendix \ref{sec_gl})
\cite{chandrabook}
\begin{equation}
\label{pc1}
M_c^{(n-1)/n}R_c^{(3-n)/n}=\frac{K(n+1)}{G(4\pi)^{1/n}}\omega_n^{(n-1)/n}.
\end{equation}
The internal energy and the gravitational
energy of
the polytropic core are given by (see Appendices \ref{sec_vt} and \ref{sec_br})
\begin{equation}
\label{pc2}
U_c=\frac{n}{5-n}\frac{GM_c^2}{R_c},\qquad
W_c=-\frac{3}{5-n}\frac{GM_c^2}{R_c}.
\end{equation}
Therefore, its total energy $E_c=U_c+W_c$ is
\begin{eqnarray}
\label{pc3}
E_c=-\frac{3-n}{n}U_c=\frac{3-n}{3}W_c=-\frac{3-n}{5-n}\frac{GM_c^2}{R_c
}.
\end{eqnarray}
Combined with Eq. (\ref{pc1}), we obtain
\begin{eqnarray}
\label{pc4}
E_c=-\frac{3-n}{5-n}\left\lbrack
\frac{(4\pi)^{1/n}}{n+1}\right\rbrack^{n/(3-n)}\frac{1}{\omega_n^{
(n-1)/(3-n)}}\nonumber\\
\times\frac{G^ {3/(3-n)} M_c^{(5-n)/(3-n)}}{K^{n/(3-n)}}.
\end{eqnarray}

\subsection{Isothermal atmosphere of uniform density}
\label{sec_ih}

We modelize the halo by an isothermal atmosphere of mass
$M_a=M-M_c$ contained between the spheres of radius
$R_c$ and $R$. The internal energy of a gas with the isothermal equation of 
state (\ref{ih1}) is \cite{ggp} 
\begin{eqnarray}
\label{ih2b}
U&=&\frac{k_B T}{m} \int \rho\ln\rho\, d{\bf r}-\frac{3}{2}Nk_B T
\ln\left
(\frac{2\pi
k_B T}{m}\right )\nonumber\\
&-&Nk_B T- Nk_{B}T\ln \left (\frac{2m^4}{h^3}\right ).
\end{eqnarray}
It can be rewritten as
\begin{eqnarray}
\label{}
U=k_B T\int \frac{\rho}{m} \left \lbrack \ln\left (\frac{\lambda^3}{2}
\frac{\rho}{m}\right )-1\right
\rbrack\, d{\bf r},
\end{eqnarray}
where $\lambda=h/(2\pi m k_B T)^{1/2}$ is the de Broglie wavelength.
Treating the atmosphere as a gas with a uniform 
density, we obtain 
\begin{eqnarray}
\label{ih3}
U_a=\frac{k_B T}{m}(M-M_c)\Biggl\lbrack \ln\left (\frac{M-M_c}{V}\right
)\nonumber\\
-\frac{3}{2}\ln\left (\frac{2\pi k_B T}{m}\right )-1-\ln \left
(\frac{2m^4}{h^3}\right )\Biggr\rbrack,
\end{eqnarray}
where $V=(4/3)\pi R^3$ is the total volume of the system. On the other hand,
the gravitational energy of the uniform atmosphere in the presence of the
``external''
polytropic
core is given by (see Appendix \ref{sec_ge})
\begin{equation}
\label{ih4}
W_a=-\frac{3GM_c(M-M_c)}{2R}-\frac{3G(M-M_c)^2}{5R}.
\end{equation}
To obtain these results, we have assumed that $R_c\ll
R$ which is a very good approximation in all cases of physical interest.

\subsection{Free energy}
\label{sec_fe}

Using the foregoing results, the total free energy of the system
(core $+$ halo)
is 
\begin{eqnarray}
\label{ana4}
F&=&-\frac{3-n}{5-n}\frac{GM_c^2}{R_c}\nonumber\\
&+&\frac{k_B T}{m}(M-M_c)\Biggl\lbrack \ln\left (\frac{M-M_c}{V}\right
)\nonumber\\
&-&\frac{3}{2}\ln\left (\frac{2\pi k_B T}{m}\right )-1-\ln \left
(\frac{2m^4}{h^3}\right )\Biggr\rbrack,
\nonumber\\
&-&\frac{3GM_c(M-M_c)}{2R}-\frac{3G(M-M_c)^2}{5R}.
\end{eqnarray}
For a given value of $M$, $R$ and $T$, the free energy is a function $F(M_c)$ of
the core mass. The extrema of this function determine the possible 
equilibrium states of the system. More precisely, they determine the
possible equilibrium
core masses $M_c^{(i)}$ as a function of $M$, $R$ and $T$. We recall that the
equilibrium states are the same in the
canonical and in the microcanonical ensembles. Indeed, the extrema of
$F(M_c)$ at fixed mass coincide with the extrema of $S(M_c)$ at fixed
mass and energy. However, their stability may be different in the canonical and
in the microcanonical ensembles. This is the important notion of ensembles
inequivalence for systems with long-range interactions \cite{ijmpb}. In the
canonical
ensemble,
a minimum of
$F(M_c)$ at fixed mass corresponds to a stable equilibrium state (most probable
state) while
a maximum of $F(M_c)$ at fixed mass
corresponds to an unstable equilibrium state (less probable state). In the
microcanonical
ensemble,
a maximum of
$S(M_c)$ at fixed mass and energy corresponds to a stable equilibrium state
(most
probable state) while
a minimum of $S(M_c)$ at fixed mass and energy
corresponds to an unstable equilibrium state (less probable state).
We first consider the canonical ensemble. The
microcanonical ensemble is treated in Sec. \ref{sec_mce}.

It is convenient to introduce the dimensionless variables
\begin{eqnarray}
\label{ana5}
x=\frac{M_c}{M},\qquad \eta=\frac{\beta GMm}{R},
\end{eqnarray}
\begin{eqnarray}
\label{ana6}
f(x)=\frac{F(M_c)R}{GM^2},
\end{eqnarray}
\begin{eqnarray}
\label{nu}
\nu=\left\lbrack
\frac{G}{K(n+1)}\right\rbrack^{n/(3-n)}(4\pi)^{1/(3-n)}\frac{1}{\omega_n^{
(n-1)/(3-n)}}\nonumber\\
\times RM^{(n-1)/(3-n)},
\end{eqnarray}
\begin{eqnarray}
\label{defmu}
\mu=\frac{2m^4}{h^3}\sqrt{512\pi^4G^3MR^3},
\end{eqnarray}
and
\begin{eqnarray}
\label{defc}
C(\eta,\mu)=\frac{3}{2}\ln\eta-\ln\mu+\ln \left (\frac{6}{\sqrt{\pi}}\right ),
\end{eqnarray}
so that Eq. (\ref{ana4}) can be rewritten as
\begin{eqnarray}
\label{ana7}
f(x)&=&-\frac{3-n}{5-n}\nu
x^{(5-n)/(3-n)}\nonumber\\
&+&\frac{1}{\eta}(1-x)\left\lbrack
\ln(1-x)+C(\eta,\mu)-1\right\rbrack\nonumber\\
&-&\frac{3}{2}x(1-x)-\frac{3}{5}(1-x)^2
\end{eqnarray}
with $0\le x\le 1$. On the other hand, introducing
\begin{eqnarray}
\label{ana7b}
y=\frac{R_c}{R},
\end{eqnarray}
the core
mass-radius relation (\ref{pc1}) can be written as 
\begin{eqnarray}
\label{ana7c}
y\, x^{\frac{n-1}{3-n}}=\frac{1}{\nu}.
\end{eqnarray}
The condition that $y\le 1$ ($R_c\le R$) when $x=1$ ($M_c=M$) implies $\nu\ge
1$.

\subsection{Connection to DM halos}
\label{sec_c}

Before going further, let us connect the dimensionless variables introduced
previously to the parameters of the DM halos. The variable $x$ represents the
normalized core mass. Using Eq. (\ref{def}) it can be written as 
\begin{eqnarray}
\label{c1}
x=\frac{M_c}{M_h}.
\end{eqnarray}
The variable $\eta$ represents the normalized inverse temperature. For DM
halos, using Eqs. (\ref{iso1}) and (\ref{def}), we get
\begin{eqnarray}
\label{c2}
\eta=\frac{\beta GM_h m}{r_h}=1.84.
\end{eqnarray}
We see that the normalized inverse temperature is of order $1 - 2$. This is
essentially a consequence of the virial theorem. Since our approach is
approximate, we will allow $\eta$ to vary  slightly around this value.
To fix the ideas we take $\eta=1$. Finally,
the variable $\nu$ characterizes the mass $M_h$ of the DM halos. For fermionic
DM halos, we have\footnote{This parameter is related to the ``degeneracy
parameter'' $\mu$
introduced in
\cite{ijmpb} and given by Eq. (\ref{defmu}). We have
\begin{eqnarray}
\nu_{\rm F}=\left (\frac{4}{9\omega_{3/2}}\right
)^{1/3}\mu^{2/3}=0.149737\, \mu^{2/3}=\lambda \, \mu^{2/3}.
\label{foo}
\end{eqnarray}
} 
\begin{eqnarray}
\label{c3}
\nu_{\rm F}=8\left (\frac{4\pi^2}{3}\right
)^{2/3}\frac{1}{\omega_{3/2}^{1/3}}\frac{Gm^{8/3}}{h^2}RM^{1/3}.
\end{eqnarray}
Using Eqs. (\ref{iso1})
and (\ref{def}), we get
\begin{eqnarray}
\label{c4}
\nu_{\rm F}=8\left (\frac{4\pi^2}{3}\right
)^{2/3}\frac{1}{\omega_{3/2}^{1/3}}\frac{Gm^{8/3}}{h^2}\frac{1}{\sqrt{1.76\,
\Sigma_0}}M_h^{5/6}.
\end{eqnarray}
Normalizing the halo mass by the minimum halo mass from Eq. (\ref{fpv3}) we
obtain
\begin{eqnarray}
\label{c4b}
\nu_{\rm F}=0.582\, \left (\frac{M_h}{(M_h)_{\rm min}}\right )^{5/6}.
\end{eqnarray}
For noninteracting BECDM halos, we have
\begin{eqnarray}
\label{c5}
\nu_{\rm B}=\frac{2}{\omega_2}\frac{Gm^2}{\hbar^2}RM.
\end{eqnarray}
Using Eqs. (\ref{iso1}) and (\ref{def}), we
get
\begin{eqnarray}
\label{c6}
\nu_{\rm B}=\frac{2}{\omega_2}\frac{Gm^2}{\hbar^2}\frac{1}{\sqrt{1.76\,
\Sigma_0}} M_h^{3/2}.
\end{eqnarray}
Normalizing the halo mass by the minimum halo mass from Eq. (\ref{nipv3})
we
obtain
\begin{eqnarray}
\label{c6b}
\nu_{\rm B}=0.293\, \left (\frac{M_h}{(M_h)_{\rm min}}\right
)^{3/2}.
\end{eqnarray}
For self-interacting BECDM halos in the TF limit, we have\footnote{This
parameter is related to the parameter $\mu_{\rm TF}=Gm^3R^2/a_s\hbar^2$
introduced in
\cite{modeldm}. We have $\nu_{\rm TF}=\mu_{\rm TF}^{1/2}$.} 
\begin{eqnarray}
\label{c7}
\nu_{\rm TF}=\left (\frac{Gm^3}{a_s\hbar^2}\right )^{1/2}R.
\end{eqnarray}
Using Eqs. (\ref{iso1}) and
(\ref{def}), we
get
\begin{eqnarray}
\label{c8}
\nu_{\rm TF}=\left (\frac{Gm^3}{a_s\hbar^2}\right )^{1/2}\frac{1}{\sqrt{1.76\,
\Sigma_0}} M_h^{1/2}.
\end{eqnarray}
Normalizing the halo mass by the minimum halo mass from Eq. (\ref{sipv3})
we
obtain
\begin{eqnarray}
\label{c8b}
\nu_{\rm TF}=2.72\, \left (\frac{M_h}{(M_h)_{\rm min}}\right
)^{1/2}.
\end{eqnarray}

\subsection{Equilibrium states}
\label{sec_es}

\begin{figure}[!h]
\begin{center}
\includegraphics[clip,scale=0.3]{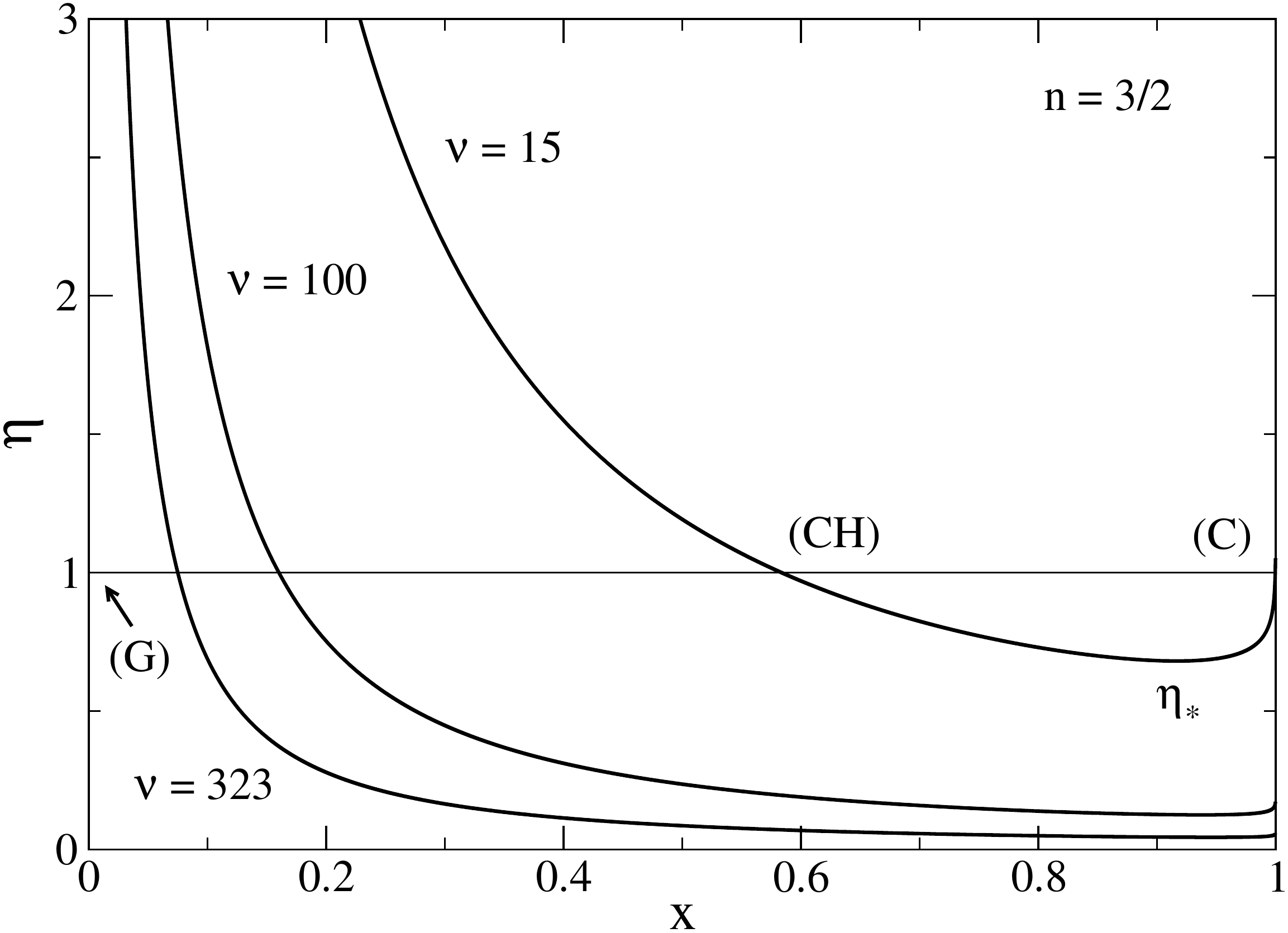}
\caption{The function $\eta(x)$ for different values of
$\nu>\nu_{\rm CCP}$
and for $n=3/2$
(the figure for $n=2$ is similar). We have
indicated the
gaseous phase (G), the condensed phase (C) and the core-halo phase (CH) on the
curve corresponding to $\nu=15$ (for $\eta=1$, the core-halo solution
corresponds to $x_{\rm CH}=0.583$). We have not represented the
region $x\rightarrow 0$ and $\eta>\eta_c\sim 2.52$ which is unphysical as
explained in the text.}
\label{xetaN1p5}
\end{center}
\end{figure}

The equilibrium states of DM halos, corresponding to $f'(x)=0$, are the
solutions of the
equation 
\begin{equation}
\label{ana8}
\ln(1-x)-\frac{9}{5}x\eta+\frac{3}{10}\eta+C(\eta,\mu)+\nu\eta
x^{2/(3-n)}=0.
\end{equation}
This equation determines the normalized core mass $x=M_c/M$ as a function of
$\eta$, $\nu$
and $\mu$. It is convenient to introduce the notation
\begin{equation}
\label{ana9b}
\eta_0=-\frac{10}{3}C(\eta,\mu),
\end{equation}
so
we can rewrite Eq.
(\ref{ana8}) as
\begin{equation}
\label{ana9}
\ln(1-x)+\nu\eta x^{2/(3-n)}-\frac{9}{5}x\eta+\frac{3}{10}(\eta-\eta_0)=0.
\end{equation}
We note that $\eta_0$ depends weakly (logarithmically) on $\eta$
and $\mu$ so, in a first approximation, it can be treated as a
constant. The solutions of Eq. (\ref{ana8}) can be easily found by
studying the
inverse function
\begin{equation}
\label{ana10}
\eta(x)=\frac{\eta_0-\frac{10}{3}\ln(1-x)}{1+\frac{10}{3}\left
\lbrack \nu x^{2/(3-n)}-\frac{9}{5} x \right \rbrack}
\end{equation}
for a given value of $\nu$ and $\mu$ (see Fig. \ref{xetaN1p5}). Our analytical
model is
valid for sufficiently large values of $\nu$ and $\mu$ (corresponding to large
DM halos).
On the other hand, the results depend on the value of the polytropic index $n$.
In the following, we
assume $1<n<3$.\footnote{The index $n=1$ is special and has been treated in
\cite{modeldm}. The condition $n<3$ is required in order to have a stable
core \cite{chandrabook}.} For $x\rightarrow 0$, we
get
\begin{equation}
\label{ana11}
\eta(x)=\eta_0+\left (6\eta_0+\frac{10}{3}\right )x+...
\end{equation}
Close to $x=0$, the curve $\eta(x)$
is always increasing. For $x\rightarrow 1$, we get
\begin{eqnarray}
\label{ana12}
\eta\sim \frac{-\ln(1-x)}{\nu-\frac{3}{2}}\rightarrow
+\infty,
\end{eqnarray}
where we have assumed
$\nu>3/2$ in order to avoid unphysical results due to the invalidity of our
model for small values of $\nu$.

We note that the inverse temperature $\eta(x)$ becomes infinite
at some $x_i\neq 1$ when the denominator in Eq. (\ref{ana10}) vanishes, i.e.,
when
\begin{equation}
\label{ana10b}
1+\frac{10}{3}\left
\lbrack \nu x_i^{2/(3-n)}-\frac{9}{5} x_i \right \rbrack=0.
\end{equation}
Instead of solving Eq. (\ref{ana10b}) for $x_i$ as a function of $\nu$, it is
simpler to study the inverse function
\begin{equation}
\label{ana10c}
\nu(x_i)=\frac{\frac{9}{5}x_i-\frac{3}{10}}{x_i^{2/(3-n)}}.
\end{equation}
This function (not represented) has the following properties: (i) $\nu(x_i)$ is
positive provided
that $x_i\ge 1/6$; (ii) $\nu(1)=3/2$; (iii) there is a maximum
\begin{equation}
\label{ana10d}
\nu_{\rm CCP}^{\rm app}=\frac{3(3-n)}{10(n-1)}[3(n-1)]^{2/(3-n)}
\end{equation}
at $(x_i)_*=1/[3(n-1)]$. In order to avoid unphysical results related to the 
divergence of the inverse temperature at some $x_i\neq 1$, we assume that
$\nu>\nu_{\rm CCP}^{\rm app}$. We find $\nu_{\rm CCP}^{\rm app}=2.7$ for
$n=2$ and $\nu_{\rm CCP}^{\rm app}=1.545$ for $n=3/2$. In a
sense, this critical value $\nu_{\rm CCP}^{\rm app}$ is the counterpart of
the canonical critical point that appears in the exact caloric
curve of
self-gravitating fermions and bosons (see Fig. 32 of \cite{ijmpb} for
fermions) although it manifests itself in a singular manner in our
simple analytical model. Its exact
value for fermions is  $\nu^{\rm F}_{\rm CCP}=2.85$ corresponding to $\mu_{\rm
CCP}=83$ \cite{ijmpb}. When $\nu>\nu_{\rm CCP}$ several
equilibrium states exist for the same value of the temperature leading to
canonical phase transitions (see below).

When $\nu>\nu_{\rm CCP}^{\rm app}$ the curve $\eta(x)$ presents  a
maximum at $(x_c^{\rm artifact}(\nu),\eta_c^{\rm artifact}(\nu))$ and a minimum
at
$(x_*(\nu),\eta_*(\nu))$. They are determined by Eq. 
(\ref{ana8}) and by the equation
\begin{equation}
\label{ana13}
\frac{2}{3-n}\nu\eta x_e^{(n-1)/(3-n)}-\frac{1}{1-x_e}-\frac{9}{5}\eta=0
\end{equation}
obtained by differentiating Eq. (\ref{ana8}) with respect to $x$ and writing
$d\eta/dx=0$.

For
$\nu,\mu\rightarrow
+\infty$, we find
that 
\begin{equation}
\label{ana16b}
x_c^{\rm artifact}\sim \left\lbrack
\frac{9}{10}(3-n)\frac{1}{\nu}\right\rbrack^{(3-n)/(n-1)} \rightarrow 0
\end{equation}
and
\begin{equation}
\label{ana16bb}
\eta_c^{\rm artifact}\sim \frac{10}{3}\ln\mu\rightarrow +\infty.
\end{equation}
We know that the maximum   inverse
temperature of a classical isothermal self-gravitating gas confined within
a box is $\eta_c^{\rm class}=2.52$ \cite{emden}. For self-gravitating fermions
and bosons, the   maximum   inverse
temperature $\eta_c(\nu,\mu)$  of the gaseous phase is close to  $\eta_c^{\rm
class}$ and tends to this value when $\nu,\mu\rightarrow
+\infty$ (see Fig. 14 of \cite{ijmpb} for fermions). Therefore, the above
results indicate that our simple
analytical model is not valid for $x\rightarrow 0$ and $\eta>\eta_c\sim 2.52$.
In particular, the maximum at $(x_c^{\rm artifact}(\nu),\eta_c^{\rm
artifact}(\nu))$ is an artifact of our model. The true maximum is at
$(x_c(\nu),\eta_c(\nu))\sim (0,2.52)$.

On the other hand, for
$\nu,\mu\rightarrow +\infty$, we find
that
\begin{eqnarray}
\label{ana17}
\frac{1}{1-x_*}&\sim& \frac{2}{3-n}\Biggl
\lbrack\frac{3}{2}\ln\nu+\ln\mu\nonumber\\
&+&\ln\left (\frac{2}{3-n}\right
)-\ln\left(\frac{6}{\sqrt{\pi}}\right )\Biggr \rbrack\rightarrow 0
\end{eqnarray}
and
\begin{equation}
\label{ana18}
\eta_*\sim \Biggl
\lbrack\frac{3}{2}\ln\nu+\ln\mu
+\ln\left (\frac{2}{3-n}\right
)-\ln\left(\frac{6}{\sqrt{\pi}}\right )\Biggr \rbrack \frac{1}{\nu}\rightarrow
0.
\end{equation}
For fermions ($n=3/2$), using Eq. (\ref{foo}), we obtain at leading order
$1/(1-x_*)\sim
(8/3)\ln\mu$ and $\eta_*\sim (2/\lambda)(\ln\mu/\mu^{2/3})$ \cite{pt}. We
note that $\eta_*(\nu)$ 
decreases as $\nu$ increases. This is consistent with the properties of the
exact
caloric curve of self-gravitating fermions (see Fig. 34 in \cite{ijmpb}).

\subsection{Stability of the equilibrium states}
\label{sec_stabana}

\begin{figure}[!h]
\begin{center}
\includegraphics[clip,scale=0.3]{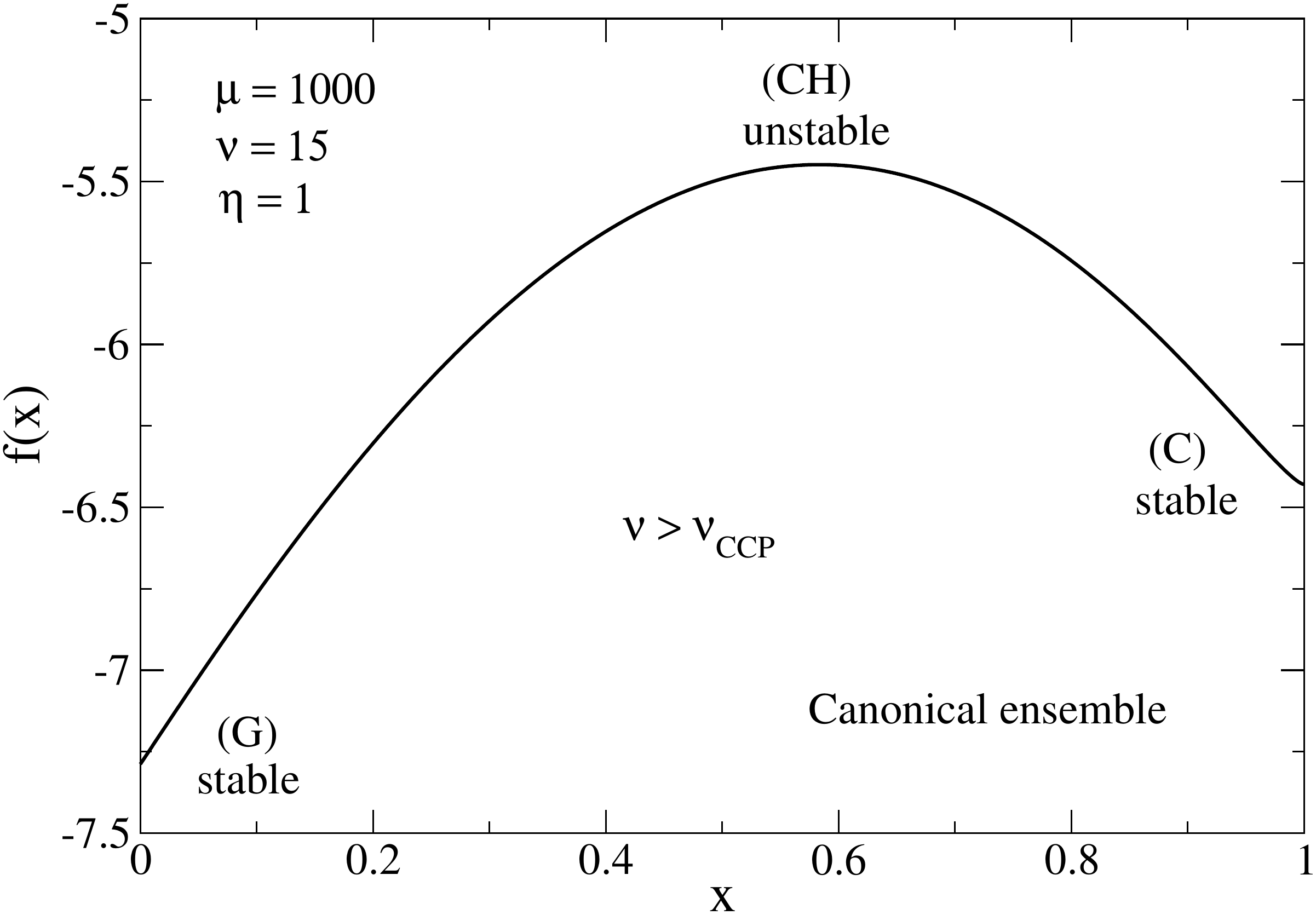}
\caption{Free energy $f(x)$ as a function of the core mass
$x$ for self-gravitating fermions ($n=3/2$) with $\nu=15$ (corresponding
to $\mu=10^3$) at $\eta=1$. The core-halo  solution (CH) at $x_*=0.583$ is a
maximum of
free energy at fixed mass. Therefore, it is unstable in the canonical
ensemble.}
\label{xf1000}
\end{center}
\end{figure}

Let us now consider more specifically the function
$f(x)$ giving the free
energy of the system as a function of the core mass $x$ for a given value of
$\nu$, $\mu$ and
$\eta$. Using Eq. (\ref{ana9b}), we can rewrite Eq.
(\ref{ana7}) as
\begin{eqnarray}
\label{ana19}
f(x)&=&-\frac{3-n}{5-n}\nu x^{(5-n)/(3-n)}\nonumber\\
&+&\frac{1}{\eta}
(1-x)\left\lbrack -\frac{3}{10}\eta_0+\ln(1-x)-1\right\rbrack\nonumber\\
&-&\frac{3}{2}x(1-x)-\frac{3}{5}(1-x)^2.
\end{eqnarray}
Its first derivative is 
\begin{equation}
\label{ana20b}
f'(x)=-\nu x^{2/(3-n)}+\frac{3}{10}\frac{\eta_0}{\eta}-\frac{1}{\eta}
\ln(1-x)-\frac{3}{10}+\frac{9}{5}x.
\end{equation}
The condition $f'(x)=0$ determines the possible equilibrium states of the system
as we have
just
seen. The stability of these equilibrium states in the canonical ensemble is
then
determined by the sign of the second derivative of the free energy: 
\begin{eqnarray}
\label{ana20}
f''(x)=-\nu \frac{2}{3-n} x^{(n-1)/(3-n)}+\frac{1}{\eta(1-x)}+\frac{9}{5}.
\end{eqnarray}
In the canonical ensemble an equilibrium state is stable when $f''(x)>0$,
corresponding to a minimum of
free energy, and  unstable when $f''(x)<0$, corresponding to a maximum of free
energy.  In Fig. \ref{xf1000} we have plotted the curve $f(x)$ in the case
$\nu>\nu_{\rm CCP}$ and $\eta_*<\eta<\eta_c$ where the system has three
equilibrium states as
detailed in the following section.

The values of the function $f(x)$ at
$x=0$ and $x=1$ are
\begin{eqnarray}
\label{ana21}
f(0)=-\frac{1}{\eta}\left (\frac{3}{10}\eta_0+1\right
)-\frac{3}{5}
\end{eqnarray}
and
\begin{eqnarray}
\label{ana22}
f(1)= -\frac{3-n}{5-n}\nu.
\end{eqnarray}
For $x\rightarrow 0$, we find that 
\begin{equation}
\label{ana23}
f(x)=f(0)+\frac{3}{10}\left (
\frac{\eta_0}{\eta}-1\right ) x+...
\end{equation}
In practice $\eta<\eta_c<\eta_0$ so 
the term in parenthesis is positive.  Since
the function $f(x)$ is
defined for $x\ge 0$, and since the slope of the function $f(x)$ at $x=0$ is
positive, the solution $x=0$ (gaseous phase)  is a local
minimum of $f(x)$ even though $f'(0)\neq 0$. We shall therefore
consider that the
solution $x=0$ is a stable equilibrium state.

\subsection{The different equilibrium states}
\label{sec_des}

After these mathematical preliminaries, we are now ready to perform the complete
analysis of the equilibrium states of
our simple analytical model. As explained previously we assume $\nu>\nu_{\rm
CCP}$.

The curve $\eta(x)$ is made of a vertical 
branch at $x=0$ up to $\eta=\eta_c$, then it decreases, reaches a minimum
$\eta_*$ at $x_*$,
and finally increases up to infinity when $x\rightarrow 1$ (see Fig.
\ref{xetaN1p5}). When 
$\eta<\eta_*$, there is a unique equilibrium state ($x=0$). It corresponds
to
the gaseous phase (G). It is stable (minimum of free
energy). When 
$\eta>\eta_c$, there is a unique equilibrium state ($x\simeq 1$). It
corresponds to the condensed phase (C). It is stable (minimum of free
energy).
When $\eta_*<\eta<\eta_c$ there are three equilibrium states (see
Figs. \ref{xetaN1p5} and
\ref{xf1000}): (i) a gaseous
phase (G); (ii) a core-halo phase (CH);
(iii) a condensed phase (C). Let us analyze these solutions in more detail 
in the limit $\nu,\mu\rightarrow
+\infty$:

(i) The gaseous solution (G) corresponds to a purely isothermal  halo
without core. The core mass is equal to zero: $x_{\rm G}=0$. This solution is
stable, being a minimum of free energy,
although the derivative of $f(x)$ does not vanish  at $x=0$ as
explained above.

\begin{figure}[!h]
\begin{center}
\includegraphics[clip,scale=0.3]{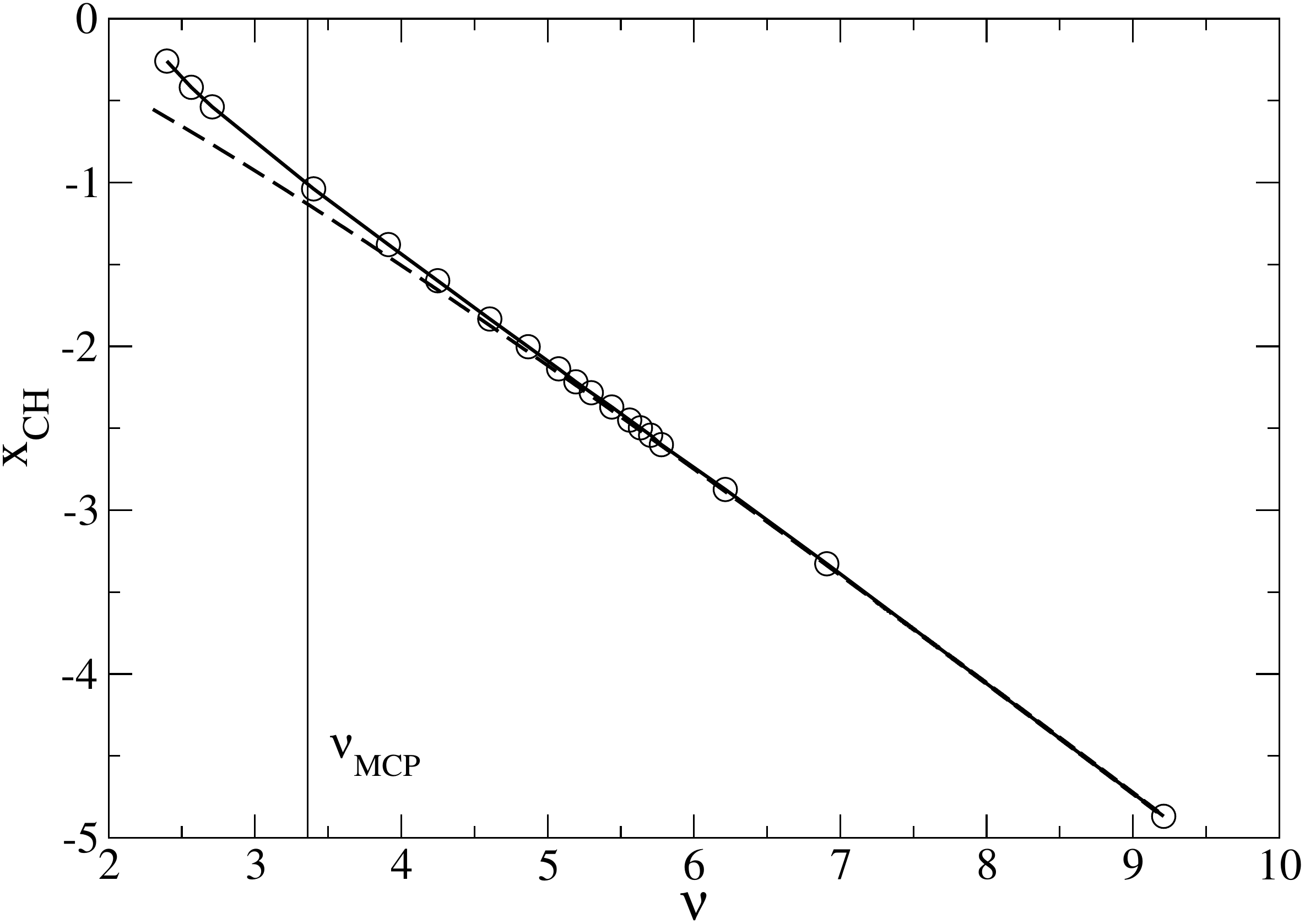}
\caption{Core mass ($x$) as a function of the halo mass ($\nu$)
for self-gravitating fermions ($n=3/2$). The bullets are obtained by solving Eq.
(\ref{ana8})
exactly (including
all the terms) for $\eta=1$. The dashed line corresponds to Eq. (\ref{ana24})
taking into account the fact that $\eta_0$ depends on $\mu$ hence on $\nu$. For
$\nu$ large we find an effective exponent $0.667$ instead of $(3-n)/2=3/4$
because of
logarithmic corrections in $\eta_0$. Coincidently, in the region of interest
$\nu<\nu_{\rm MCP}$ (see Sec. \ref{sec_mce}) we find an effective exponent very
close to $3/4$.}
\label{nuxch}
\end{center}
\end{figure}

(ii) The core-halo solution (CH) corresponds to an isothermal  halo
harboring a core with a small mass ($x_{\rm CH}\ll 1$). From Eq. (\ref{ana9}),
we find
that the normalized core
mass scales as
\begin{eqnarray}
\label{ana24}
x_{\rm CH}\sim \left\lbrack
\frac{3}{10}\frac{\eta_0-\eta}{\eta}\frac{1}{\nu}\right
\rbrack^{(3-n)/2}.
\end{eqnarray}
For fermions, using Eq. (\ref{foo}), we obtain at leading
order $x_{\rm CH}\sim (\ln\mu/\lambda\eta)^{3/4}\mu^{-1/2}$
\cite{pt}. This asymptotic formula is compared with the
exact value of $x_{\rm CH}$ in Fig.
\ref{nuxch}. Substituting Eq. (\ref{ana24}) into Eq. (\ref{ana20}) we find
that
\begin{equation}
f''(x_{\rm CH})\sim -\frac{2}{3-n} \left\lbrack
\frac{3}{10}\frac{\eta_0-\eta}{\eta}\right
\rbrack^{(n-1)/2}\nu^{(3-n)/2}\rightarrow -\infty.
\end{equation}
Therefore, the core-halo solution is
unstable in the canonical ensemble being a maximum of free energy. 

(iii) The condensed solution (C) corresponds to a quantum core surrounded
by a tenuous atmosphere ($x_{\rm C}\sim 1$). From Eq. (\ref{ana9}), we find
that the normalized
core mass
scales as  
\begin{eqnarray}
\label{ana30}
1-x_{\rm C}\sim \frac{\sqrt{\pi}}{6}\mu \frac{e^{3\eta/2}}{\eta^{3/2}}
e^{-\eta\nu},
\end{eqnarray}
showing that the quantum core contains almost all the mass. Substituting Eq.
(\ref{ana30}) into Eq. (\ref{ana20}) we find that
\begin{equation}
f''(x_{\rm C})\sim
\frac{6}{\sqrt{\pi}}\frac{1}{\mu} \frac{\eta^{1/2}}{e^{3\eta/2}} e^{\eta
\nu}\rightarrow +\infty.
\end{equation}
Therefore, the condensed
solution is stable, being a minimum of free energy.

The occurence of three equilibrium states at the same
temperature reveals the existence of a canonical phase transition associated
with an isothermal collapse \cite{aaiso}. Such
gravitational phase transitions are analyzed in detail in \cite{ijmpb}. The
gaseous (G)
and condensed (C) solutions are stable while the core-halo (CH) solution is
unstable. It has a negative specific heat (see
Sec. \ref{sec_mce}) which is forbidden in the canonical ensemble. It plays the
role of a ``germ'' or a ``critical droplet'' in the langage of phase
transitions and nucleation (the quantum core is the analogue of
the droplet). It creates a
barrier of free energy (see Fig. \ref{xf1000}) that the
system must cross to pass from the gaseous phase to the condensed phase (or the
converse). Depending on the
value of the temperature $T$ with respect to a temperature of transition $T_t$
\cite{ijmpb},
the gasous and the condensed states may
be either fully stable (global minimum of free energy) or metastable (local
minimum of
free
energy). However, for systems with long-range interactions, metastable states
have very long
lifetimes scaling as $e^N$ where $N\gg 1$
is the number of particles in the system \cite{lifetime}, so they are
strongly stable in practice.\footnote{The system initially in  
the metastable gaseous phase (G) must spontaneously  form a quantum core of mass
$(M_c)_{\rm CH}$ to overcome the barrier of free energy and 
collapse in the condensed phase (C). The probability to spontaneously
form such a core, thanks to energy
fluctuations, is extremely weak, scaling as $e^{-N}$. This is a such a rare
event that the metastable gaseous phase is stable in practice.} By contrast,
the core-halo state  (CH) is
unstable in the canonical ensemble, being a maximum (not a
minimum) of free energy at fixed mass. This looks
like a bad news since this core-halo structure is the
most interesting structure from a physical point of view. Fortunately, we show
in the next section that this core-halo structure is stable in the 
microcanonical ensemble (if the halo mass is not too large), being a maximum of
entropy at fixed mass and energy. This is a manifestation of ensembles
inequivalence
for systems with long-range
interactions \cite{ijmpb}. Since the core-halo structure seems to
appear in observations and numerical
simulations we suggest, following our previous paper \cite{modeldm},  that the
microcanonical ensemble is more
relevant than the canonical ensemble in the physics of DM halos.

\subsection{Microcanonical ensemble}
\label{sec_mce}

In the microcanonical ensemble, a stable equilibrium state is obtained by 
maximizing the entropy $S$ at fixed mass $M$ and
energy $E$. Let us first compute the energy and the entropy of the system
(quantum core $+$ isothermal atmosphere) by using the same analytical model as
in the previous sections.

The energy of the
quantum core is given by Eq.
(\ref{pc3}).  On the other hand, the energy of an isothermal self-gravitating
gas is
\begin{eqnarray}
\label{mce1}
E=\frac{3}{2}Nk_B T+W,
\end{eqnarray}
where the first term is the kinetic (thermal) energy and the second term is the
gravitational energy. Treating the atmosphere as a gas with a uniform density,
we get
\begin{equation}
\label{mce2}
E_a=\frac{3}{2}(M-M_c)\frac{k_B T}{m}+W_a,
\end{equation}
where $W_a$ is the gravitational energy of the atmosphere in the presence of the
core. It is given by Eq. (\ref{ih4}). Therefore, the total energy of the system
(core $+$
atmosphere) is
\begin{eqnarray}
\label{mce3}
E&=&-\frac{3-n}{5-n}\frac{GM_c^2}{R_c}+\frac{3}{2}(M-M_c)\frac{k_B
T}{m}\nonumber\\
&-&\frac{3GM_c(M-M_c)}{2R}-\frac{3G(M-M_c)^2}{5R}.
\end{eqnarray}
In the microcanonical ensemble, the total energy $E$ is fixed. Therefore, Eq.
(\ref{mce3})
determines the temperature $T(M_c)$ of the halo as a function of the core mass 
$M_c$.

The entropy of an isothermal self-gravitating gas is
\begin{eqnarray}
\label{mce4}
S=-k_B\int\frac{\rho}{m}\ln\rho\, d{\bf r}+\frac{3}{2}Nk_B \ln\left
(\frac{2\pi
k_B T}{m}\right )\nonumber\\
+\frac{5}{2}Nk_B+Nk_{B}\ln \left (\frac{2m^4}{h^3}\right ).
\end{eqnarray}
Treating the atmosphere as a gas with a
uniform
density,
we get
\begin{eqnarray}
\label{mce5}
S=-\frac{k_B}{m}(M-M_c)\ln\left
(\frac{M-M_c}{V}\right )\nonumber\\
+\frac{3}{2}\frac{k_B}{m}(M-M_c)\ln\left
(\frac{2\pi
k_B T}{m}\right )
+\frac{5}{2}\frac{k_B}{m}(M-M_c)\nonumber\\
+\frac{k_{B}}{m}(M-M_c)\ln \left
(\frac{2m^4}{h^3}\right ).\qquad
\end{eqnarray}
Since the entropy of the core is equal to zero,
Eq. (\ref{mce5}) represents the total entropy of the system. 

From the above expressions, we note that the free energy $F=E-TS$ of an
isothermal self-gravitating gas (which is the relevant thermodynamic potential
in the canonical ensemble) is
\begin{eqnarray}
\label{mce6}
F=k_B T \int\frac{\rho}{m}\ln\rho\, d{\bf r}-\frac{3}{2}Nk_B T
\ln\left (\frac{2\pi
k_B T}{m}\right )\nonumber\\
-Nk_B T- Nk_{B}T\ln \left (\frac{2m^4}{h^3}\right )+W.
\end{eqnarray}
This justifies the expression of the internal energy $U=F-W$ given by
Eq. (\ref{ih2b}). On the other hand, the free energy $F=E-TS$ calculated
with Eqs. (\ref{mce3}) and (\ref{mce5})
returns Eq. (\ref{ana4}) of our analytical model.

Introducing the dimensionless variables
\begin{eqnarray}
\label{mce7}
\Lambda=-\frac{ER}{GM^2},\qquad s(x)=\frac{S}{Nk_B},
\end{eqnarray}
we get
\begin{eqnarray}
\label{mce8}
\Lambda&=&\frac{3-n}{5-n}\nu
x^{(5-n)/(3-n)}-\frac{3}{2\eta}(1-x)\nonumber\\
&+&\frac{3}{2}x(1-x)+\frac{3}{5}
(1-x)^2
\end{eqnarray}
and
\begin{eqnarray}
\label{mce9}
s(x)=-(1-x)\left\lbrack\ln(1-x)+C(\eta,\mu)-\frac{5}{2}
\right\rbrack.
\end{eqnarray}
The first equation determines the inverse temperature $\eta(x)$ as a function of
the core mass $x$ for a fixed value
of the energy
$\Lambda$. Then, the maximization of the entropy $s(x)$ (at fixed energy
$\Lambda$) 
determines the equilibrium state(s) in
the microcanonical ensemble. From the above expressions, we note that the
dimensionless free
energy (relevant in the canonical ensemble) is given by
\begin{eqnarray}
\label{mce10}
f(x)=-\Lambda(x)-\frac{s(x)}{\eta},
\end{eqnarray}
where  now the inverse temperature $\eta$ is fixed and the energy $\Lambda(x)$
depends on the core mass
$x$. This returns Eq.
(\ref{ana7}).

One can easily check that the condition $s'(x)=0$ for a fixed value of
$\Lambda$ yields Eq. (\ref{ana8}) which was previously obtained from the
condition
$f'(x)=0$ for a fixed value of
$\eta$. As a result, the equilibrium states (extrema of entropy at fixed mass
and energy and extrema of free energy at fixed mass) are the same in the
microcanonical and canonical ensembles. However, their stability may be
different in the microcanonical and canonical ensembles. This is the notion
of ensembles inequivalence for systems with long-range interactions
\cite{ijmpb}.

\begin{figure}[!h]
\begin{center}
\includegraphics[clip,scale=0.3]{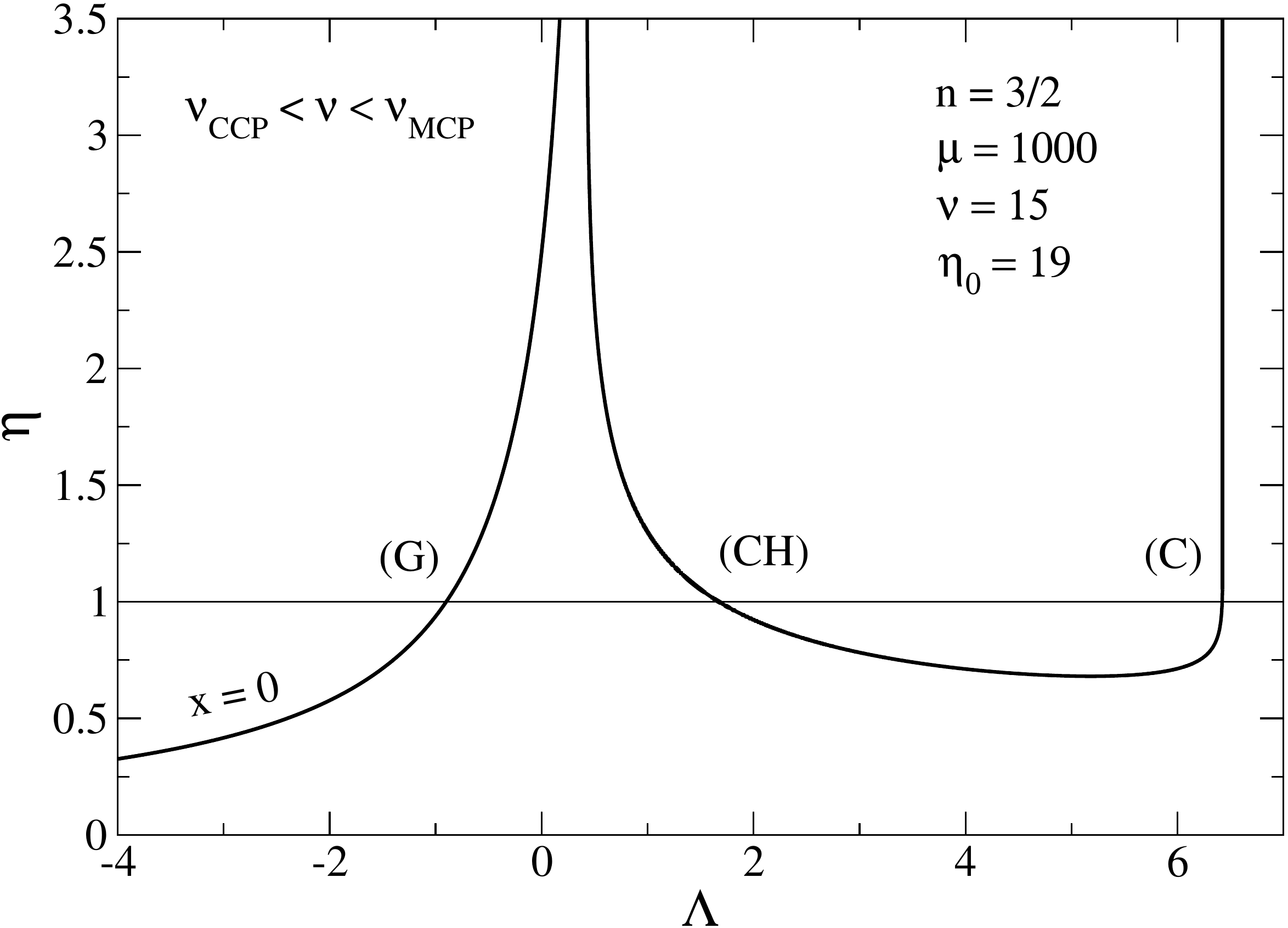}
\caption{Caloric curve of self-gravitating fermions $n=3/2$ with $\nu=15$
(corresponding to $\mu=1000$) in the framework of our analytical model. It is
in good agreement with the exact caloric curve represented in Fig. 10 of
\cite{pt}.}
\label{lambdaeta1000}
\end{center}
\end{figure}

\begin{figure}[!h]
\begin{center}
\includegraphics[clip,scale=0.3]{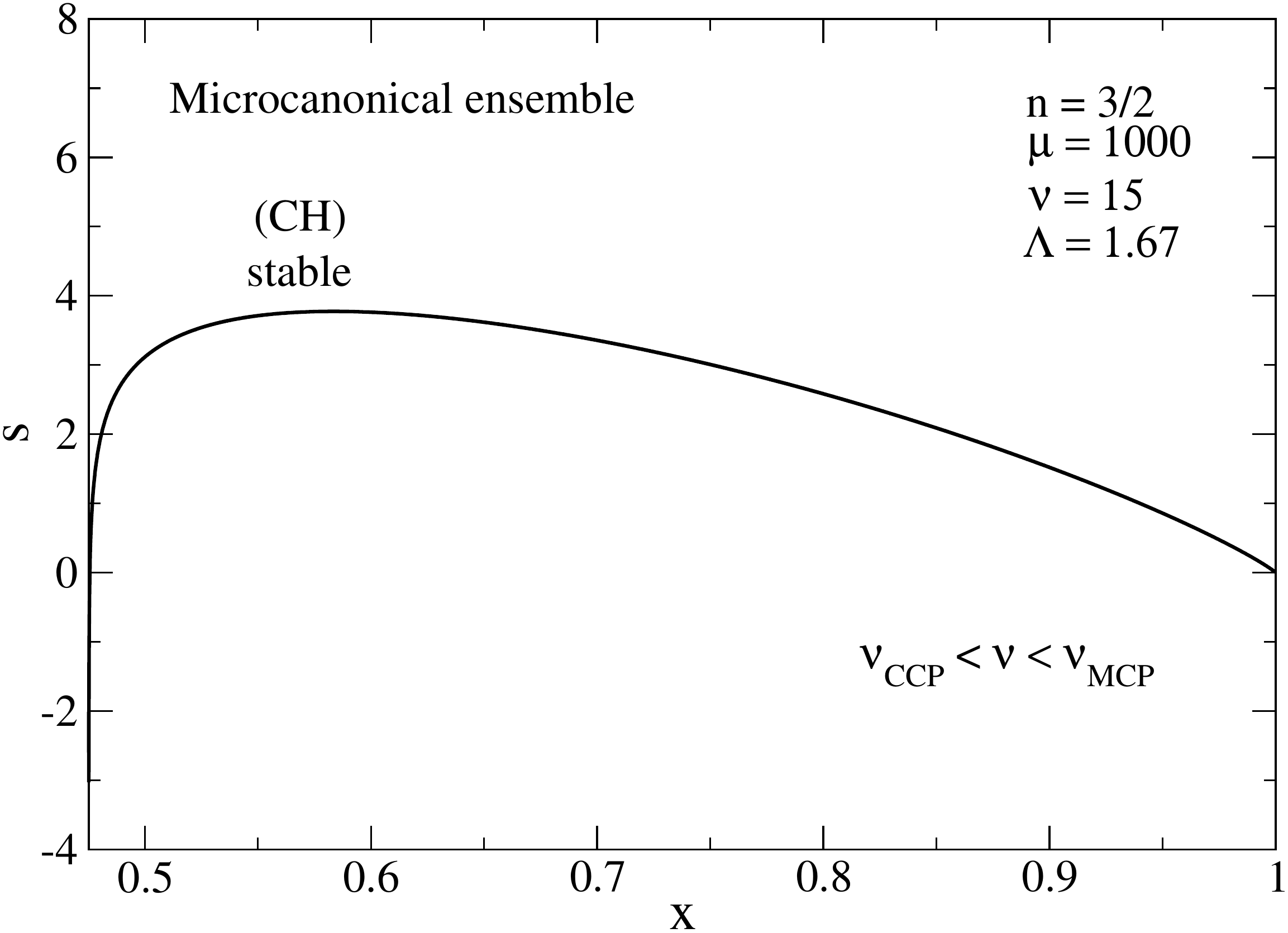}
\caption{Entropy $s(x)$ as a function of the core mass $x$ for  self-gravitating
fermions  $n=3/2$ with $\nu=15$ (corresponding to $\mu=1000$) at $\Lambda=1.67$.
The core-halo
 solution (CH) at $x_*=0.583$ is a maximum of entropy at fixed mass and energy.
Therefore, it is stable in the microcanonical ensemble.}
\label{xs1000}
\end{center}
\end{figure}

Let us first consider all the possible equilibrium states determined by
Eq. (\ref{ana8}).
For each of them, we can compute the inverse temperature $\eta(x)$ and
the energy  $\Lambda(x)$ and plot the caloric curve 
$\eta(\Lambda)$, relating the temperature to the energy, parametrized by the
core mass $x$. The minimum energy, corresponding to $x=1$, is
$\Lambda_{\rm max}=[(3-n)/(5-n)]\nu$. We can also add the caloric
curve of the
gaseous phase ($x=0$) which is simply given by $\Lambda=3/5-3/(2\eta)$. The
complete caloric curve
for $n=3/2$ (fermions) and $\nu=15$ (corresponding to $\mu=1000$) is represented
in Fig. \ref{lambdaeta1000}. For $\eta=1$, we
recover the three solutions (G), (CH) and (C) studied in Sec. \ref{sec_des}.
We note for future reference that
the core-halo solution has an energy $\Lambda_{\rm CH}=1.67$ and a core mass
$x_{\rm CH}=0.583$. We clearly see that the core-halo
phase has a negative specific heat $C=dE/dT<0$ implying that it is unstable in
the canonical ensemble as we have found. However, it is known that negative
specific heats are
allowed in the microcanonical ensemble for systems with long-range interactions
\cite{ijmpb}.
In the present case,  the
core-halo solution (CH) turns out to be stable in the microcanonical ensemble.
This is confirmed in Fig. \ref{xs1000} by plotting the entropy $s(x)$ versus the
core mass
at the energy $\Lambda=1.67$. We see that the core-halo solution at $x_{\rm
CH}=0.583$ is a maximum of
entropy at fixed energy. {\it Therefore, the core-halo  solution (CH), with a
negative
specific heat, is unstable in the canonical ensemble while it is stable in
the microcanonical ensemble.} This makes the core-halo
solution extremely important since it is now justified as being the ``most
probable'' configuration of the system (maximum entropy state).

\begin{figure}[!h]
\begin{center}
\includegraphics[clip,scale=0.3]{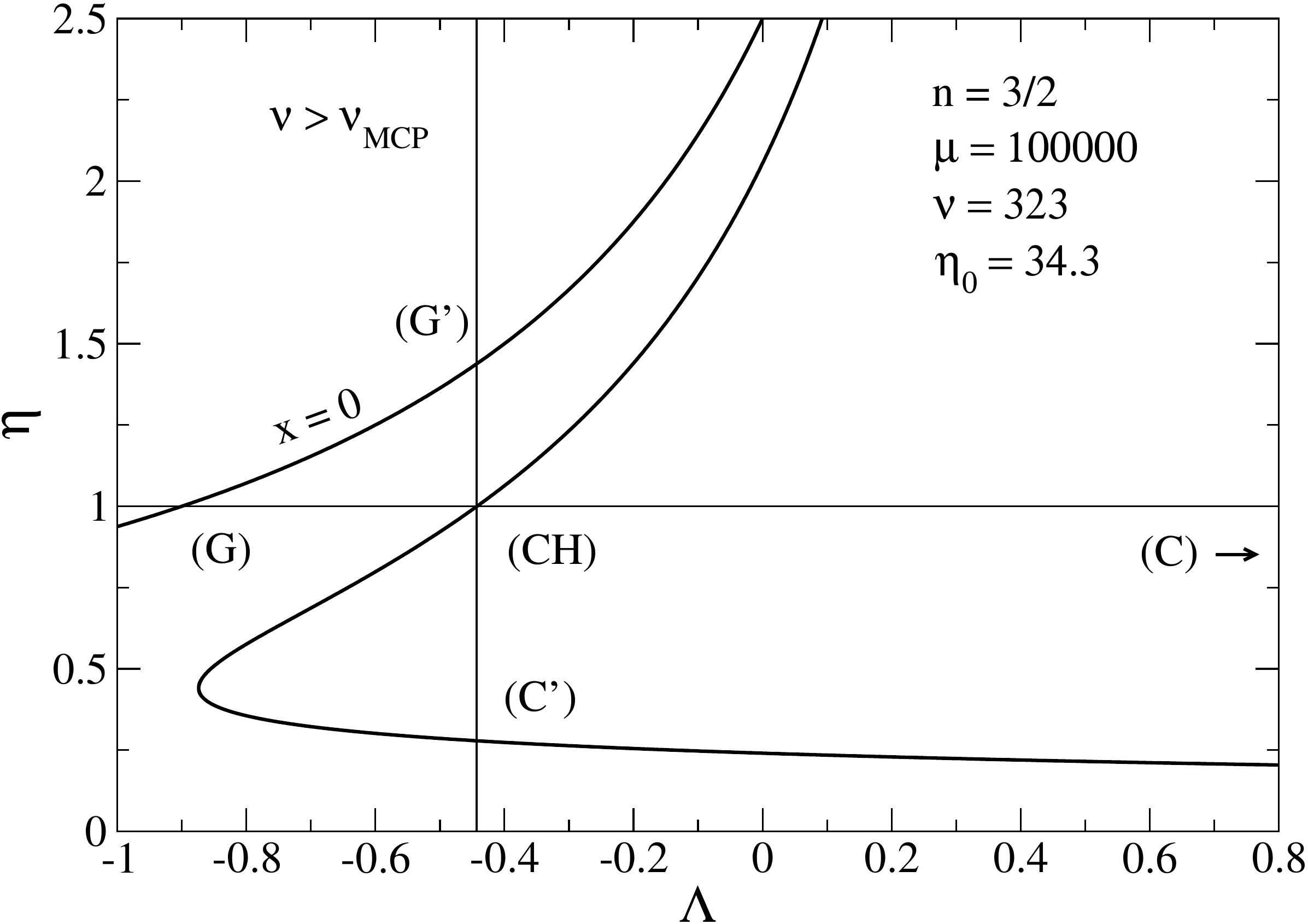}
\caption{Caloric curve of self-gravitating fermions $n=3/2$ with $\nu=323$
(corresponding to $\mu=10^5$) in the framework of our analytical model. It is
in good agreement with the exact caloric curve represented in Fig. 7 of
\cite{pt}.}
\label{lambdaeta100000}
\end{center}
\end{figure}

\begin{figure}[!h]
\begin{center}
\includegraphics[clip,scale=0.3]{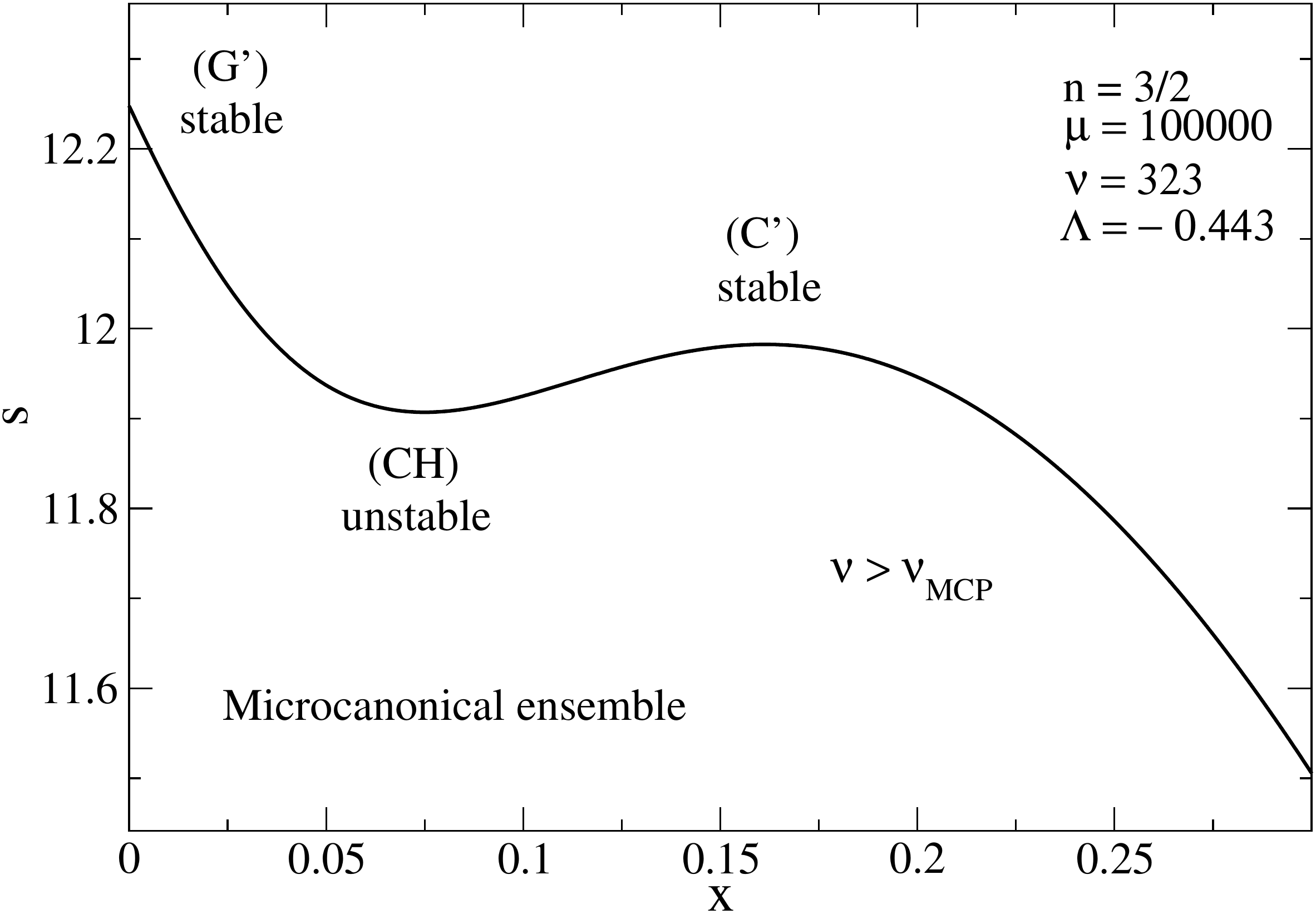}
\caption{Entropy $s(x)$ as a function of the core mass $x$ for  self-gravitating
fermions $n=3/2$ with $\nu=323$ (corresponding to $\mu=10^5$) at
$\Lambda=-0.443$. The
 solution (CH) at $x_{\rm CH}=0.0750$ is a minimum of entropy at fixed mass
and energy.
Therefore, it is unstable in the microcanonical ensemble.}
\label{xs100000}
\end{center}
\end{figure}

This result is valid only when $\nu<\nu_{\rm MCP}$, where $\nu_{\rm MCP}$ is the
microcanonical critical point (its exact value for fermions is $\nu^{\rm F}_{\rm
MCP}=28.8$  corresponding to $\mu_{\rm MCP}=2670$ \cite{ijmpb}). When
$\nu>\nu_{\rm MCP}$ the caloric curve
$\eta(\Lambda)$ becomes multivalued (see Fig. \ref{lambdaeta100000}) leading
to microcanonical phase
transitions associated with the gravothermal catastrophe \cite{lbw,ijmpb}. In
that case, we can
have a microcanonical phase transition from the gaseous phase (G') to the 
condensed phase (C') that we do not analyze in detail here (see \cite{ijmpb}
for a detailed discussion). We just point out that the core-halo solution (CH)
is now both canonically {\it and} microcanonically unstable (for $\eta=1$ we
have
$\Lambda_{\rm CH}=-0.443$ and $x_{\rm CH}=0.0750$).
This is confirmed in Fig. \ref{xs100000} by plotting the entropy $s(x)$ versus
the
core mass at the energy $\Lambda=-0.443$. We see that the core-halo solution is
a
minimum of entropy at fixed energy. {\it Therefore, the core-halo  solution (CH)
is
unstable in the  canonical  and in the microcanonical ensembles}.

\subsection{General scenario of DM halos}
\label{sec_scenario}

Recalling that $\nu$ is a measure of the size of a DM halo (see
Sec.
\ref{sec_c}), and that the virial theorem fixes the value of
the normalized temperature to $\eta\sim 1$ [see
Eq. (\ref{c2})], the preceding results confirm and refine the scenario
developed
in \cite{modeldm}:  

(i) There is a ``minimum halo''  of mass $(M_h)_{\rm min}\sim 10^{8}\,
M_{\odot}$
(value obtained from the observations)
corresponding to a purely quantum core without isothermal atmosphere (ground
state).

(ii) For $(M_h)_{\rm min}<M_h<(M_h)_{\rm CCP}$, the caloric curve
$\eta(\Lambda)$ is monotonic. As a result, the virial condition $\eta\sim 1$
determines a unique solution: a quantum phase (Q). It corresponds to a DM halo
with a quantum core and a tenuous isothermal atmosphere.

(iii) For $(M_h)_{\rm CCP}<M_h<(M_h)_{\rm MCP}$, the caloric curve
$\eta(\Lambda)$ has an $N$-shape structure (see Fig. \ref{lambdaeta1000} of
this paper and Fig. 10 of \cite{pt}). As a result, the virial condition
$\eta\sim 1$
determines three solutions: a gaseous phase (G), a core-halo phase (CH)
and a condensed phase (C). Among these three solutions, the  core-halo state
(CH) corresponding to a DM halo with a quantum core and an isothermal atmosphere
is the most relevant.
This solution is unstable in the canonical ensemble but it is stable in the
microcanonical ensemble. It has a negative
specific heat.  The quantum
core may mimic a large bulge (not a black hole) as argued in
\cite{modeldm}. The core mass -- halo mass relation $M_c(M_h)$, which is of
prime interest, is studied in the following sections. 

(iv) For $M_h>(M_h)_{\rm MCP}$, the caloric curve
$\eta(\Lambda)$ has a $Z$-shape structure (see Fig.
\ref{lambdaeta100000} of
this paper and Fig. 7 of \cite{pt}). As before,
the virial condition $\eta\sim 1$
determines three solutions: (G),  (CH)
and (C). However, the  core-halo solution
(CH) is now unstable in both canonical and microcanonical ensembles. In that
case, the system
collapses and
undergoes a gravothermal catastrophe \cite{lbw}. A possibility is that it
collapses from the gaseous phase (G') to the condensed phase (C') by forming a
quantum core of mass $M_c \sim (1/3)M_h$ \cite{acf} and expulsing a hot 
atmosphere at large distances.\footnote{As argued in \cite{clm2,modeldm}, this
scenario may not be relevant for DM halos. However, in another context, it
may be relevant to explain the supernova phenomenon of massive stars
\cite{pomeau2,acf}.} Alternatively, the core of the system may become
relativistic
during
the gravothermal catastrophe and finally undergo a dynamical instability of
general
relativistic origin leading ultimately to a  supermassive black hole (this is
the scenario of Balberg {\it et al.} \cite{balberg}  advocated in
\cite{clm2,modeldm}).
We stress that the process of gravothermal catastrophe is generally very long
(secular) but it may be relevant in galactic nuclei or if the DM
particle has a self-interaction \cite{balberg}.

In conclusion, we predict that DM halos with a mass $10^8\,
M_{\odot}<M_h<(M_h)_{\rm MCP}$ harbor a quantum core (bosonic soliton or fermion
ball) while  DM halos with a mass $M_h>(M_h)_{\rm MCP}$ harbor a supermassive
black hole. This result is connected to the fundamental existence of a
microcanonical
critical point in the statistical mechanics of self-gravitating systems
\cite{ijmpb}. If our scenario turns out to be correct, it would
provide a beautiful illustration of the curious thermodynamics of
self-gravitating systems in a cosmological context. The value of $(M_h)_{\rm
MCP}$ depends on the type of particles that compose DM. For self-interacting
bosons in TF limit we found $(M_h)^{\rm TF}_{\rm MCP}\sim 2\times 10^{12}\,
M_{\odot}$ (corresponding to $\mu_{\rm MCP}^{\rm TF}\sim 10^5$ and $\nu_{\rm
MCP}^{\rm TF}\sim 316$) and for fermions we found
$(M_h)^{\rm TF}_{\rm MCP}= 9.81\times 10^{9}\,
M_{\odot}$ (corresponding to $\mu_{\rm MCP}^{\rm F}=2670$ and $\nu_{\rm
MCP}^{\rm F}=28.8$). The case of noninteracting bosons is under
investigation \cite{modeldm}.\footnote{In comparison, the canonical critical
point is $(M_h)^{\rm TF}_{\rm CCP}= 3.27\times 10^{9}\,
M_{\odot}$ (corresponding to $\mu_{\rm CCP}^{\rm TF}=130$ and $\nu_{\rm
CCP}^{\rm TF}=11.4$) for self-interacting
bosons in TF limit and $(M_h)^{\rm F}_{\rm CCP}=6.12\times 10^{8}\,
M_{\odot}$ (corresponding to $\mu_{\rm CCP}^{\rm F}=83$ and $\nu_{\rm
CCP}^{\rm F}=2.85$) for fermions } Presently, the obtained value of
$(M_h)_{\rm
MCP}$ is a rough
estimate which should be improved in the future if our scenario is relevant. On
the other hand, we do not the preclude the presence of supermassive black holes
in DM
halos of mass $M_h<(M_h)_{\rm MCP}$ if they are formed by a mechanism different
from the one \cite{balberg} that we have advocated.

{\it Remark:} The gaseous solutions (G)  constitute
the lower branch of the generic phase
diagram $M_c(M_h)$ reported in Fig. 49 of  \cite{modeldm}. The core-halo 
solutions (CH)
constitute
the upper branch of this phase diagram that appears above the
canonical critical point (bifurcation point) $(M_h)_{\rm CCP}$.
This
is the branch of
most physical interest in the physics of DM halos. The core mass -- halo mass
relation on the core-halo  branch (CH) is studied in detail in
the following sections. This branch becomes unstable above $(M_h)_{\rm MCP}$
where the quantum core (bulge) is replaced by a supermassive black hole.

\section{Justification of the velocity
dispersion tracing relation and determination of the $M_c(M_v)$ relation}
\label{sec_vdt}

\subsection{The velocity
dispersion tracing relation}

We consider the core-halo solution (CH) of Sec. \ref{sec_des}. The
normalized core mass is given by
Eq. (\ref{ana24}). We first show that this result is equivalent to the
``velocity
dispersion tracing'' relation \cite{mocz,bbbs,modeldm}
\begin{eqnarray}
\label{ana31b}
v_c^2\sim v_h^2 \quad {\rm or} \quad M_c\sim \frac{R_c}{r_h}M_h,
\end{eqnarray}
stating that the
velocity dispersion in the core $v_c^2\sim GM_c/R_c$ is of the same order as the
velocity dispersion in the halo $v_h^2\sim GM_h/r_h$. Using Eq. (\ref{pc1}),
this relation can be rewritten as
\begin{eqnarray}
\label{dim1}
M_c\sim \left\lbrack
\frac{K(n+1)}{G}\right\rbrack^{n/2}
\frac{\omega_n^{(n-1)/2}}{(4\pi)^{1/2}} \left
(\frac{M_h}{r_h}\right )^{(3-n)/2}.
\end{eqnarray}
On the other hand, using Eqs. (\ref{ana5}) and (\ref{nu}), we find that Eq.
(\ref{ana24}) is equivalent to
\begin{eqnarray}
\label{dim2}
M_c\sim \left\lbrack \frac{3}{10}\frac{\eta_0-\eta}{\eta}\right
\rbrack^{(3-n)/2} \left\lbrack
\frac{K(n+1)}{G}\right\rbrack^{n/2}\frac{\omega_n^{
(n-1)/2}}{(4\pi)^{1/2}}\nonumber\\
\times \left (\frac{M_h}{r_h}\right )^{(3-n)/2}.
\end{eqnarray}
The formulae (\ref{dim1}) and (\ref{dim2}) are consistent with each other up
to a multiplicative factor  of order unity containing a logarithmic
correction.
As a result, Eq. (\ref{ana24}) is equivalent to the ``velocity
dispersion tracing'' relation from Eq. (\ref{ana31b}). Our study provides
therefore a
justification of this relation
from (effective) thermodynamical arguments.

\subsection{The $M_c(M_h)$ relation}

If we now substitute the relation $M_h=1.76\,
\Sigma_0
r_h^2$  from Eq. (\ref{iso1}) into Eq. (\ref{dim1}) we obtain
\begin{eqnarray}
\label{dim3}
M_c\sim \left\lbrack
\frac{K(n+1)}{G}\right\rbrack^{n/2}
\frac{\omega_n^{(n-1)/2}}{(4\pi)^{1/2}}(1.76\, \Sigma_0 M_h)^{(3-n)/4}.\quad
\end{eqnarray}
This equation gives the relation between the core mass $M_c$ and the halo mass
$M_h$. It displays the fundamental scaling
\begin{eqnarray}
\label{dim3b}
M_c\propto M_h^{(3-n)/4}.
\end{eqnarray}
Combining Eq. (\ref{dim3}) with the minimum halo mass from Eq.
(\ref{p7}), we get 
\begin{eqnarray}
\label{sat1}
\frac{M_c}{(M_h)_{\rm min}}\sim A_n \left\lbrack \frac{M_h}{(M_h)_{\rm
min}}\right\rbrack^{(3-n)/4},
\end{eqnarray}
where
\begin{eqnarray}
\label{sat2}
A_n= \left (\frac{1.76}{4\pi}\right
)^{(3-n)/4}\frac{\xi_1^{(n+1)/2}(-\theta'_1)^{(n-1)/2}}{\xi_h^{(3n-1)/4}
(-\theta'_h)^{(n+1)/4}}
\end{eqnarray}
is a constant that depends on the polytropic index
$n$. We
find below that the prefactor $A_n$ in Eq. (\ref{sat1}) is of order unity.
Therefore, the ratio between the core mass and the halo mass is
$M_c/M_h\sim  [M_h/(M_h)_{\rm min}]^{-(n+1)/4}$. Accordingly, the core mass
becomes negligible in front of the halo mass when $M_h\gtrsim 10-100\,
(M_h)_{\rm min}$ with $(M_h)_{\rm min}\sim 10^8\, M_{\odot}$.

\subsection{The $M_c(M_v)$ relation}
\label{sec_mcmv}

Following our previous work \cite{modeldm}, we have defined the halo mass
$M_h$ and the halo radius $r_h$ such that $r_h$ represents the distance at
which the central density is divided by $4$ (see Appendix \ref{sec_p}). However,
Schive
{\it et al.} \cite{ch3} use another definition of
the halo mass $M_v$ and
halo radius $r_v$. They are connected by
\begin{eqnarray}
\label{sc1}
M_v=\frac{4}{3}\pi r_v^3 \zeta(0)\rho_{m,0}
\end{eqnarray}
where $\rho_{m,0}=\Omega_{m,0}\epsilon_0/c^2$ is the
present background matter density in the Universe and $\zeta(0)$ is a
prefactor of order $\sim 350$. For the numerical applications, we shall take
$\Omega_{m,0}=0.3089$ and $\epsilon_0/c^2=8.62\times 10^{-24}\, {\rm
g \, m^{-3}}$ giving $\rho_{m,0}=2.66\times 10^{-24}\, {\rm
g \, m^{-3}}$ \cite{planck2016}. 
Using
\begin{eqnarray}
\label{sc2}
\frac{GM_v}{r_v}\sim \frac{GM_h}{r_h},
\end{eqnarray}
in consistency with Eq. (\ref{ana31b}), and Eqs. (\ref{iso1}) and (\ref{sc1}),
we obtain 
\begin{eqnarray}
\label{sc3}
M_h\sim \frac{1}{1.76\, \Sigma_0}\left\lbrack \frac{4}{3}\pi
\zeta(0)\rho_{m,0}\right\rbrack^{2/3} M_v^{4/3}.
\end{eqnarray}
The scaling $M_h\propto M_v^{4/3}$ was previously noted in \cite{modeldm}.
Normalizing the halo mass by the minimum halo mass, we get
\begin{eqnarray}
\label{sc4}
\frac{M_h}{(M_h)_{\rm min}}\sim B_n \left\lbrack \frac{M_v}{(M_h)_{\rm
min}}\right\rbrack^{4/3}
\end{eqnarray}
with
\begin{eqnarray}
\label{sc5}
B_n=\frac{1}{1.76\, \Sigma_0}\left\lbrack \frac{4}{3}\pi
\zeta(0)\rho_{m,0}\right\rbrack^{2/3} (M_h)_{\rm min}^{1/3}.
\end{eqnarray}
Combining Eqs. (\ref{sat1}) and (\ref{sc4}), we
finally  obtain the core mass -- halo mass relation
\begin{eqnarray}
\label{sc6}
\frac{M_c}{(M_h)_{\rm min}}\sim A_n B_n^{(3-n)/4} \left\lbrack
\frac{M_v}{(M_h)_{\rm
min}}\right\rbrack^{(3-n)/3}.
\end{eqnarray}
It exhibits the fundamental scaling $M_c\sim M_v^{(3-n)/3}$. 

{\it Remark:} Since $B_n$ is a dimensioness constant of order
$10^{-3}$ (see below), Eq. (\ref{sc5}) provides a
relation
between the DM particle parameters $(m,a_s)$ [via $(M_h)_{\rm
min}$], the universal DM surface
density $\Sigma_0$ and the present density of matter in the Universe
$\rho_{\rm m,0}$. Expressing $\Sigma_0$ and $\rho_{\rm m,0}$ in terms of the
cosmological constant $\Lambda$, we
will be able (see Sec. \ref{sec_app}) to obtain the DM particle parameters
$(m,a_s)$ in terms of the fundamental constants of physics.

\section{Application to quantum models of DM halos}
\label{sec_app}

We now apply these results to quantum models of DM halos made of fermions,
noninteracting bosons, and self-interacting bosons as described in Sec.
\ref{sec_qm}. For reasons that will become clear below,
we treat the case of noninteracting bosons first.

\subsection{Noninteracting bosons}
\label{sec_qmni}

A noninteracting self-gravitating BEC ($a_s=0$) is equivalent to
a polytrope of index $n=2$  with a
polytropic constant given by Eq. (\ref{t4}) (see Sec. \ref{sec_ni} and
Appendix \ref{sec_comp}). Using Eqs. (\ref{sat1}), (\ref{sat2})
and the results of
Appendix \ref{sec_p}, we
get
\begin{eqnarray}
\label{sat5}
\frac{M_c}{(M_h)_{\rm min}}\sim 1.84\, 
\left\lbrack \frac{M_h}{(M_h)_{\rm
min}}\right\rbrack^{1/4}.
\end{eqnarray}
On the other hand, using Eqs. (\ref{nipv3}) and (\ref{sc5}), we find that
\begin{eqnarray}
\label{sc10}
B_{2}&=&\frac{1}{1.76\, \Sigma_0} \left\lbrack \frac{4}{3}\pi
\zeta(0)\rho_{m,0}\right\rbrack^{2/3} (1.61)^{1/3}\left
(\frac{\hbar^4\Sigma_0}{G^2m^4}\right )^{1/9}\nonumber\\
&=& 2.04\times 10^{-3}.
\end{eqnarray}
Therefore, Eq. (\ref{sc4}) takes the form
\begin{eqnarray}
\label{sc4ni}
\frac{M_h}{(M_h)_{\rm min}}\sim  2.04\times 10^{-3} \left\lbrack
\frac{M_v}{(M_h)_{\rm
min}}\right\rbrack^{4/3}.
\end{eqnarray}
Combining Eqs. (\ref{sat5}) and (\ref{sc4ni}), we obtain the core mass -- halo
mass relation
\begin{eqnarray}
\label{sat6}
\frac{M_c}{(M_h)_{\rm min}}\sim 0.391\, 
\left\lbrack \frac{M_v}{(M_h)_{\rm
min}}\right\rbrack^{1/3}.
\end{eqnarray}
It exhibits the fundamental scaling $M_c\propto M_v^{1/3}$.
This theoretical scaling is consistent with the scaling found
numerically by Schive {\it et al.} \cite{ch3} (see also \cite{veltmaat}). These
authors also presented an
heuristic argument to justify this relation. As discussed in
Refs. \cite{mocz,bbbs,modeldm}, their argument is equivalent to assuming the 
velocity dispersion
tracing relation (\ref{ana31b}). We stress, however, that this
relation is not obvious {\it a priori} and that other relations, such as the
energy tracing relation corresponding to $GM_c^2/R_c\sim GM_h^2/r_h$, could be
contemplated as well
 \cite{mocz}. They would lead to different results. The fact
that the velocity
dispersion
tracing relation
(\ref{ana31b}) can be justified from a maximum entropy principle (most probable
state), as
shown in the present paper, may provide a
physical basis for it.

On the other hand, reversing Eq. (\ref{sc10}) following the remark at the end
of Sec. \ref{sec_mcmv}, we get
\begin{eqnarray}
\label{sc11}
m=\frac{(1.61)^{3/4}}{(1.76)^{9/4}} \left\lbrack \frac{4}{3}\pi
\zeta(0)\rho_{\rm
m,0}\right\rbrack^{3/2}\frac{\hbar}{\Sigma_0^2G^{1/2}}\frac{1}{{B_2}^{9/4}}.
\end{eqnarray}
The present matter density in the Universe is given by
$\rho_{\rm m,0}=\Omega_{\rm m,0}\epsilon_0/c^2$ and the density of dark energy  
is given by $\rho_{\Lambda}=\Lambda/{8\pi G}=\Omega_{\rm de,0}\epsilon_0/c^2$ 
where $\Omega_{\rm de,0}\simeq 1-\Omega_{\rm m,0}=0.6911$ is the present
fraction
of dark energy and $\Lambda=1.00\times 10^{-35}\, {\rm s^{-2}}$ is the
cosmological constant.  Therefore, we
can write the  present DM density in terms of the cosmological constant
as
\begin{eqnarray}
\label{sc1b}
\rho_{\rm m,0}=\frac{\Omega_{\rm m,0}}{\Omega_{\rm de,0}}\frac{\Lambda}{8\pi
G}=0.0178\, \frac{\Lambda}{G}.
\end{eqnarray}
On the other hand, in the framework of the logotropic
model \cite{epjp,lettre,ouf}, we have theoretically predicted that the surface
density of the DM halos is constant and that its universal value is given in
terms of an
effective cosmological constant (whose value is the same as Einstein's
cosmological constant) by\footnote{We emphasize that there is no free parameter
in the logotropic model  \cite{epjp,lettre,ouf}. In particular, the prefactor
$0.0207$ in Eq. (\ref{sc12}) is predicted by our model.}
\begin{eqnarray}
\label{sc12}
\Sigma_0=0.0207\, \frac{c\sqrt{\Lambda}}{G}.
\end{eqnarray}
Now that this formula has been isolated, we can use it independently
from the theory developed in
\cite{epjp,lettre,ouf}.
Combining Eqs. (\ref{sc11}), (\ref{sc1b}) and (\ref{sc12}) we find that the mass
of the
noninteracting bosonic particle is given by  
\begin{eqnarray}
\label{sc14}
m=1.41\times 10^{11} \, m_{\Lambda},
\end{eqnarray}
where
\begin{equation}
m_{\Lambda}= \frac{\hbar\sqrt{\Lambda}}{c^2}=2.08\times 10^{-33}\,
{\rm eV/c^2}.
\label{sc15}
\end{equation}
This mass  scale is often interpreted as the
smallest mass of the bosons predicted by string theory \cite{axiverse} or as
the upper bound on the mass of the graviton
\cite{graviton}.\footnote{It is
simply
obtained by equating the Compton wavelength of the particle $\lambda_C=\hbar/mc$
with the Hubble radius $R_\Lambda=c/H_0\sim c/\sqrt{\Lambda}$ (the typical size
of the visible
Universe) giving $m_{\Lambda}=\hbar H_0/c^2\sim \hbar \sqrt{\Lambda}/c^2$. 
By comparison, if we identify the Compton wavelength 
$\lambda_C=\hbar/mc$ with the Schwarzschild radius $r_S\sim Gm/c^2$ we get the
Planck mass $M_P=(\hbar c/G)^{1/2}$.} This is also the mass of an hypothetical
particle called the cosmon \cite{ouf}. We see that it fixes the mass scale of
the DM
particle in the case where it is a noninteracting boson. Nevertheless, there
is a huge proportionality factor between
them, of the order of $10^{11}$. We have also found this result in Appendix F
of \cite{ouf} from considerations based on the Jeans
instability. These considerations are further developed in Appendix
\ref{sec_mdm} and generalized to the case of self-interacting bosons and
fermions.

{\it Remark:} Returning to original variables, and using Eqs. (\ref{iso1}),
(\ref{ana31b}) and (\ref{t6}), the core mass -- halo mass relation of DM
halos made of noninteracting bosons can be written as
\begin{eqnarray}
\label{ana33}
M_c\sim 2.29 \left (\frac{\hbar^2M_h}{Gm^2r_h}\right )^{1/2}\sim 2.64 \left
(\frac{\hbar^4\Sigma_0M_h}{G^2m^4}\right )^{1/4}
\end{eqnarray}
leading to $M_c\propto M_h^{1/4}\propto
M_v^{1/3}$.

\subsection{Self-interacting bosons}
\label{sec_qmi}

A self-gravitating BEC with a repulsive self-interaction in
the TF limit ($\hbar=0$) is equivalent to a polytrope of index $n=1$ with a
polytropic constant given by Eq. (\ref{tf2}) (see Sec. \ref{sec_tf}). Using Eqs.
(\ref{sat1}), (\ref{sat2}) and the results of
Appendix \ref{sec_p}, we
get
\begin{eqnarray}
\label{sat7}
\frac{M_c}{(M_h)_{\rm min}}\sim 1.16\, 
\left\lbrack \frac{M_h}{(M_h)_{\rm
min}}\right\rbrack^{1/2}.
\end{eqnarray}
On the other hand, using Eqs. (\ref{sipv3}) and (\ref{sc5}), we find that
\begin{eqnarray}
\label{sat9}
B_{1}&=&\frac{1}{1.76\, \Sigma_0} \left\lbrack \frac{4}{3}\pi
\zeta(0)\rho_{m,0}\right\rbrack^{2/3} (13.0)^{1/3}\left
(\frac{a_s\hbar^{2}\Sigma_0}{Gm^{3}}\right )^{1/3}\nonumber\\
&=& 3.43\times 10^{-3}.
\end{eqnarray}
Therefore, Eq. (\ref{sc4}) takes the form
\begin{eqnarray}
\label{sc4si}
\frac{M_h}{(M_h)_{\rm min}}\sim  3.43\times 10^{-3} \left\lbrack
\frac{M_v}{(M_h)_{\rm
min}}\right\rbrack^{4/3}.
\end{eqnarray}
Combining Eqs. (\ref{sat7}) and
(\ref{sc4si}),  we obtain the core mass -- halo
mass relation
\begin{eqnarray}
\label{sat8}
\frac{M_c}{(M_h)_{\rm min}}\sim 6.79\times 10^{-2}\, 
\left\lbrack \frac{M_v}{(M_h)_{\rm
min}}\right\rbrack^{2/3}.
\end{eqnarray}
It exhibits the fundamental scaling $M_c\propto M_v^{2/3}$. This is a new
theoretical prediction \cite{modeldm} that still has to be tested with direct
numerical simulations of self-interacting bosons.

On the other hand, reversing Eq. (\ref{sat9}) following the remark at the end
of Sec. \ref{sec_mcmv}, we get
\begin{eqnarray}
\frac{m^3}{a_s}=\frac{13.0}{(1.76)^{3}} \left\lbrack \frac{4}{3}\pi
\zeta(0)\rho_{\rm m,0}\right\rbrack^{2}\frac{\hbar^{2}}{G\Sigma_0^{2}}
\frac{1}{B_{1}^{3}}.
\end{eqnarray}
Using Eqs.  (\ref{sc1b}) and (\ref{sc12}) we obtain
\begin{eqnarray}
\frac{a_s}{m^3}=5.34\times 10^{-15}\,  \frac{r_{\Lambda}}{m_{\Lambda}^3},
\end{eqnarray}
where
\begin{eqnarray}
\label{rlml3}
\frac{r_{\Lambda}}{m_{\Lambda}^3}=\frac{2Gc^2}{\Lambda\hbar^2}=6.11\times
10^{17}\, {\rm fm\, (eV/c^2)^{-3}}.
\end{eqnarray}
In this expression, $m_{\Lambda}$ is the mass of the cosmon given by Eq. 
(\ref{sc15}) and 
\begin{eqnarray}
r_{\Lambda}=\frac{2Gm_{\Lambda}}{c^2}=5.51\times 10^{-96}\, {\rm m}
\end{eqnarray}
is the gravitational radius of the cosmon \cite{ouf}.

{\it Remark:} Returning to original variables, and using Eqs.
(\ref{tf3}), (\ref{iso1}) and 
(\ref{ana31b}), the core mass -- halo mass relation of DM
halos made of self-interacting bosons in the TF limit can be written as
\begin{eqnarray}
\label{ana32}
M_c\sim \pi \left (\frac{a_s\hbar^2M_h^2}{Gm^3r_h^2}\right )^{1/2}\sim
4.17 \left (\frac{a_s\hbar^2\Sigma_0
M_h}{Gm^3}\right )^{1/2}
\end{eqnarray}
leading to $M_c\propto M_h^{1/2}\propto
M_v^{2/3}$.

\subsection{Fermions}
\label{sec_qmf}

A fermionic core is equivalent to a polytrope of index
$n=3/2$ with a polytropic constant given by Eq. (\ref{fdm2}) (see Sec.
\ref{sec_fdm}). Using Eqs. (\ref{sat1}), (\ref{sat2}) and the results of
Appendix \ref{sec_p}, we
get 
\begin{eqnarray}
\label{sat3}
\frac{M_c}{(M_h)_{\rm min}}\sim 1.50\, 
\left\lbrack \frac{M_h}{(M_h)_{\rm
min}}\right\rbrack^{3/8}.
\end{eqnarray}
On the other hand, using Eqs. (\ref{fpv3}) and (\ref{sc5}), we find that
\begin{eqnarray}
\label{sc6ff}
B_{3/2}&=&\frac{1}{1.76\, \Sigma_0} \left\lbrack \frac{4}{3}\pi
\zeta(0)\rho_{m,0}\right\rbrack^{2/3} (4.47)^{1/3}\left
(\frac{\hbar^{12}\Sigma_0^3}{G^6m^{16}}\right )^{1/15}\nonumber\\
&=&2.70\times 10^{-3}.
\end{eqnarray}
Therefore, Eq. (\ref{sc4}) takes the form
\begin{eqnarray}
\label{sc4f}
\frac{M_h}{(M_h)_{\rm min}}\sim  2.70\times 10^{-3} \left\lbrack
\frac{M_v}{(M_h)_{\rm
min}}\right\rbrack^{4/3}.
\end{eqnarray}
Combining Eqs. (\ref{sat3}) and (\ref{sc4f}), we obtain the core mass -- halo
mass
relation 
\begin{eqnarray}
\label{sat4}
\frac{M_c}{(M_h)_{\rm min}}\sim 0.163\, 
\left\lbrack \frac{M_v}{(M_h)_{\rm
min}}\right\rbrack^{1/2}.
\end{eqnarray}
It  exhibits the fundamental scaling $M_c\propto M_v^{1/2}$. This theoretical
scaling, previously
given in the form of Eq. (\ref{sat3}) in Appendix H of \cite{clm2},
is
consistent with the scaling found numerically by  Ruffini {\it et al.}
\cite{rar} (they find an exponent equal to $0.52$
instead of $1/2$).

On the other hand, reversing Eq. (\ref{sc6ff}) following the remark at the end
of Sec. \ref{sec_mcmv}, we get
\begin{equation}
\label{sc7}
m=\frac{(4.47)^{5/16}}{(1.76)^{15/16}} \left\lbrack \frac{4}{3}\pi
\zeta(0)\rho_{m,0}\right\rbrack^{5/8}\frac{\hbar^{3/4}}{\Sigma_0^{3/4}G^{3/8}}
\frac{1}{B_{3/2}^{15/16}}.
\end{equation}
Using Eqs. (\ref{sc1b}) and (\ref{sc12}) we obtain
\begin{eqnarray}
\label{sc8}
m=3.38\times 10^4 \, m_{\Lambda}^*,
\end{eqnarray}
where
\begin{eqnarray}
\label{sc9}
m_{\Lambda}^*=\left (\frac{\Lambda\hbar^3}{Gc^3}\right
)^{1/4}=\sqrt{m_{\Lambda}M_P}=5.04\times 10^{-3}\, {\rm eV/c^2}.\nonumber\\
\end{eqnarray}
To our knowledge, this mass scale has not
been introduced before. It is the geometric mean of the cosmon mass
$m_{\Lambda}$ given by Eq. 
(\ref{sc15}) and the Planck mass $M_P=(\hbar c/G)^{1/2}=2.18\times 10^{-5}\,
{\rm g}$.\footnote{In comparison, using the results of
\cite{ouf}, the mass of
the electron
may be written in terms of the fundamental constants of physics  as
$m_e=1.03\alpha(m_{\Lambda} M_P^2)^{1/3}=9.11\times 10^{-28}\, {\rm g}$ where
$\alpha\simeq 1/137$ is the fine structure constant.}

{\it Remark:} Comparing Eqs. (\ref{sc10}), (\ref{sat9}) and
(\ref{sc6ff}), we note
that the value of $B_n$ defined by Eq. (\ref{sc5}) is relatively insensitive on
the value of the polytropic index $n$ for the cases
contemplated. This is because the parameters have been chosen so that
$(M_h)_{\rm min}$ typically has a fixed value ($\sim 10^{8} M_{\odot}$).

{\it Remark:} Returning to original variables, and using Eqs.
(\ref{fdm3}), (\ref{iso1}) and 
(\ref{ana31b}), the core mass -- halo mass relation of DM
halos made of fermions may be written as
\begin{equation}
\label{ana34}
M_c\sim 3.10\frac{\hbar^{3/2}}{m^2}\left (\frac{M_h}{Gr_h}\right
)^{3/4}\sim 3.83 \frac{\hbar^{3/2}}{m^2} \left (\frac{M_h
\Sigma_0}{G^2}\right )^{3/8}
\end{equation}
leading to $M_c\propto M_h^{3/8}\propto
M_v^{1/2}$.

\subsection{Semiclassical limit}
\label{sec_sl}

It is interesting to study how the mass $M_c$, the radius $R_c$,
the velocity dispersion $GM_c/R_c$ and the energy $GM_c^2/R_c$  in the core
behave in the semiclassical limit $\hbar\rightarrow 0$.  For noninteracting
bosons,
using Eq.
(\ref{ana33}), we find $M_c\sim R_c\sim GM_c^2/R_c\sim
\hbar\rightarrow 0$ and $GM_c/R_c\sim 1$. For self-interacting
bosons, using Eq. (\ref{ana32}), we find
$M_c\sim R_c\sim GM_c^2/R_c\sim
\hbar\rightarrow 0$ and $GM_c/R_c\sim 1$. 
For fermions, using Eq.
(\ref{ana34}), we find $M_c\sim
R_c\sim GM_c^2/R_c\sim
\hbar^{3/2}\rightarrow 0$ and $GM_c/R_c\sim 1$. Therefore, in the semiclassical
limit $\hbar\rightarrow 0$, the mass $M_c$, the size $R_c$ and the energy
$GM_c^2/R_c$ of the quantum core go to zero while the velocity dispersion
$GM_c/R_c$ remains finite.

\section{Gaussian ansatz}
\label{sec_ga}

In this section, we obtain the core mass -- halo mass relation of DM halos by
combining the
velocity dispersion tracing relation (\ref{ana31b}) with the approximate
core mass-radius relation of a self-gravitating BEC at $T=0$ obtained in
\cite{prd1}
from a Gaussian ansatz.
This allows us to recover the preceding results and to generalize them to
the case of a repulsive
self-interaction ($a_s>0$), without making the TF approximation, and to the case
of an attractive self-interaction ($a_s<0$). Throughout this section, we
introduce appropriate normalizations in order to clearly see the physical origin
of the parameters and be able to refine the numerical applications when more
precise data will be available.

\subsection{Core mass-radius relation}
\label{sec_mrr}

Using a Gaussian ansatz, it is found in \cite{prd1} that the approximate
mass-radius relation of a self-gravitating BEC at $T=0$ (ground state) is given
by
\begin{equation}
\label{mrr1}
M_c=\frac{2\sigma}{\nu}\frac{\frac{\hbar^2}{Gm^2R_c}}{1-\frac{6\pi\zeta
a_s\hbar^2}{\nu Gm^3 R_c^2}}
\end{equation}
with the coefficients $\sigma=3/4$, $\zeta=1/(2\pi)^{3/2}$ and
$\nu=1/\sqrt{2\pi}$. Inversely, the radius can be expressed in terms of the
mass as
\begin{equation}
\label{mrr2}
R_c=\frac{\sigma}{\nu}\frac{\hbar^2}{GM_cm^2}\left
(1\pm\sqrt{1+\frac{6\pi\zeta\nu}{\sigma^2}\frac{GmM_c^2a_s}{\hbar^2}}\right )
\end{equation}
with $+$ when $a_s>0$ and with $\pm$ when $a_s<0$. The
results of \cite{prd1} describe the ``minimum halo'' (ground state) or the
quantum core of larger halos. For noninteracting BECs
($a_s=0$), the mass-radius relation (\ref{mrr1}) reduces to
\begin{equation}
\label{mrr1b}
M_c=\frac{2\sigma}{\nu}\frac{\hbar^2}{Gm^2R_c}.
\end{equation}
In the
repulsive case ($a_s>0$) the mass-radius relation is monotonic (see Fig. 2 of
\cite{prd1}). There is a minimum radius 
\begin{equation}
\label{mrr1c}
(R_c)_{\rm min}=\left (\frac{6\pi\zeta}{\nu}\right )^{1/2}\left
(\frac{a_s\hbar^2}{Gm^3}\right )^{1/2}
\end{equation}
corresponding to $M_c\rightarrow +\infty$ (TF limit).
Measuring the DM particle  mass in units of
$10^{-22}{\rm eV/c^2}$ and the scattering length in units of $10^{-62}\, {\rm
fm}$, we get $(R_c)_{\rm min}=963\, a_s^{1/2}m^{-3/2}\, {\rm pc}$. For $R\gg
(R_c)_{\rm min}$
we are in the noninteracting
limit (\ref{mrr1b}). In the attractive case
($a_s<0$)  the mass-radius relation is nonmonotonic
(see Fig. 3 of
\cite{prd1}).  There is a maximum mass
\begin{equation}
\label{mrr3}
(M_{c})_{\rm max}=\left (\frac{\sigma^2}{6\pi\zeta\nu}\right
)^{1/2}\frac{\hbar}{\sqrt{Gm|a_s|}}
\end{equation}
corresponding to the radius
\begin{equation}
\label{mrr4}
(R_c)_*=\left (\frac{6\pi\zeta}{\nu}\right )^{1/2}\left
(\frac{|a_s|\hbar^2}{Gm^3}\right )^{1/2}.
\end{equation}
Measuring the DM particle  mass in units of
$10^{-22}{\rm eV/c^2}$ and the scattering length in units of $10^{-62}\, {\rm
fm}$, we get $(M_{c})_{\rm max}=1.67\times 10^8\, m^{-1/2}|a_s|^{-1/2}\,
M_{\odot}$ and $(R_c)_{\rm min}=963\, \, |a_s|^{1/2}m^{-3/2}\, {\rm pc}$.

No equilibrium state exists with a mass $M_c>(M_{c})_{\rm max}$. For
$M_c<(M_{c})_{\rm max}$ the branch $R_c>(R_c)_*$ (corresponding to the solutions
(\ref{mrr2}) with the sign $+$) is stable
and the branch $R_c<(R_c)_*$ (corresponding to the solutions (\ref{mrr2}) with
the sign $-$) is 
unstable. For $R\gg (R_c)_*$ we are in the noninteracting
limit (\ref{mrr1b}) and for $R\ll (R_c)_*$ we are in the (unstable)
nongravitational limit where
\begin{equation}
\label{mrr4b}
M_c=\frac{\sigma}{3\pi\zeta}\frac{mR_c}{|a_s|}.
\end{equation}

\subsection{The minimum halo mass}
\label{sec_mhm}

\begin{figure}[!h]
\begin{center}
\includegraphics[clip,scale=0.3]{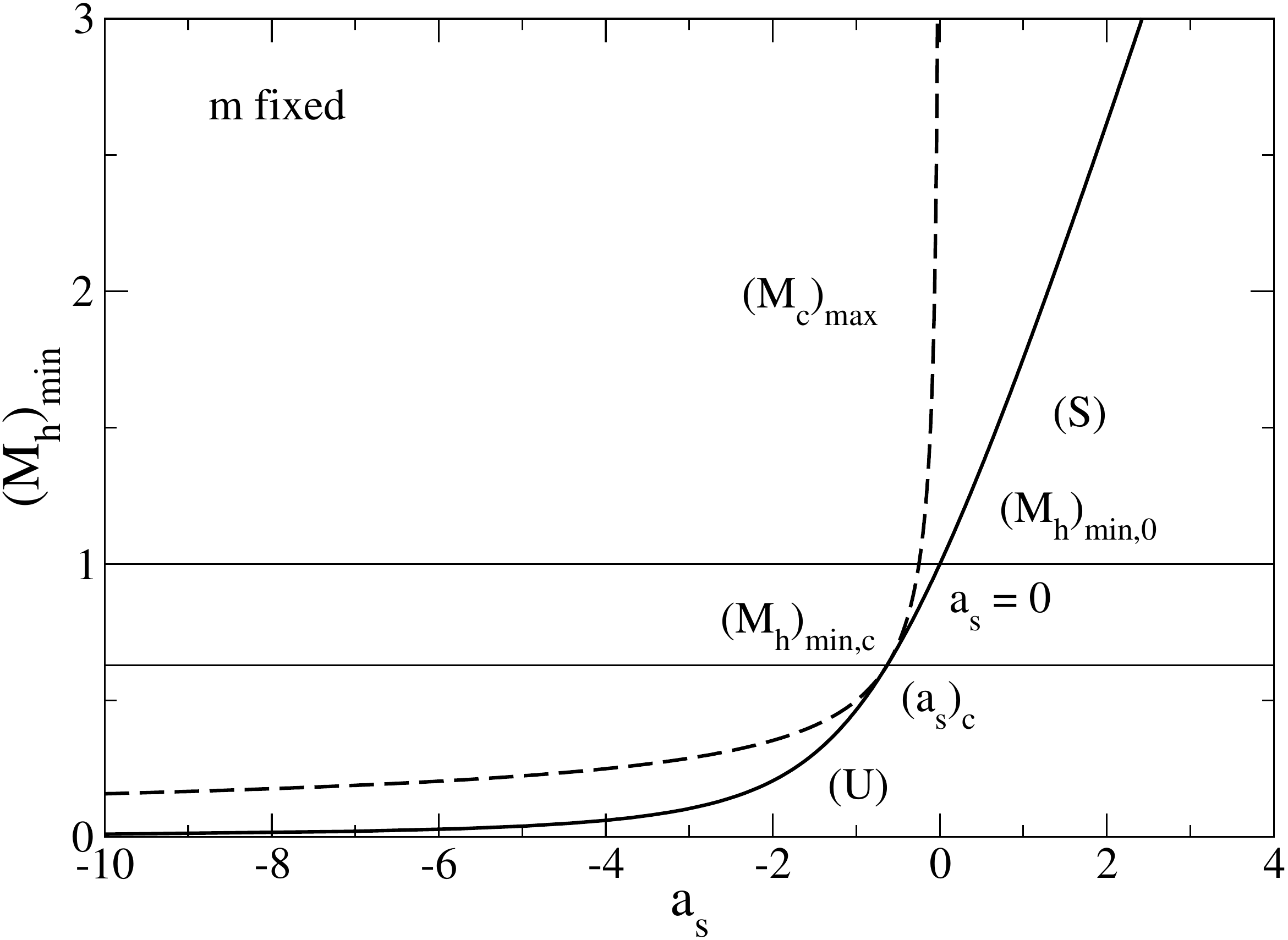}
\caption{Minimum halo mass $(M_{h})_{\rm min}$ as a function of the scattering
length $a_s$ for a fixed mass $m$ of the DM particle (solid line). We have also
plotted the maximum mass $(M_c)_{\rm
max}$ \cite{prd1} as a function of $a_s<0$ (dashed line). The intersection
between these curves determines the critical minimum halo mass $(M_{h})_{\rm
min,c}$. The mass is normalized
by
$(M_{h})_{\rm min,0}$ and the scattering length by $a_*$. The stable part of
the curve starts at the critical minimum halo point ($(a_s)_c,(M_{h})_{\rm
min,c}$).}
\label{am}
\end{center}
\end{figure}

We first determine the minimum halo mass $(M_h)_{\rm min}$. As explained
previously, the minimum halo corresponds to the
ground state ($T=0$) of the self-gravitating BEC. In our approximate approach
we
write the surface density as
\begin{equation}
\label{mrr5}
\Sigma_0=\alpha\frac{M_c}{R_c^2},
\end{equation}
where $\alpha$ is a constant of order unity (in the numerical applications we
take $\alpha=1/1.76$ for the reason explained in footnote 17). Eliminating
$R_c$ between
Eqs.
(\ref{mrr1}) and (\ref{mrr5}), and treating $\Sigma_0$ as a universal constant,
we get the minimum halo mass $(M_{h})_{\rm min}$ as a function of $m$ and $a_s$
(we recall that $M_h=M_c$ for the ground state since there is no isothermal
atmosphere by definition). For noninteracting BECs 
($a_s=0$) we find that the minimum halo mass is 
\begin{equation}
\label{mrr6}
(M_h)_{\rm min,0}=\frac{2^{2/3}\sigma^{2/3}}{\nu^{2/3}\alpha^{1/3}}\left
(\frac{\hbar^4\Sigma_0}{G^2m^4}\right )^{1/3}.
\end{equation}
The prefactor is $2.92$. This result can be compared with Eq.
(\ref{nipv3}). Measuring the DM particle  mass in units of
$10^{-22}{\rm eV/c^2}$, we get $(M_h)_{\rm min,0}=2.95\times 10^8 m^{-4/3}
M_{\odot}$. In the
general case,
valid for an arbitrary value of $a_s$, we find that the minimum halo mass
$(M_{h})_{\rm min}$ is determined by the equation
\begin{equation}
\label{mrr7}
\frac{a_s}{a_*}=\frac{(M_{h})_{\rm min}}{(M_{h})_{\rm
min,0}}-\sqrt{\frac{(M_{h})_{\rm min,0}}{(M_{h})_{\rm min}}},
\end{equation}
where we have introduced the appropriate scattering length scale
\begin{equation}
\label{mrr8}
a_*=\frac{2^{2/3}\sigma^{2/3}\alpha^{2/3}\nu^{1/3}}{6\pi\zeta}\left
(\frac{Gm^5}{\hbar^2\Sigma_0^2}\right )^{1/3}.
\end{equation}
The prefactor is $0.553$.  Measuring the DM particle  mass in
units of
$10^{-22}{\rm eV/c^2}$, we get $a_*=1.28\times 10^{-62} m^{5/3}\, {\rm fm}$.
For $a_s=0$ we recover $(M_{h})_{\rm
min}=(M_{h})_{\rm min,0}$. More generally, the noninteracting limit is valid
for $|a_s|\ll a_*$. The relation $(M_{h})_{\rm min}/(M_h)_{\rm min,0}$ vs
$a_s/a_*$ is plotted in Fig. \ref{am}. For a given mass $m$, we see that
$(M_{h})_{\rm min}$ is larger than $(M_{h})_{\rm min,0}$ when $a_s>0$ and
smaller  when $a_s<0$.

In the repulsive case, for $a_s\gg a_*$, we have
\begin{equation}
\label{mrr9}
\frac{(M_{h})_{\rm min}}{(M_{h})_{\rm
min,0}}\sim \frac{a_s}{a_*}.
\end{equation}
This corresponds to the TF limit. Returning to the original variables, we
obtain 
\begin{equation}
\label{mrr9b}
(M_{h})_{\rm min}\sim
\frac{6\pi\zeta}{\alpha\nu}\frac{ a_s\hbar^2\Sigma_0}{Gm^3}.
\end{equation}
The prefactor is $5.28$. This result can be compared with Eq.
(\ref{sipv3}). Measuring the DM particle  mass in units of
$10^{-22}{\rm eV/c^2}$ and the scattering length in units of $10^{-62}\, {\rm
fm}$, we get $(M_{h})_{\rm min}=2.30\times 10^8\, a_s m^{-3}\,
M_{\odot}$.

In the attractive case, for $|a_s|\gg a_*$, we have
\begin{equation}
\label{mrr10}
\frac{(M_{h})_{\rm min}}{(M_{h})_{\rm
min,0}}\sim \left (\frac{a_*}{a_s}\right )^2.
\end{equation}
This corresponds to the nongravitational limit in which the configurations are
unstable. Returning to the original variables, we obtain 
\begin{equation}
\label{mrr10b}
(M_{h})_{\rm min}\sim
\frac{4\sigma^2\alpha}{(6\pi\zeta)^2}\frac{m^2}{\Sigma_0 a_s^2}.
\end{equation}
The prefactor is $0.892$. On the other hand, the normalized
maximum mass (\ref{mrr3}) can be written as
\begin{equation}
\label{mrr11}
\frac{(M_{c})_{\rm max}}{(M_h)_{\rm min,0}}=\frac{1}{2}\left
(\frac{a_*}{|a_s|}\right )^{1/2}.
\end{equation}
We find that the minimum halo is critical (i.e. $(M_h)_{\rm min}=(M_c)_{\rm
max}$) for
\begin{equation}
\label{mrr11a}
\frac{(a_s)_c}{a_*}=-\frac{1}{2^{2/3}},\qquad \frac{(M_{h})_{\rm
min,c}}{(M_h)_{\rm min,0}}=\frac{1}{2^{2/3}}.
\end{equation}
These relations determine the minimum scattering length $(a_s)_c$ of the DM
particle and the minimum mass $(M_{h})_{\rm
min,c}$ of the minimum halo. 
Returning to the
original variables, we obtain
\begin{equation}
\label{mrr11b}
(a_s)_c=-\frac{\sigma^{2/3}\alpha^{2/3}\nu^{1/3}}{6\pi\zeta}
\left (\frac{Gm^5}{\hbar^2\Sigma_0^2}\right )^{1/3},
\end{equation}
\begin{equation}
\label{mrr11c}
(M_{h})_{\rm
min,c}=\frac{\sigma^{2/3}}{\alpha^{1/3}\nu^{2/3}}\left (\frac{\hbar^{4}
\Sigma_0}{G^{2}m^{4}}\right )^{1/3}.
\end{equation}
The prefactors are $0.348$ and $1.84$. Measuring the DM
particle mass in units of
$10^{-22}{\rm eV/c^2}$, we get $(a_s)_c=-8.06\times 10^{-63} m^{5/3}\, {\rm
fm}$ and $(M_h)_{\rm min,c}=1.86\times 10^8 m^{-4/3}
M_{\odot}$. These
relations can be directly obtained by writing $\Sigma_0=\alpha (M_c)_{\rm
max}/(R_c)^2_*$ and using Eqs. (\ref{mrr3}) and (\ref{mrr4}). The minimum
halo is stable only for $a_s\ge (a_s)_c$. It has a mass 
 $(M_{h})_{\rm min}\ge (M_{h})_{\rm min,c}$. When $a_s< (a_s)_c$ the minimum
halo is unstable (it lies on the branch $R_c< (R_c)_*$ of the
mass-radius relation).  We note that $(M_{h})_{\rm min,c}$ is relatively close
to $(M_{h})_{\rm min,0}$. Therefore, when $(a_s)_c<a_s<0$, the minimum halo
mass $(M_{h})_{\rm min}$ is always of the order of $(M_{h})_{\rm min,0}$ (see
the stripe in Fig.
\ref{am}).

\begin{figure}[!h]
\begin{center}
\includegraphics[clip,scale=0.3]{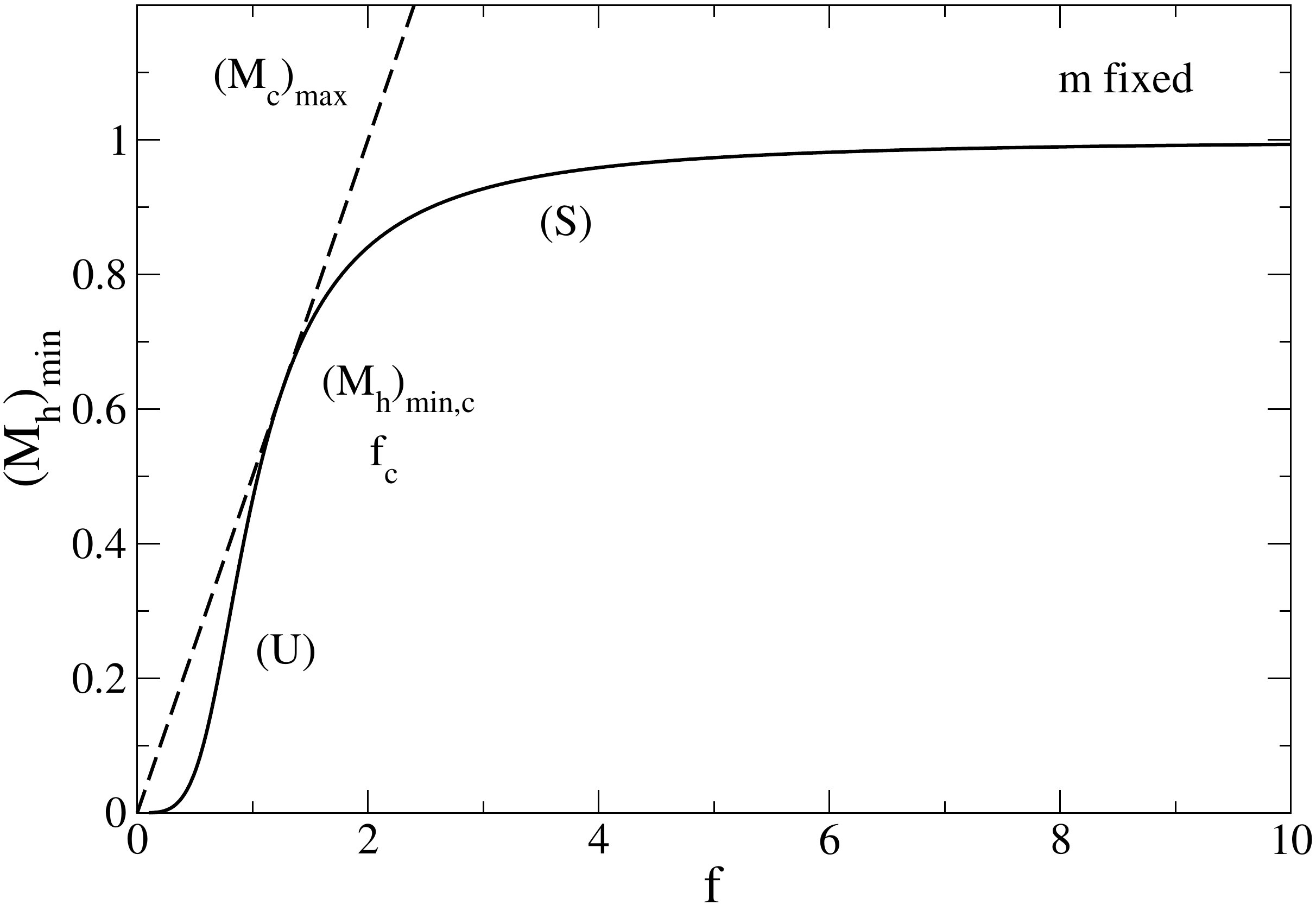}
\caption{Minimum halo mass $(M_{h})_{\rm min}$ as a function of the decay
constant $f$ for a fixed mass $m$ of the DM particle (solid line). We have
also
plotted the maximum mass $(M_c)_{\rm
max}$ \cite{prd1} as a function of $f$ (dashed line). The intersection
between these curves determines the critical minimum halo mass $(M_{h})_{\rm
min,c}$.  The mass is normalized
by
$(M_{h})_{\rm min,0}$ and the decay constant by $f_*$. The stable part of
the curve starts at the critical minimum halo point ($f_c,(M_{h})_{\rm
min,c}$).}
\label{fmh}
\end{center}
\end{figure}

For bosons with an attractive self-interaction, like the axion
\cite{marshrevue}, it is more
convenient to express the results in terms of the decay constant
(see, e.g., \cite{phi6})
\begin{equation}
\label{mrr19}
f=\left (\frac{\hbar c^3 m}{32\pi |a_s|}\right )^{1/2},
\end{equation}
rather than the scattering length $a_s$. Measuring the DM
particle  mass in units of
$10^{-22}{\rm eV/c^2}$ and the scattering length in units of $10^{-62}\, {\rm
fm}$, we get $f=1.40\times 10^{14} m^{1/2}|a_s|^{-1/2}\, {\rm GeV}$. We can
write
\begin{equation}
\label{lmrr19a}
\frac{f}{f_*}=\left (\frac{a_*}{|a_s|}\right
)^{1/2}
\end{equation}
with
\begin{equation}
\label{lmrr19b}
f_*=\frac{(6\pi\zeta)^{1/2}}{(32\pi)^{1/2}2^{1/3}\sigma^{1/3}\alpha^{1/3}\nu^{1/
6}} \frac{\hbar^{5/6}\Sigma_0^{1/3}c^{3/2}}{G^{1/6}m^{1/3}}.
\end{equation}
The prefactor is $0.134$. Measuring the DM particle  mass in
units of
$10^{-22}{\rm eV/c^2}$, we get
$f_*=1.24\times 10^{14} m^{-1/3}\, {\rm GeV}$. Eq. (\ref{mrr7}) can
be rewritten as
\begin{equation}
\label{lmrr7}
\left (\frac{f_*}{f}\right )^2=\sqrt{\frac{(M_{h})_{\rm min,0}}{(M_{h})_{\rm
min}}}-\frac{(M_{h})_{\rm min}}{(M_{h})_{\rm
min,0}}.
\end{equation}
It determines the minimum halo mass $(M_h)_{\rm min}$ in terms of $m$
and $f$. This relation is
plotted in Fig. \ref{fmh}. The 
maximum mass (\ref{mrr3}) can be written as 
\begin{equation}
\label{slmrr11}
(M_c)_{\rm max}=\left (\frac{32\pi\sigma^2}{6\pi\zeta\nu}\right
)^{1/2}\left (\frac{\hbar}{Gc^3}\right )^{1/2}\frac{f}{m}
\end{equation}
or, in normalized form, as
\begin{equation}
\label{lmrr11}
\frac{(M_{c})_{\rm max}}{(M_h)_{\rm min,0}}=\frac{1}{2}\frac{f}{f_*}.
\end{equation}
Using Eqs. (\ref{mrr11a}) and (\ref{lmrr19a}), the
minimum
decay constant
corresponding to the critical minimum halo is
\begin{equation}
\label{lmrr20b}
\frac{f_c}{f_*}=2^{1/3}.
\end{equation}
Returning to the original variables, we find 
\begin{equation}
\label{lmrr19bz}
f_c=\frac{(6\pi\zeta)^{1/2}}{(32\pi)^{1/2}\sigma^{1/3}\alpha^{
1/3 } \nu^ { 1/
6}} \frac{\hbar^{5/6}\Sigma_0^{1/3}c^{3/2}}{G^{1/6}m^{1/3}}.
\end{equation}
The prefactor is $0.169$. Measuring the DM particle  mass in
units of $10^{-22}{\rm eV/c^2}$, we get $f_c=1.56\times 10^{14} m^{-1/3}\, {\rm
GeV}$. Only the upper part of the curve $(M_h)_{\rm min}(f)$
starting from the point ($f_{c},(M_h)_{\rm min,c}$) is stable. The
noninteracting limit corresponds to  $f\gg f_*$.

\subsection{The $m(a_s)$ and $m(f)$ relations}
\label{sec_mas}

\begin{figure}[!h]
\begin{center}
\includegraphics[clip,scale=0.3]{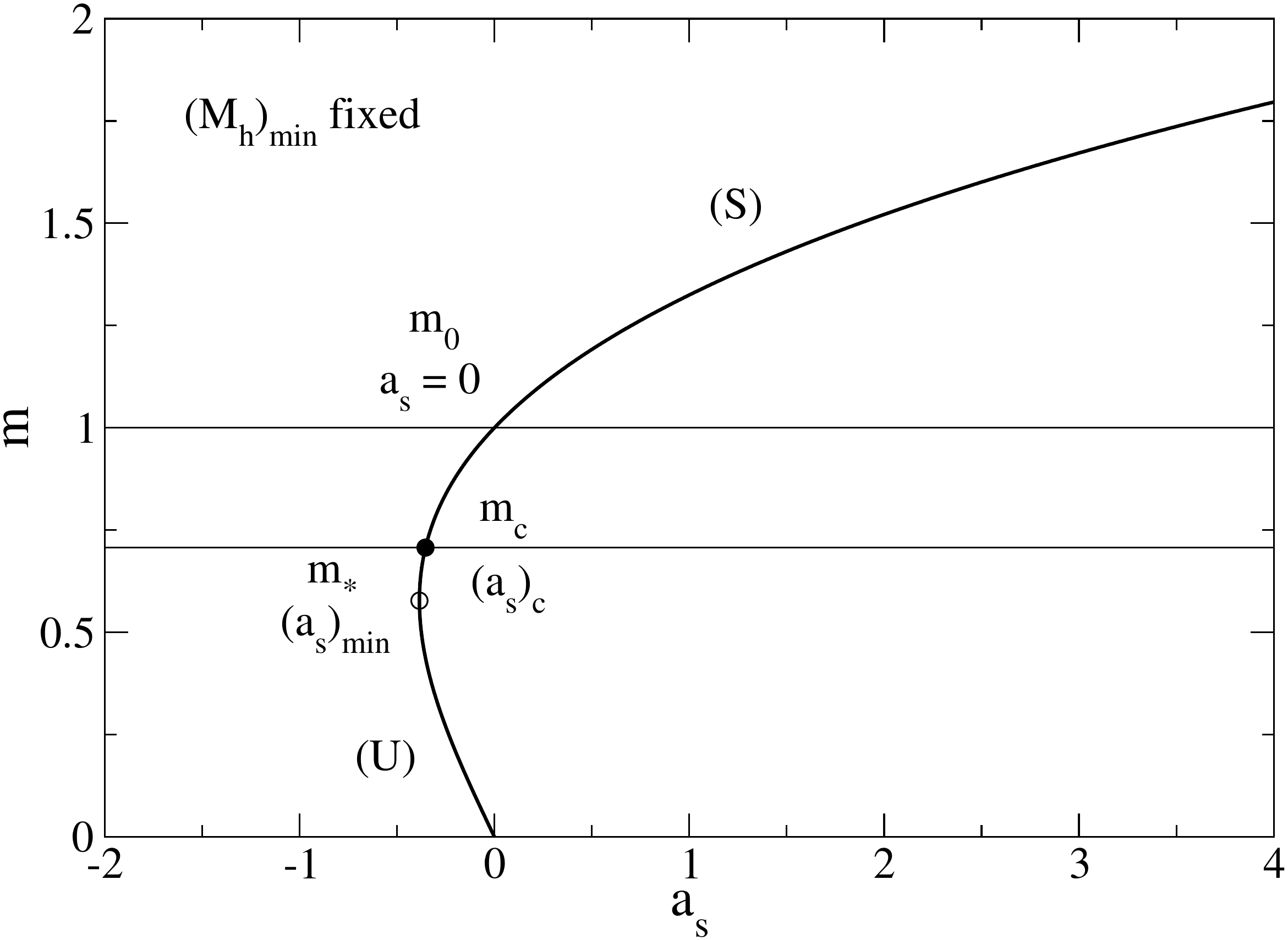}
\caption{Mass $m$ of the DM particle as a function of its scattering
length $a_s$  for a fixed minimum halo mass $(M_h)_{\rm min}$ (solid line). The
mass is normalized by
$m_0$ and the scattering length by $a'_*$. The stable part of
the curve starts at the critical minimum halo point ($(a_s)_c,m_{c}$).}
\label{abism}
\end{center}
\end{figure}

If we consider that the minimum halo mass $(M_h)_{\rm min}$ is known from the
observations, and take $(M_h)_{\rm min}\sim 10^8\, M_{\odot}$ (Fornax) to fix
the ideas, we can use the relation (\ref{mrr7}) to determine the mass $m$ that
the DM particle must have as a function of its scattering length $a_s$ in order
to match the value of
the minimum halo mass $(M_h)_{\rm min}$. For $a_s=0$ we find from Eq.
(\ref{mrr6}) that
\begin{equation}
\label{mas1}
m_0=\frac{2^{1/2}\sigma^{1/2}}{\nu^{1/2}\alpha^{1/4}}\frac{\hbar\Sigma_0^{1/4}}{
G^ {1/2}(M_h)_{\rm min}^{3/4}}.
\end{equation}
The prefactor is $2.23$. In that case we obtain $m_0=2.25\times
10^{-22}\, {\rm eV}/c^2$ which can be compared to Eq. (\ref{ni8})
and Eq. (\ref{mbni}). We can
then write
\begin{equation}
\label{mas2}
\frac{(M_{h})_{\rm min}}{(M_{h})_{\rm
min,0}}=\left (\frac{m}{m_0}\right)^{4/3}.
\end{equation}
On the other hand, we can write
\begin{equation}
\label{mas3}
\frac{a_s}{a_*}=\frac{a_s}{a'_*}\left (\frac{m_0}{m}\right)^{5/3},
\end{equation}
where we have introduced the appropriate scattering length scale
\begin{equation}
\label{mas4}
a'_*=\frac{2^{3/2}\sigma^{3/2}\alpha^{1/4}}{\nu^{1/2}6\pi\zeta}\frac{\hbar}{G^{
1/2}\Sigma_0^{1/4}(M_h)_{\rm min}^{5/4}}.
\end{equation}
The prefactor is $2.11$. We find
$a'_*=4.95\times 10^{-62}\, {\rm fm}$.
Substituting Eqs. (\ref{mas2}) and (\ref{mas3}) into Eq. (\ref{mrr7}), we obtain
\begin{equation}
\label{mas5}
\frac{a_s}{a'_*}=\left (\frac{m}{m_0}\right )^3-\frac{m}{m_0}.
\end{equation}
This relation determines the mass $m$ of the DM particle as a function of
its scattering length $a_s$ in order to yield a minimum halo of mass $(M_h)_{\rm
min}$. It is plotted in Fig. \ref{abism}. For $a_s=0$, we recover $m=m_0$ which
is the mass of a noninteracting boson.
More generally, the noninteracting limit corresponds to $|a_s|\ll a'_*$. We
see that $m$ is larger than $m_0$ when $a_s>0$ and smaller when  $a_s<0$.
Therefore, we can increase the DM particle mass by allowing for a repulsive
self-interaction between the bosons. As
discussed in Appendix D.4 of \cite{suarezchavanis3} this could alleviate some
tensions   with observations of the Lyman-$\alpha$ forest
encountered in the noninteracting model \cite{hui}. By contrast, an attractive
self-interaction implies a (slightly) smaller DM particle mass and may therefore
be even
more in conflict with observations of the Lyman-$\alpha$ forest. As a result, a
repulsive self-interaction ($a_s>0$) is priviledged over an attractive
self-interaction ($a_s<0$). In this respect, we recall that
theoretical models of particle physics  usually lead to particles with
an attractive self-interaction (e.g., the QCD axion). However, some authors
\cite{fan,reig} have pointed out the possible existence of particles with a
repulsive self-interaction (e.g. the light majoron).

In the repulsive case, for $a_s\gg a'_*$, we have
\begin{equation}
\label{mas6}
\frac{m}{m_0}\sim \left (\frac{a_s}{a'_*}\right )^{1/3}.
\end{equation}
This corresponds to the TF limit. Returning to the
original variables, we obtain 
\begin{equation}
\label{mas7}
\frac{a_s}{m^3}\sim \frac{\nu\alpha}{6\pi\zeta}\frac{G(M_h)_{\rm
min}}{\hbar^2\Sigma_0}.
\end{equation}
The prefactor is $0.189$. We find
$a_s/m^3=4.35\times 10^3\, {\rm fm}/({\rm eV}/c^2)^3$ which can be compared
with Eq. (\ref{tf9}) and Eq. (\ref{mbft}).

In the attractive case, the curve $m(a_s)$ presents a turning point at
\begin{equation}
\label{mas8}
\frac{(a_s)_{\rm min}}{a'_*}=-\frac{2}{3\sqrt{3}},\qquad
\frac{m_*}{m_0}=\frac{1}{\sqrt{3}}.
\end{equation}
However, this turning
point does {\it not} correspond to the critical minimum halo for which [see Eqs.
(\ref{mrr11a}), (\ref{mas2}) and (\ref{mas3})]
\begin{equation}
\label{mas9}
\frac{(a_s)_{c}}{a'_*}=-\frac{1}{2^{3/2}},\qquad
\frac{m_c}{m_0}=\frac{1}{\sqrt{2}}.
\end{equation}
Returning to the
original variables, we obtain
\begin{equation}
\label{mas10}
(a_s)_{c}=-\frac{\sigma^{3/2}\alpha^{1/4}}{6\pi\zeta\nu^{1/2}}\frac{\hbar}{G^{
1/2}\Sigma_0^{1/4}(M_h)_{\rm min}^{5/4}},
\end{equation}
\begin{equation}
\label{mas10b}
m_{c}=\frac{\sigma^{1/2}}{\alpha^{1/4}\nu^{1/2}}\frac{\hbar\Sigma_0^{1/4}}{G^{
1/2}(M_h)_{\rm min}^{3/4}}.
\end{equation}
The prefactors are $0.746$ and $1.58$. We find $(a_s)_{c}=-1.75\times 10^{-62}\,
{\rm fm}$ and $m_c=1.59\times
10^{-22}\, {\rm eV}/c^2$ which can be compared with Eq. (D19) in Appendix D of
\cite{suarezchavanis3}.  Only the upper part of the curve $m(a_s)$
starting from the point ($(a_s)_{c},m_c$) is stable. The existence of a
stable minimum
halo in the Universe implies that $a_s\ge
(a_s)_c$.  In that case, the DM particle mass satisfies $m\ge m_c$. We note that
$m_c$ is relatively close
to $m_0$. Therefore, when $(a_s)_c<a_s<0$, the minimum DM particle
mass $m$ is always of the order of $m_0$ (see
the stripe in Fig.
\ref{abism}).

\begin{figure}[!h]
\begin{center}
\includegraphics[clip,scale=0.3]{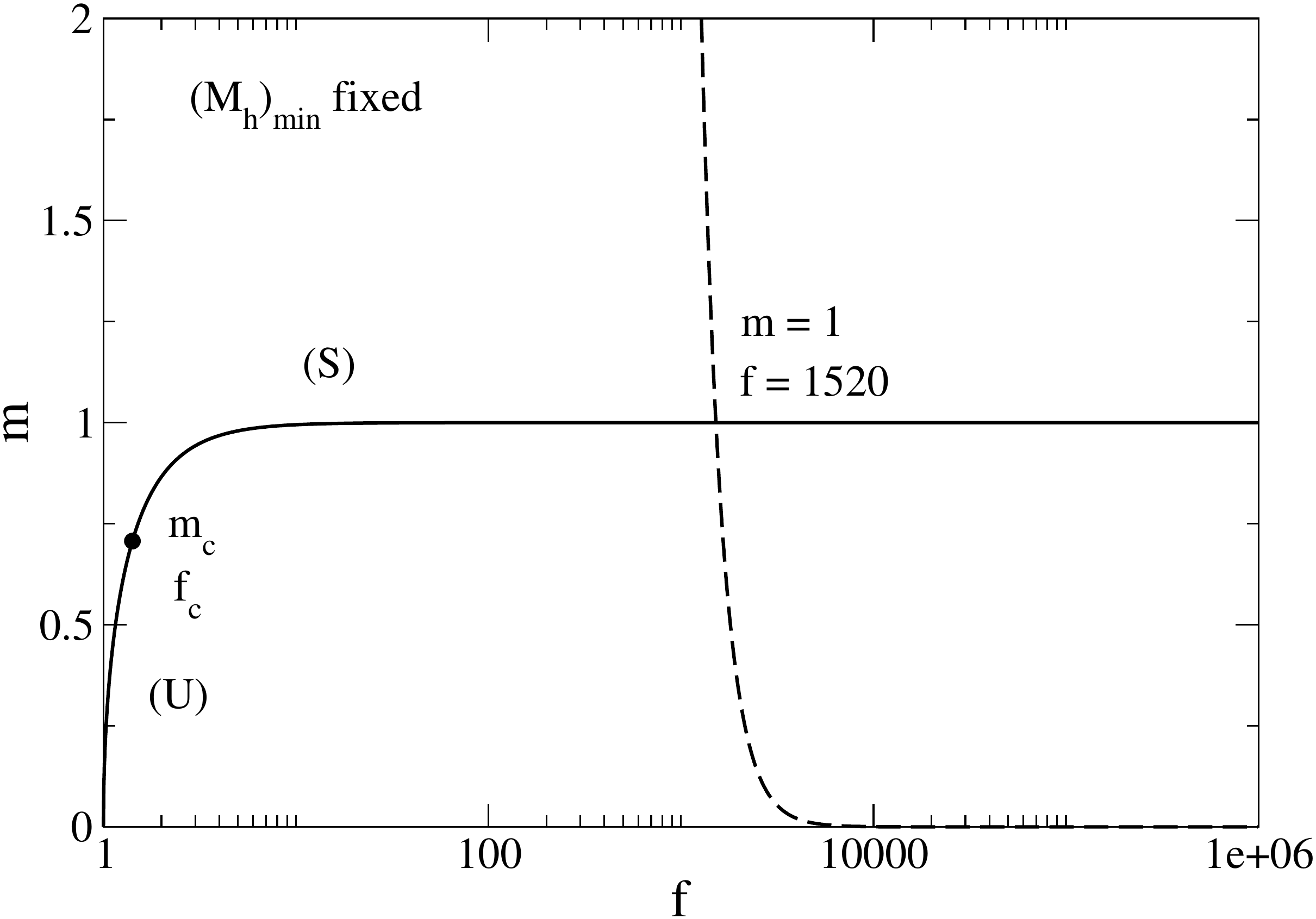}
\caption{Mass $m$ of the DM particle as a function of its decay constant $f$ 
for a fixed minimum halo mass $(M_h)_{\rm min}$ (solid line). The
mass is normalized by
$m_0$ and the decay constant by $f'_*$. The stable part of
the curve starts at the critical minimum halo point ($f_c,m_{c}$). The dashed
line corresponds to Eq. (\ref{hui2}) predicted by particle physics and
cosmology. The intersection between the two curves determines the mass and the 
decay constant of the DM particle.}
\label{fm}
\end{center}
\end{figure}

For bosons with an attractive self-interaction, like the axion
\cite{marshrevue}, it is more
convenient to express the results in terms of the decay constant
(\ref{mrr19}). We can write
\begin{equation}
\label{mrr19a}
\frac{f}{f'_*}=\left (\frac{m}{m_0}\right )^{1/2}\left (\frac{a'_*}{|a_s|}\right
)^{1/2}
\end{equation}
with
\begin{equation}
\label{mrr19b}
f'_*=\frac{(6\pi\zeta)^{1/2}}{8\pi^{1/2}\sigma^{1/2}\alpha^{1/4}}\hbar^{1/2}
\Sigma_0^{1/4}(M_h)_{\rm min}^{1/4}c^{3/2}.
\end{equation}
The prefactor is $0.103$. We find
$f'_*=9.45\times 10^{13}\, {\rm GeV}$. Eq. (\ref{mas5}) can
be rewritten as
\begin{equation}
\label{mrr19c}
\frac{m}{m_0}=\sqrt{1-\left (\frac{f'_*}{f}\right )^2}.
\end{equation}
It determines the relation between $m$ and $f$ in order to have a minimum halo
(ground state) of mass $(M_h)_{\rm min}$. This relation is
plotted in Fig. \ref{fm} (note that $f\ge f'_*$). Using Eqs.
(\ref{mas9}) and (\ref{mrr19a}), the
minimum
decay constant
corresponding to the critical minimum halo is
\begin{equation}
\label{mrr20b}
\frac{f_c}{f'_*}=\sqrt{2}.
\end{equation}
Returning to the original variables, we obtain
\begin{equation}
\label{mrr19bz}
f_c=\frac{\sqrt{2}(6\pi\zeta)^{1/2}}{8\pi^{1/2}\sigma^{1/2}\alpha^{1/4}}\hbar^{
1/2 }
\Sigma_0^{1/4}(M_h)_{\rm min}^{1/4}c^{3/2}.
\end{equation}
The prefactor is $0.146$. We find
$f_c=1.34\times 10^{14}\, {\rm GeV}$. Only the upper part of the curve $m(f)$
starting from the point ($f_{c},m_c$) is stable. The
existence of a stable minimum
halo in the Universe implies that $f\ge f_c$.  In that case, the DM particle
mass satisfies $m\ge m_c$. The noninteracting limit corresponds to $f\gg f'_*$.

There is an interesting by-product of our analysis. Indeed, particle physics and
cosmology lead to the following relation between $f$ and $m$ \cite{hui}:
\begin{equation}
\label{hui1}
\Omega_{\rm axion}\sim 0.1 \left (\frac{f}{10^{17}\, {\rm GeV}}\right )^2\left
(\frac{m}{10^{-22}\, {\rm eV}}\right )^{1/2}.
\end{equation}
Taking $\Omega_{\rm axion}\sim \Omega_{\rm m,0}=0.3089$ and $(M_h)_{\rm
min}\sim 10^{8}\, M_{\odot}$, this relation can be rewritten as
\begin{equation}
\label{hui2}
\frac{m}{m_0}\sim 5.32\times 10^{12}\left (\frac{f}{f'_*}\right )^{-4}.
\end{equation}
This relation is independent from Eq. (\ref{mrr19c}). Equating Eqs.
(\ref{mrr19c}) and (\ref{hui2}), we obtain $f=1520\,
f'_*=1.44\times 10^{17}\, {\rm GeV}$ and $m=m_0=2.25\times
10^{-22}\, {\rm eV}/c^2$. Therefore, we can
determine $f$ and $m$ {\it individually}. We note that $m$ has the same value
as in the noninteracting case while $f$ has a finite value $f=1520\,
f'_*=1.44\times 10^{17}\, {\rm GeV}$. It corresponds to
$a_s=-2.14\times 10^{-68}\, {\rm fm}$. Interestingly, $f$ lies in the range
$10^{16}\, {\rm GeV}\le f\le 10^{18}\, {\rm GeV}$ expected in particle physics
\cite{hui} (we stress that the value of $f$ has been {\it deduced} from our
model based on the core mass-radius relation (\ref{mrr1})). Since $f\gg
f'_*$, we are essentially in the noninteracting regime.

{\it Remark:} we note that for the critical minimum halo the ratio
\begin{equation}
\label{rew}
\frac{(a_s)_c}{m_c^{5/3}}=-\frac{\sigma^{2/3}\alpha^{2/3}\nu^{1/3}}{6\pi\zeta}
\left (\frac{G}{\hbar^2\Sigma_0^2}\right )^{1/3}
\end{equation}
is independent of $(M_h)_{\rm min}$. The prefactor is $0.348$. We find
$(a_s)_c/m_c^{5/3}=-3.75\times 10^{-26}\, {\rm fm}/({\rm eV}/c^2)^{5/3}$.

\subsection{The $M_c(M_h)$ relation}
\label{sec_mcmh}

To obtain the core mass --  halo mass relation $M_c(M_h)$ we use the
velocity dispersion tracing relation (\ref{ana31b}), the core mass-radius
relation
(\ref{mrr2}) and
the halo-mass radius relation $M_h=1.76\, \Sigma_0
r_h^2$  from Eq. (\ref{iso1}). We obtain
\begin{equation}
\label{mrr12}
\frac{M_c}{(M_h)_{\rm min,0}}=\left (\frac{M_h}{(M_h)_{\rm min,0}}\right
)^{1/4}\sqrt{1+\frac{a_s}{a_*}\left (\frac{M_h}{(M_h)_{\rm min,0}}\right
)^{1/2}}.
\end{equation}
For convenience, we have taken $\alpha=1/1.76$. In this manner, when
$M_c=M_h$, we recover the condition (\ref{mrr7}) determining the minimum
halo mass.\footnote{This can be understood as follows. Combining
Eqs. (\ref{ana31b}) and
(\ref{iso1}) we get
\begin{equation}
\label{mrr12b}
\frac{M_c}{R_c}=\sqrt{1.76\, \Sigma_0 M_h}.
\end{equation}
This equation, that uses the relation $M_h\sim 1.76\, \Sigma_0
r_h^2$,  is valid only for sufficiently large halos (see Sec. \ref{sec_iso}).
However, if we extrapolate
this equation to the minimum halo for which $M_h=M_c$, we get $M_c/R_c^2=1.76\,
\Sigma_0$. Comparing this equation with Eq. (\ref{mrr5}) we obtain
$\alpha=1/1.76$.} For $a_s=0$, we get
\begin{equation}
\label{mrr13}
\frac{M_c}{(M_h)_{\rm min,0}}=\left ( \frac{M_h}{(M_h)_{\rm min,0}}\right
)^{1/4}.
\end{equation}
This relation is also valid for $|a_s|\ll a_*$ and
$M_h/(M_h)_{\rm min,0}\ll (a_*/a_s)^2$. We recover the scaling from Eq.
(\ref{sat5}). Returning to the
original variables, we get 
\begin{equation}
\label{mrr13b}
M_c=\frac{2^{1/2}\sigma^{1/2}}{\nu^{1/2}\alpha^{1/4}}\left
(\frac{\hbar^4\Sigma_0M_h}{G^2m^4}\right )^{1/4}.
\end{equation}
The prefactor is $2.23$. For a DM halo of mass $M_h=10^{12}\,
M_{\odot}$ similar to the one that surrounds our Galaxy, we obtain a core
mass $M_c=10^{9}\,
M_{\odot}$  (we have
taken $(M_h)_{\rm min}=10^{8}\,
M_{\odot}$). The corresponding core radius is $R_c=63.5\, {\rm
pc}$ [see Eq. (\ref{mrr1b})]. The quantum core represents a bulge or a nucleus
(it cannot mimic a black hole).

\begin{figure}[!h]
\begin{center}
\includegraphics[clip,scale=0.3]{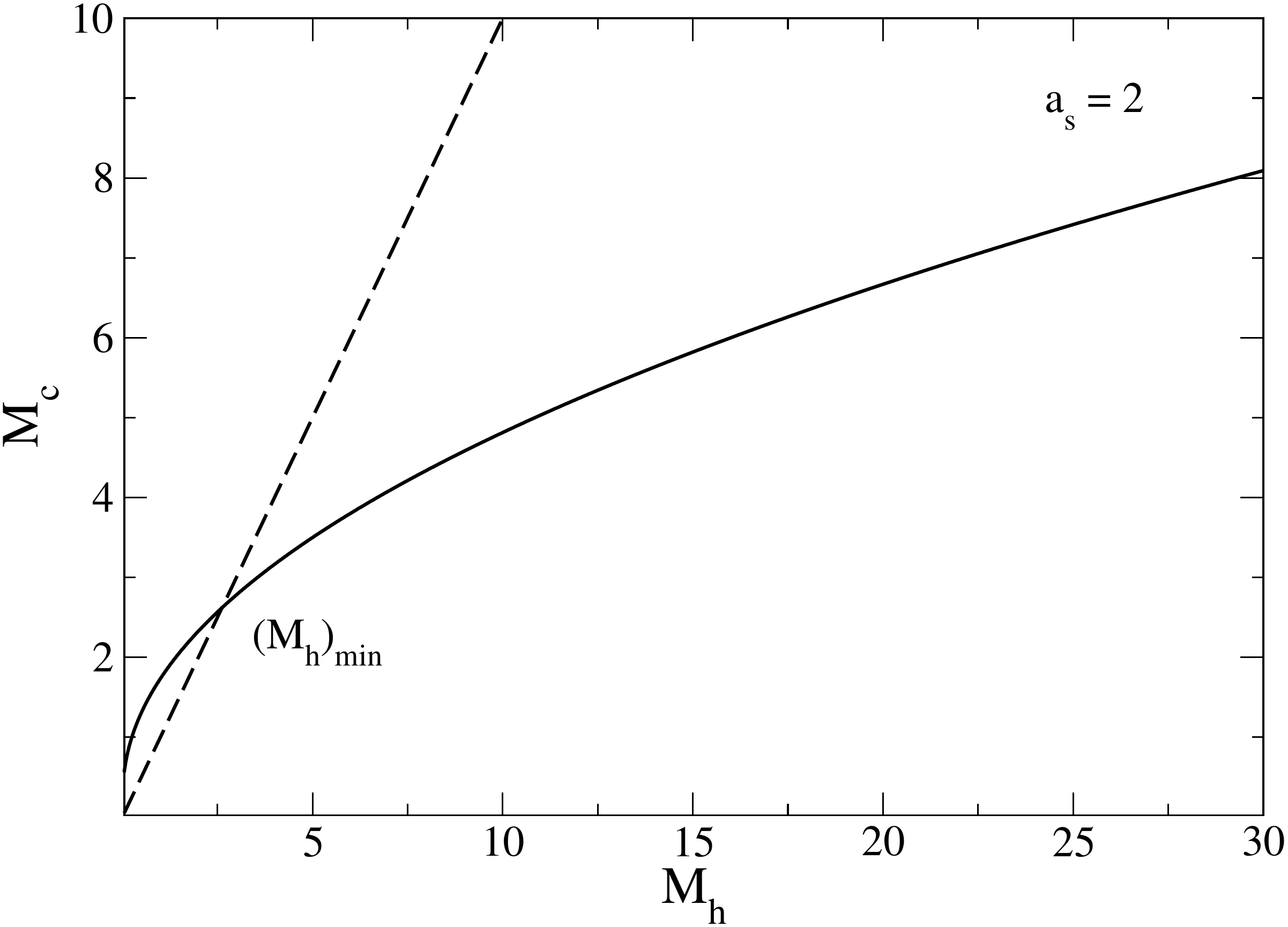}
\caption{Core mass $M_c$ as a function of the halo mass $M_h$ for a repulsive
self-interaction $a_s>0$ (solid line). We have also plotted the relation
$M_c=M_h$ (dashed line) determining the minimum halo mass $(M_{h})_{\rm min}$.
The mass is
normalized by $(M_{h})_{\rm min,0}$ and the scattering length by $a_*$.}
\label{mhmca2}
\end{center}
\end{figure}

\begin{figure}[!h]
\begin{center}
\includegraphics[clip,scale=0.3]{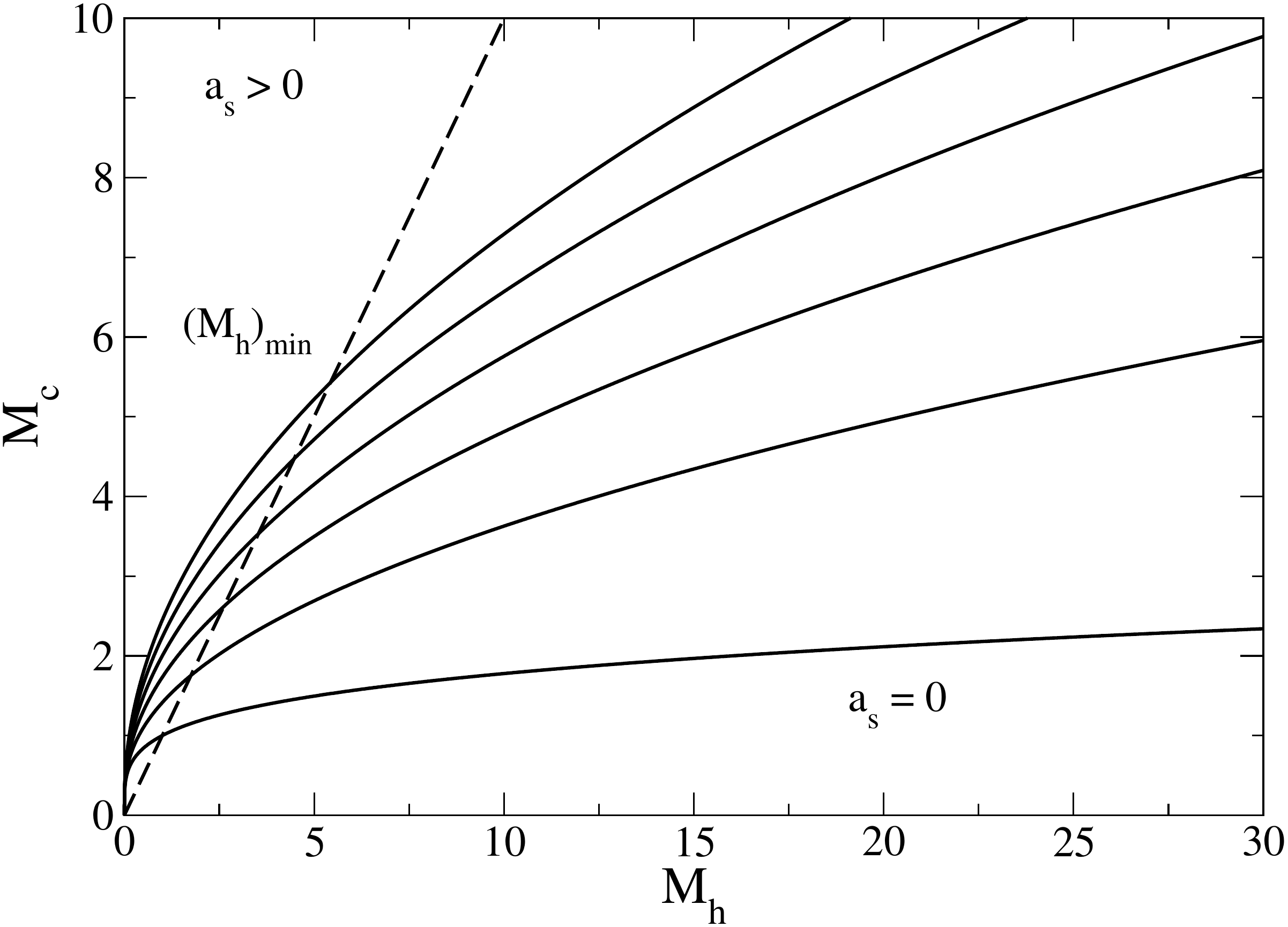}
\caption{Core mass $M_c$ as a function of the halo mass $M_h$ for
different
values of the scattering length $a_s\ge 0$ and a fixed value of the mass
$m$ (solid lines). We have
plotted the position of the minimum halo mass $(M_{h})_{\rm min}$  (dashed
line).
The mass is
normalized by $(M_{h})_{\rm min,0}$. We have indicated the curve
corresponding to the noninteracting case $a_s=0$. }
\label{multipos}
\end{center}
\end{figure}

In the repulsive case, for $a_s\gg a_*$ or $M_h/(M_h)_{\rm
min,0}\gg (a_*/a_s)^2$, we have
\begin{equation}
\label{mrr14}
\frac{M_c}{(M_h)_{\rm min,0}}=\left (\frac{a_s}{a_*}\right )^{1/2}\left
(\frac{M_h}{(M_h)_{\rm min,0}}\right
)^{1/2}.
\end{equation}
This corresponds to the TF limit. Using Eq. (\ref{mrr9}), we obtain 
\begin{equation}
\label{mrr14b}
\frac{M_c}{(M_h)_{\rm min}}=\left
(\frac{M_h}{(M_h)_{\rm min}}\right
)^{1/2}.
\end{equation}
We recover the scaling from Eq. (\ref{sat7}).  Returning to the
original variables, we get 
\begin{equation}
\label{mrr14c}
M_c=\frac{(6\pi\zeta)^{1/2}}{\nu^{1/2}\alpha^{1/2}}\left
(\frac{\hbar^2\Sigma_0a_sM_h}{Gm^3}\right )^{1/2}.
\end{equation}
The prefactor is $2.30$. For a DM halo of mass
$M_h=10^{12}\,
M_{\odot}$ similar to the one that surrounds our Galaxy, we obtain a core
mass $M_c=10^{10}\,
M_{\odot}$ (we have taken $(M_h)_{\rm min}=10^{8}\,
M_{\odot}$). The core radius is $R_c=635\, {\rm
pc}$ [see Eq. (\ref{mrr1c})]. The quantum core represents a bulge or a
nucleus (it cannot mimic a black hole). The
core mass -- halo mass relation for $a_s>0$ is
plotted in Figs. \ref{mhmca2} and \ref{multipos}. These solutions are valid for
$M_h>(M_h)_{\rm
min}$.

\begin{figure}[!h]
\begin{center}
\includegraphics[clip,scale=0.3]{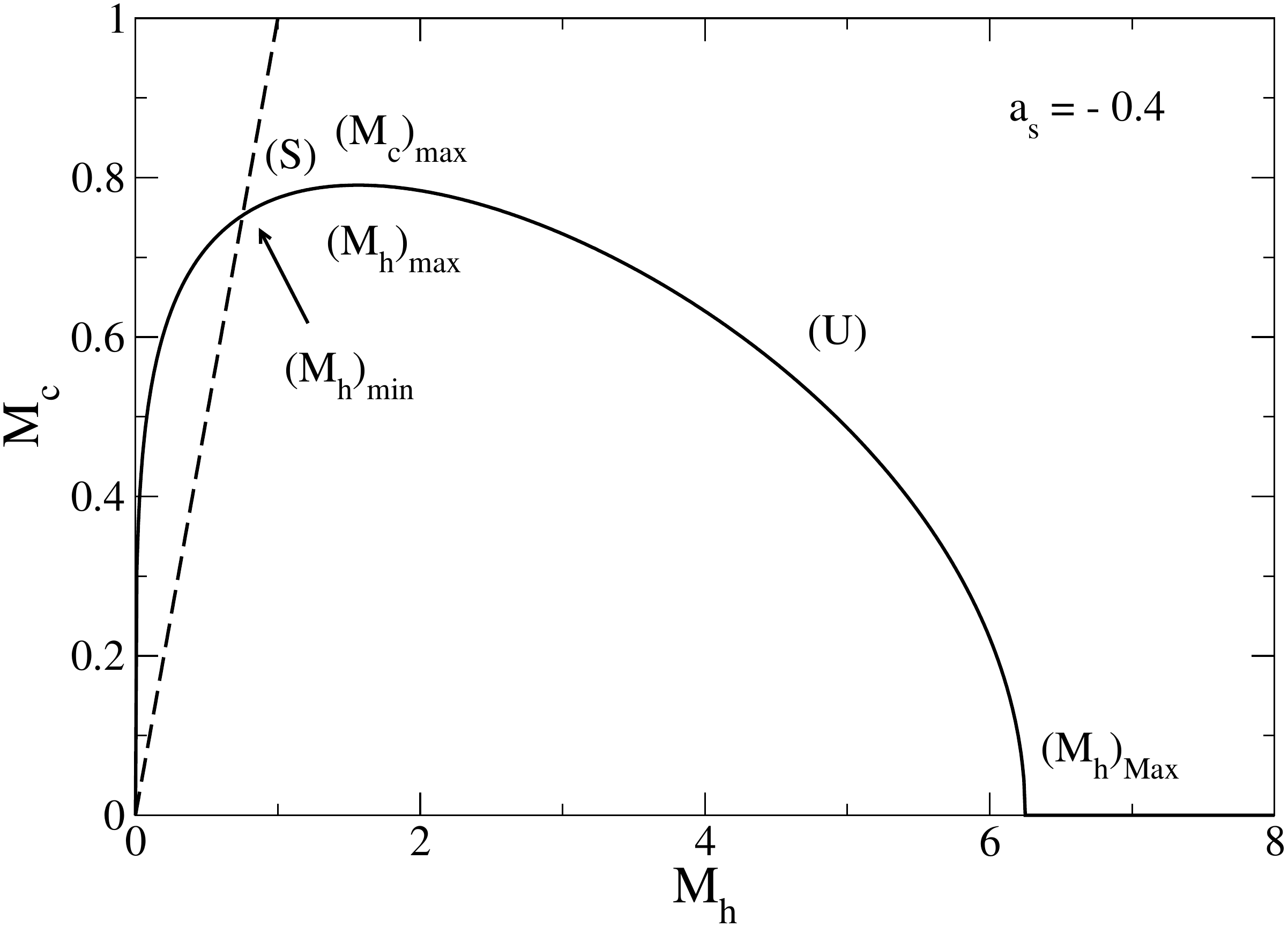}
\caption{Core mass $M_c$ as a function of the halo mass $M_h$ for an attractive
self-interaction $a_s<0$ (solid line). We have also plotted the relation
$M_c=M_h$ (dashed line) determining the minimum halo mass $(M_{h})_{\rm min}$.
The mass is
normalized by $(M_{h})_{\rm min,0}$ and the scattering length by $a_*$.}
\label{mhmcaminus0p4}
\end{center}
\end{figure}

\begin{figure}[!h]
\begin{center}
\includegraphics[clip,scale=0.3]{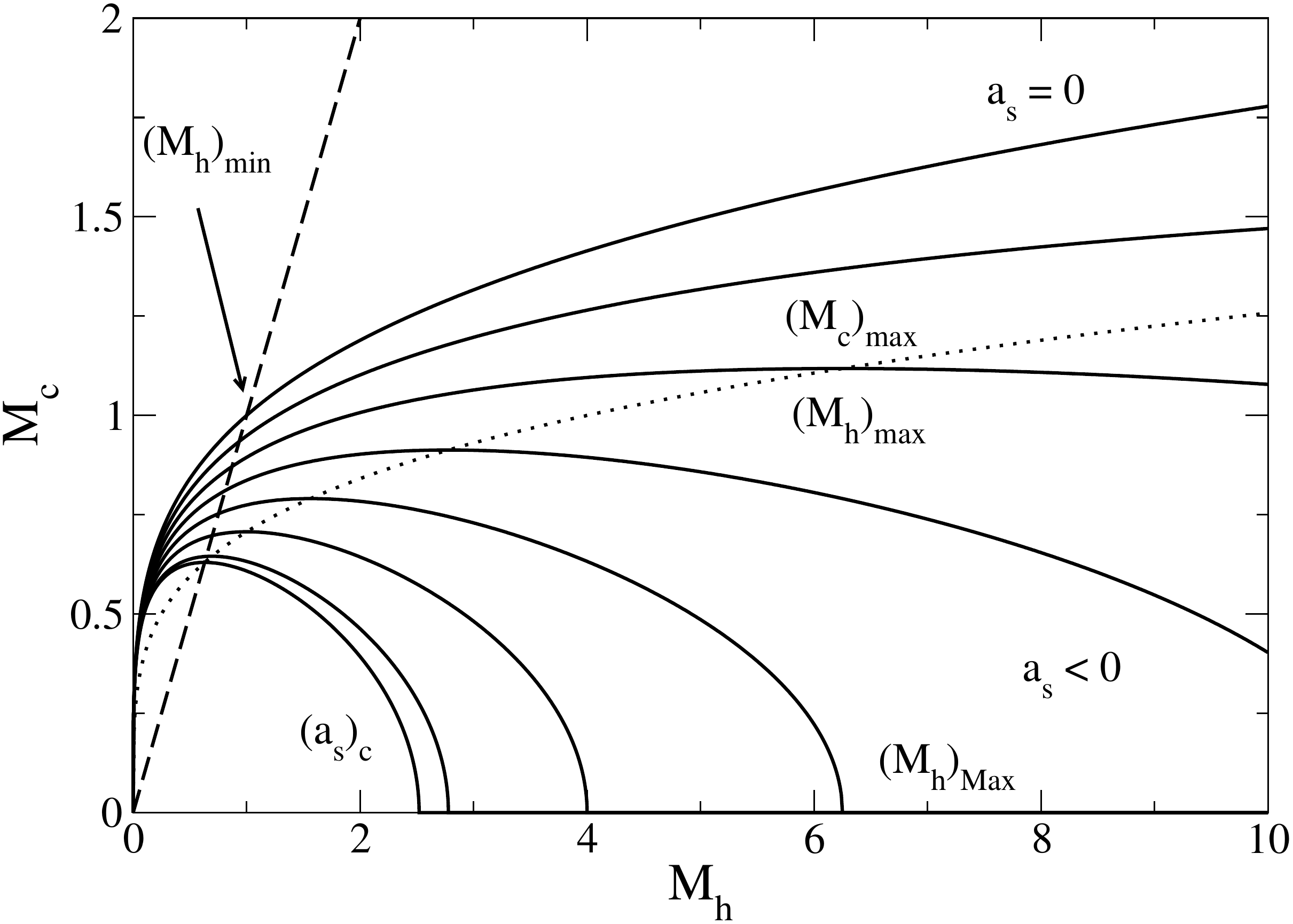}
\caption{Core mass $M_c$ as a function of the halo mass $M_h$ for different
values of the scattering length $a_s\le 0$ and a fixed value of the mass
$m$ (solid lines). We have
plotted the position of the minimum halo mass $(M_{h})_{\rm min}$  (dashed line)
and the position of the maximum halo mass $(M_{h})_{\rm max}$  (dotted line).
The mass is
normalized by $(M_{h})_{\rm min,0}$. We have indicated the curve corresponding
to the minimum scattering length $(a_s)_c$ for which $(M_{h})_{\rm
min}=(M_{h})_{\rm max}$ (so that the minimum halo is critical) and the curve
corresponding to the noninteracting case
$a_s=0$. }
\label{multineg}
\end{center}
\end{figure}

 In the attractive case, the core mass  vanishes at
\begin{equation}
\label{mrr15}
\frac{(M_h)_{\rm Max}}{(M_h)_{\rm min,0}}=\left
(\frac{a_*}{a_s}\right )^2.
\end{equation}
On the other hand the core mass is maximum at
\begin{equation}
\label{mrr16}
\frac{(M_h)_{\rm max}}{(M_h)_{\rm min,0}}=\frac{1}{4}\left
(\frac{a_*}{a_s}\right )^2
\end{equation}
with the value
\begin{equation}
\label{mrr17}
\frac{(M_c)_{\rm max}}{(M_h)_{\rm min,0}}=\frac{1}{2}\left
(\frac{a_*}{|a_s|}\right )^{1/2}.
\end{equation}
This corresponds to the maximum core mass given by Eq. (\ref{mrr11}).
The core mass --  halo mass
relation for $a_s<0$  is
plotted in Figs. \ref{mhmcaminus0p4} and \ref{multineg}. These solutions are
valid for
$M_h>(M_h)_{\rm
min}$. On the other hand, the branch $(M_h)_{\rm
max}\le M_h\le (M_h)_{\rm
Max}$ corresponds to unstable states so that only the branch
$(M_h)_{\rm
min}\le M_h\le (M_h)_{\rm
max}$, corresponding to stable states, is physical. In summary, when
$a_s<(a_s)_c$ there is no halo with a stable quantum core (see Sec.
\ref{sec_mhm}). When
$(a_s)_c<a_s<0$ a stable quantum core exists
only in the range $(M_h)_{\rm min}\le M_h\le (M_h)_{\rm max}$. It has a mass 
$(M_h)_{\rm min}\le M_c\le (M_c)_{\rm max}$. Coming back to the original
variables, the maximum halo mass is
\begin{equation}
\label{mrr18}
(M_h)_{\rm max}=\frac{\sigma^2\alpha}{(6\pi\zeta)^2} \frac{m^2}{a_s^2\Sigma_0}.
\end{equation}
The prefactor is $0.223$. Measuring the DM particle  mass in
units of
$10^{-22}{\rm eV/c^2}$ and the scattering length in units of $10^{-62}\, {\rm
fm}$, we get $(M_{h})_{\rm max}=1.21\times 10^8\, m^{2}|a_s|^{-2}\,
M_{\odot}$. Note that if we
determine the maximum halo mass $(M_h)_{\rm max}$
approximately by equating Eqs. (\ref{mrr13}) and (\ref{mrr11}) [or
equivalently Eqs. (\ref{mrr13b}) and (\ref{mrr3})], we obtain a value that
differs from the real one [Eq. (\ref{mrr16}) or equivalently (\ref{mrr18})] by
a factor $1/4$.

For bosons with an attractive self-interaction, like the axion
\cite{marshrevue}, it is more
convenient to express the results in terms of the decay constant (\ref{mrr19}).
When $f<f_c$ there is no halo with a stable
quantum core. When $f>f_c$ a stable quantum core  exists
only in the range $(M_h)_{\rm min}\le M_h\le (M_h)_{\rm max}$. The minimum halo
mass  $(M_h)_{\rm min}$ is close to $(M_h)_{\rm min,0}$ given by Eq.
(\ref{mrr6}). The maximum halo mass and the maximum core mass can be written as
\begin{equation}
\label{mrr20}
(M_h)_{\rm max}=\frac{\sigma^2\alpha (32\pi)^2}{(6\pi\zeta)^2} 
\frac{f^4}{\hbar^2 c^6 \Sigma_0},
\end{equation}
\begin{equation}
\label{mrr20c}
(M_c)_{\rm max}=\left (\frac{32\pi\sigma^2}{6\pi\zeta\nu}\right )^{1/2} 
\left (\frac{f^2\hbar}{c^3m^2G}\right )^{1/2}.
\end{equation}
We note that the maximum halo
mass depends only on $f$ while the  maximum core mass depends on $f$ and $m$.
The prefactors are $2255$ and  $10.9$. 
Measuring the DM particle  mass in units of
$10^{-22}{\rm eV/c^2}$ and the decay constant in units of $10^{15}\, {\rm GeV}$,
we get $(M_h)_{\rm max}=3.14\times 10^{11} f^4\, M_{\odot}$ and $(M_c)_{\rm
max}=1.19\times 10^{9} (f/m)\, M_{\odot}$. For $f=1.44\times 10^{17}\,
{\rm GeV}$ and $m=2.25\times
10^{-22}\, {\rm eV}/c^2$ (see Sec. \ref{sec_mas}), we find $(M_h)_{\rm
max}=1.35\times 10^{20}\, M_{\odot}$ and $(M_c)_{\rm
max}=7.62\times 10^{10}\, M_{\odot}$. Since the largest DM halos observed in
the Universe have a mass
$M_h\sim 10^{14}\, M_{\odot}\ll (M_h)_{\rm
max}$, these results suggest that the effect of an
attractive self-interaction is negligible: Everything
happens {\it as if} the bosons were not self-interacting. This favors the
consideration of a repulsive self-interaction \cite{modeldm}.

In this section, we have expressed the core mass -- halo mass relation in terms
of $M_h$. This relation  could easily be expressed in terms of $M_v$ by using
Eq. (\ref{sc4}) with $B=2.79\times 10^{-3}$ (we have taken $(M_h)_{\rm
min}=10^{8}\,
M_{\odot}$).

{\it Remark:} We can directly obtain the expression (\ref{mrr18}) of the maximum
halo mass $(M_h)_{\rm max}$ from the relation [see Eq. (\ref{mrr12b})]
\begin{equation}
\label{mrr21}
\frac{(M_c)_{\rm max}}{(R_c)_*}=\sqrt{1.76\, \Sigma_0 (M_h)_{\rm max}}
\end{equation}
with Eqs. (\ref{mrr3}) and (\ref{mrr4}). If we consider a
self-gravitating BEC with a central black hole it can be shown that the ratio
$(M_c)_{\rm max}/(R_c)_*$ is independent of the black hole
mass \cite{epjpbh}. This implies that the expression (\ref{mrr18})
of the
maximum halo mass is unchanged for a fixed central black hole. The
case where the black hole mass changes with the halo mass is treated in
\cite{mcmhbh}.

\subsection{Summary}
\label{sec_disc}

In the noninteracting case ($a_s=0$) the halos with a mass $M_h>(M_h)_{\rm
min,0}$ [see Eq. (\ref{mrr6})] contain a quantum core of mass $M_c$ given by Eq.
(\ref{mrr13}). All the configurations are stable.

For a repulsive self-interaction ($a_s>0$) the halos with a mass
$M_h>(M_h)_{\rm min}$ [see Eq. (\ref{mrr7})] contain a quantum core of mass
$M_c$ given by Eq. (\ref{mrr12}). For $a_s\ll a_*$ and $M_h$ not too
large ($M_h/(M_h)_{\rm
min,0}\ll (a_*/a_s)^2$),
we are in the noninteracting limit discussed previously. For $a_s\gg a_*$ or
 $M_h$ sufficiently large ($M_h/(M_h)_{\rm
min,0}\gg (a_*/a_s)^2$) we are in the
TF limit. In that case, the minimum halo mass $(M_h)_{\rm min}$ is given by Eq.
(\ref{mrr9b}) and the  mass $M_c$ of the quantum core is given by Eq.
(\ref{mrr14b}).  All the configurations are stable.

For an attractive self-interaction ($a_s<0$) the halos can contain a stable
quantum core only if $(a_s)_c<a_s<0$ [see Eq. (\ref{mrr11b})] or equivalently if
$f>f_c$ [see Eq. (\ref{lmrr19bz})]. When this condition is fulfilled the
quantum core is
stable for $(M_h)_{\rm min}<M_h<(M_h)_{\rm
max}$. The minimum halo mass $(M_h)_{\rm min}$ [see Eqs. (\ref{mrr7}) and
(\ref{lmrr7})] is of
the order of $(M_h)_{\rm min,0}$ [see Eq. (\ref{mrr6})]. When we reach the
maximum halo mass [see Eqs.  (\ref{mrr18}) and (\ref{mrr20})], the core reaches
its maximum limit
[see Eqs. (\ref{mrr3}) and (\ref{mrr20c})] and collapses. The result of the
collapse (dense
axion star, black hole, bosenova...) is discussed
in \cite{braaten,cotner,bectcoll,ebycollapse,tkachevprl,helfer,phi6,visinelli,
moss}.

\begin{figure}[!h]
\begin{center}
\includegraphics[clip,scale=0.3]{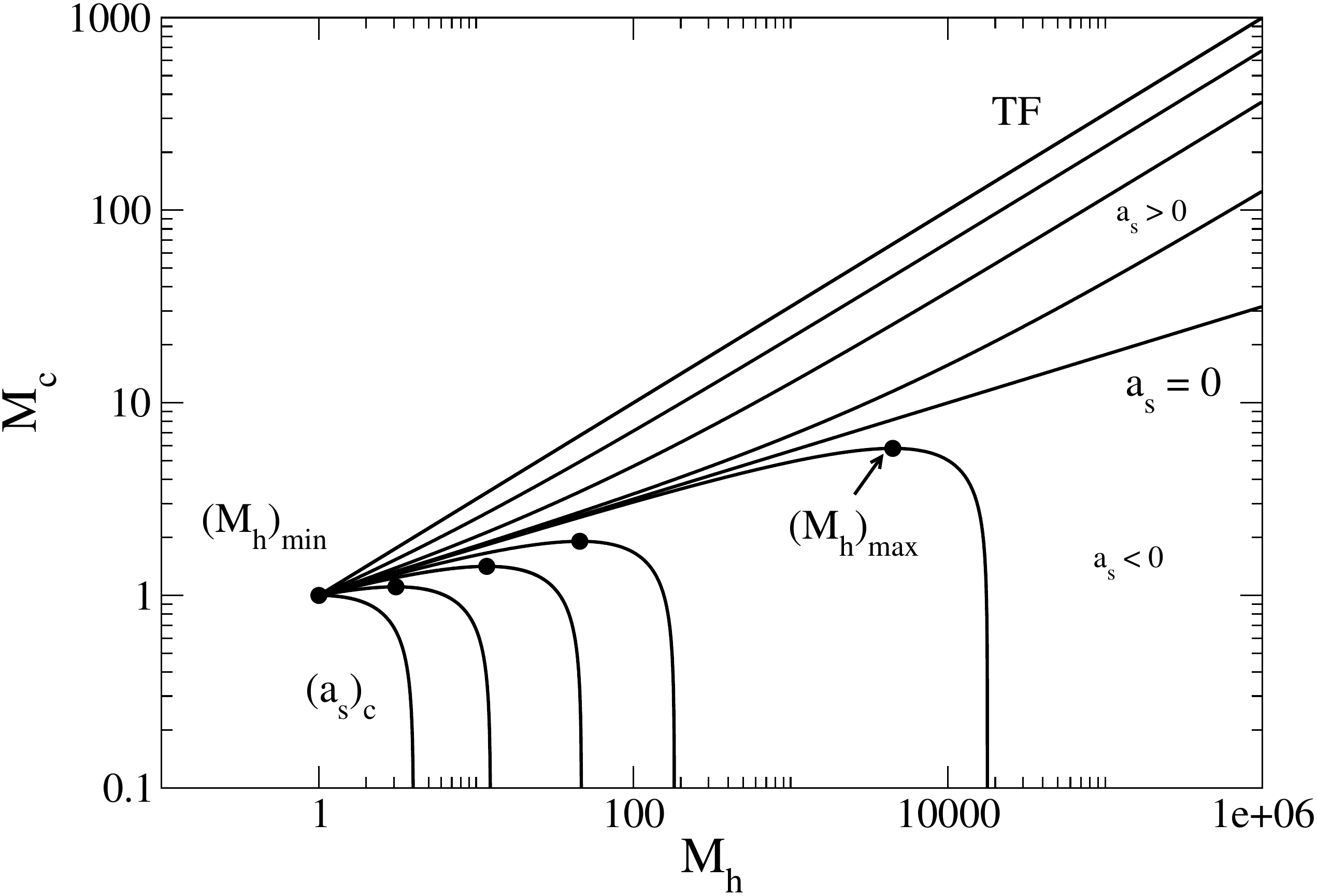}
\caption{Core mass $M_c$ as a
function
of the halo mass $M_h$ for different
values of $a_s$ and $m$ (see Table I) so that the minimum halo mass
$(M_{h})_{\rm
min}$
is fixed (as explained in the text). The mass is normalized by $(M_{h})_{\rm
min}$
(typically
$(M_{h})_{\rm min}\sim 10^8\, M_{\odot}$).  We have
plotted the position of the minimum halo mass $(M_{h})_{\rm min}$ (common
origin) and the position of the maximum halo mass $(M_{h})_{\rm max}$
(bullets) above which the quantum core becomes unstable when
$a_s<0$.
 We have indicated the curve
corresponding
to the minimum scattering length
$(a_s)_c/a'_*=-1/2^{3/2}$ for which
$(M_{h})_{\rm
min}=(M_{h})_{\rm max}$ (critical minimum halo), the curve corresponding to the
noninteracting case
$a_s=0$ [see Eq. (\ref{mrr13})], and the curve corresponding to the TF limit
$a_s/a'_*\gg 1$ [see Eq. (\ref{mrr14b})].}
\label{mhmcnorm}
\end{center}
\end{figure}

These results are summarized in Fig. \ref{mhmcnorm} representing the general
core mass --
halo mass relation with a normalization such that the minimum halo mass is
fixed to a common value
obtained from the observations (typically $(M_{h})_{\rm min}\sim 10^8\,
M_{\odot}$). The minimum halo mass determines the relation
between $m$
and $a_s$ as explained in Sec. \ref{sec_mas}. The construction
of Fig. \ref{mhmcnorm}  is detailed below and the selected values of the DM
particle  parameters are given in
Table I. On this representation, we clearly
see that, as compared to the noninteracting case ($a_s=0$), the core mass $M_c$
increases more rapidly with $M_h$ in the case of a repulsive
self-interaction ($a_s>0$) and less rapidly in
the case of an attractive
self-interaction ($a_s<0$). For $a_s\ge 0$, there
is a stable core for any
halo mass. For $(a_s)_c<a_s<0$, there is a maximum halo mass
$(M_h)_{\rm max}$ associated with the
existence of a maximum core mass $(M_c)_{\rm max}$. Above that mass, the
quantum core collapses.

\begin{table*}[t]
\centering
\begin{tabular}{|c|c|c|c|c|c|c|}
\hline
  $\frac{(M_h)_{\rm min}}{(M_h)_{\rm min,0}}$ & $\frac{a_s}{a_*}$ &
$\frac{m}{m_0}$
& $\frac{a_s}{a'_*}$ & $\frac{f}{f'_*}$ & $\frac{(M_h)_{\rm max}}{(M_h)_{\rm
min}}$ & $\frac{(M_c)_{\rm max}}{(M_h)_{\rm min}}$\\
\hline
$10^4$ &  $10^4$ & $10^3$ & $10^9$ & & &\\
\hline
$1.6$ &  $0.809$ & $1.42$ & $1.46$& & & \\
\hline
$1.1$ &  $0.1465$ & $1.07$ & $0.165$& & & \\
\hline
$1.01$ &  $0.015$ & $1.01$ & $0.01515$& & & \\
\hline
$1$ &  $0$ & $1$ & $0$& $\infty$ & & \\
\hline
$0.995$ &  $-0.00751$ & $0.996$ & $-0.00746$ & $11.55$ & $4460$ & $5.80$ \\
\hline
$0.95$ &  $-0.0760$ & $0.962$ & $-0.0713$ & $3.67$ & $45.6$ & $1.91$ \\
\hline
$0.9$ &  $-0.154$ & $0.924$ & $-0.135$ & $2.615$ & $11.7$ & $1.41$ \\
\hline
$0.8$ &  $-0.318$ & $0.846$ & $-0.241$ & $1.87$ & $3.09$ & $1.11$ \\
\hline
$0.630$ &  $-0.630$ & $0.707$ & $-0.354$ & $1.41$ & $1$ & $1$ \\
\hline
\end{tabular}
\label{table1}
\caption{Values of the DM particle parameters selected in Fig. \ref{mhmcnorm}.
The scales
$m_0$, $a'_*$ and $f'_*$ are given by Eqs. (\ref{mas1}), (\ref{mas4}) and
(\ref{mrr19b}). For $(M_h)_{\rm min}=10^8\, M_{\odot}$, we obtain
$m_0=2.25\times 10^{-22}\, {\rm eV}/c^2$, $a'_*=4.95\times 10^{-62}\, {\rm fm}$
and $f'_*=9.45\times 10^{13}\, {\rm GeV}$. The noninteracting
limit corresponds to $|a_s|\ll a'_*$ (hence $m\sim m_0$). The TF limit
corresponds to $a_s\gg a'_*$ (hence $m\gg m_0$). The minimum value of $a_s$
ensuring that the minimum halo is stable is $(a_s)_{c}=-0.354\, a'_*$ (hence
$m_c=0.707\, m_0$).}
\end{table*}

To make Fig. \ref{mhmcnorm}, corresponding to a
fixed minimum halo mass $(M_h)_{\rm min}$, we have proceeded in the following
manner. For a given value of $a_s/a'_*$ we
can obtain $m/m_0$ from Eq. (\ref{mas5}). Then, we get $a_s/a_*$ from
Eq.
(\ref{mas3}) or, equivalently, from 
\begin{equation}
\label{proc1}
\frac{a_s}{a_*}=\left (\frac{m}{m_0}\right)^{4/3}-\left
(\frac{m_0}{m}\right)^{2/3}.
\end{equation}
Finally, we get   
$(M_h)_{\rm min}/(M_h)_{\rm min,0}$ from Eq. (\ref{mas2}) or, equivalently, from
Eq.
(\ref{mrr7}). We can then plot
$M_c/(M_h)_{\rm min}$ as a function of
$M_{h}/(M_h)_{\rm min}$ by using Eq.
(\ref{mrr12}).  We stress
that this
procedure yields a ``universal'' curve for a given value of $a_s/a'_*$. The mass
$(M_h)_{\rm min}$ of the minimum halo then determines the scales
$m_0$ and $a'_*$ according to Eqs. (\ref{mas1}) and
(\ref{mas4}).\footnote{We
have assumed that the surface density of the DM halos
is universal [see Eq. (\ref{p5})]. If this were not the case, our general model
would
remain valid but the problem would depend on $M_h$ and $r_h$ instead of just
$M_h$.} For convenience, we have proceeded the other way
round. We have chosen a value of $(M_h)_{\rm min}/(M_h)_{\rm min,0}$,
determined 
$a_s/a_*$ from Eq. (\ref{mrr7}) and plotted $M_c/(M_h)_{\rm min}$ as
a
function of
$M_{h}/(M_h)_{\rm min}$ by using Eqs. (\ref{mrr7}) and
(\ref{mrr12}). We have
then used Eqs. (\ref{mas2}) and (\ref{mas5}) to obtain the
values of $m/m_0$ and  $a_s/a'_*$ corresponding to our choice of $(M_h)_{\rm
min}/(M_h)_{\rm min,0}$. The selected values of $(M_h)_{\rm min}/(M_h)_{\rm
min,0}$ and the corresponding DM particle parameters appearing in Fig.
\ref{mhmcnorm} are reported in Table I.

\begin{figure}[!h]
\begin{center}
\includegraphics[clip,scale=0.3]{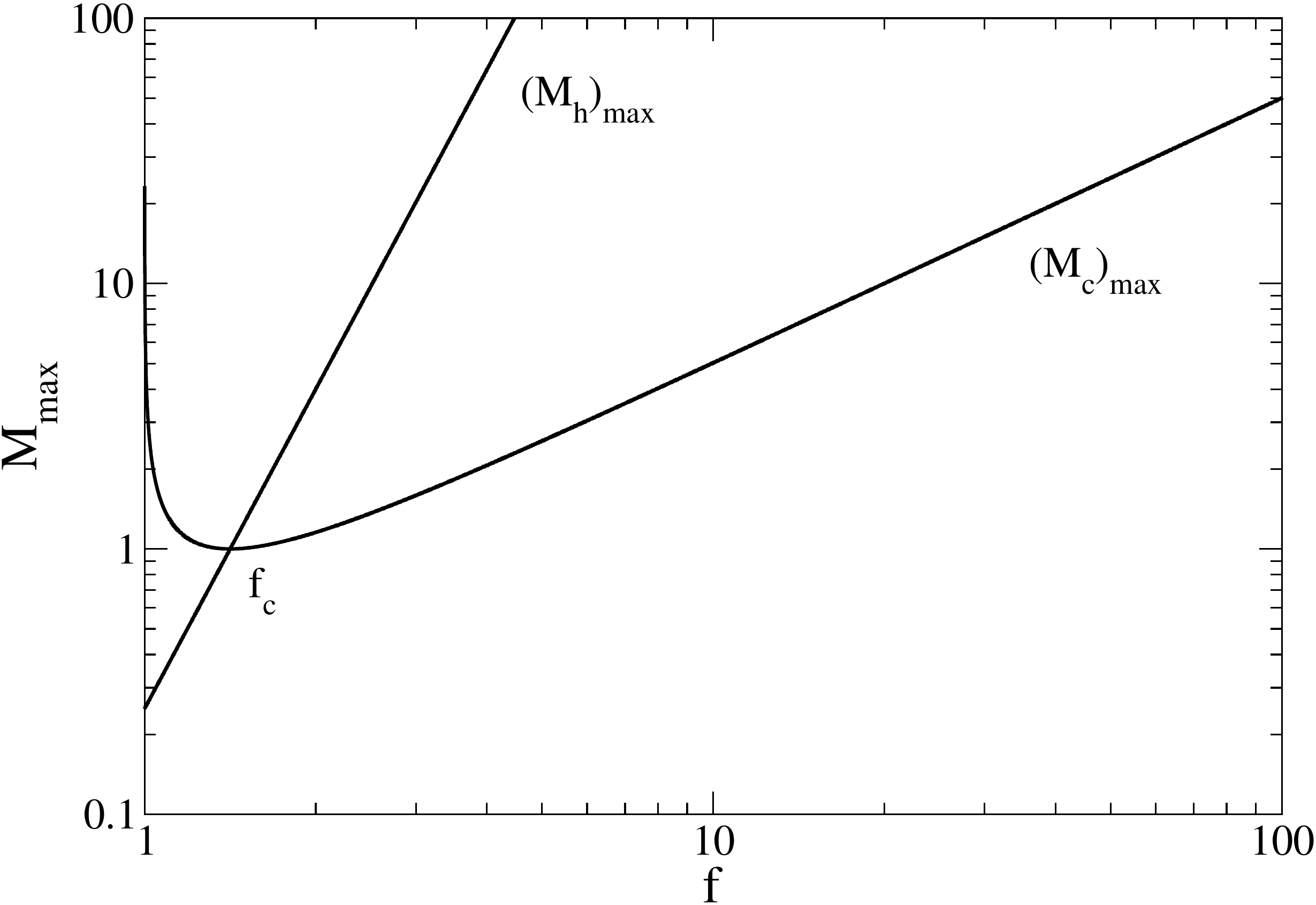}
\caption{Maximum halo mass and maximum core mass as a function of the decay
constant $f$ in the case of an attractive self-interaction. The mass is
normalized by $(M_{h})_{\rm
min}$ and the decay constant by $f'_*$.}
\label{fmmax}
\end{center}
\end{figure}

In the case of an attractive self-interaction, using Eqs. (\ref{mas2}),
(\ref{mas3}), (\ref{mrr19a}), (\ref{mrr16}) and (\ref{mrr17}), the maximum halo
mass  and the maximum core mass normalized by the minimum halo mass are given by
\begin{equation}
\label{proc2}
\frac{(M_h)_{\rm max}}{(M_h)_{\rm min}}=\frac{1}{4}\left
(\frac{a'_*}{a_s}\right )^2 \left(\frac{m}{m_0}\right )^2=\frac{1}{4}\left
(\frac{f}{f'_*}\right )^4
\end{equation}
and
\begin{equation}
\label{proc3}
\frac{(M_c)_{\rm max}}{(M_h)_{\rm min}}=\frac{1}{2}\left
(\frac{a'_*}{|a_s|}\right )^{1/2} \left(\frac{m_0}{m}\right
)^{1/2}=\frac{1}{2}\frac{f}{f'_*}\frac{m_0}{m},
\end{equation}
where $a_s/a'_*$, $f/f'_*$ and $m/m_0$ are related to each other  by Eqs.
(\ref{mas5}) and
(\ref{mrr19c}). They are plotted as a function of the
decay constant $f$ in
Fig. \ref{fmmax}. For $f\ge f_c$ (correspondingly $m_c\le m\le m_0$ and
$(a_s)_c\le a_s\le 0$), the maximum halo mass $(M_h)_{\rm max}$ and the maximum
core mass $(M_c)_{\rm max}$ increase monotonically with $f$, starting from the
value $(M_h)_{\rm min}$. For $f\gg f'_*$ (i.e. $|a_s|\ll a'_*$)
they behave as $(M_h)_{\rm max}/{(M_h)_{\rm min}}\sim (1/4)(f/f'_*)^4\sim
(1/4)(a'_*/a_s)^2$ and $(M_c)_{\rm
max}/{(M_h)_{\rm min}}\sim (1/2)(f/f'_*)\sim (1/2)(a'_*/|a_s|)^{1/2}$.
 As indicated at the end of Sec. \ref{sec_mcmh}, the
largest DM halos observed in the Universe have a mass $M_h\sim 10^{14}\,
M_{\odot}$. Therefore, if $(M_h)_{\rm max}>10^{14}\, M_{\odot}$ we will never
see the effect of an attractive self-interaction (for any DM halo). Taking
$(M_h)_{\rm min}=10^{8}\, M_{\odot}$, this condition corresponds to $f>44.7\,
f'_*=4.23\times 10^{15}\, {\rm GeV}$. Therefore, for $f>44.7\,
f'_*=4.23\times 10^{15}\, {\rm GeV}$
(correspondingly $0.99975\, m_0<m<m_0$ and $-0.0005\, a'_*=-2.47\times
10^{-65}\, {\rm fm}<a_s\le 0$) everything happens as if the bosons were
noninteracting. This is particularly true for the value  $f=1520\,
f'_*=1.44\times 10^{17}\, {\rm GeV}$ predicted in Sec. \ref{sec_mas} and, more
generally, for values of $f$ in the range
$10^{16}\, {\rm GeV}\le f\le 10^{18}\, {\rm GeV}$ expected in particle physics
\cite{hui}.

In the case of a repulsive self-interaction, for a given value of $a_s$, the
transition betwen the noninteracting limit and the TF limit occurs at a typical
halo mass  
\begin{equation}
\label{proc2b}
\frac{(M_h)_{t}}{(M_h)_{\rm min}}\sim \left
(\frac{a'_*}{a_s}\right )^2 \left(\frac{m}{m_0}\right )^2.
\end{equation}
For $(M_h)_{\rm min}<M_h\ll (M_h)_t$ we are in the  noninteracting limit and
for $M_h\gg (M_h)_t$ we are in the TF limit. Using the same argument as above,
if $(M_h)_{t}>10^{14}\, M_{\odot}$ we will not see
the effect of a repulsive self-interaction (for any DM halo). Taking $(M_h)_{\rm
min}=10^{8}\,
M_{\odot}$, this condition corresponds to $0<a_s<0.001\, a'_*=4.95\times
10^{-65}\, {\rm fm}$. Therefore, for $0<a_s<0.001\, a'_*=4.95\times
10^{-65}\, {\rm fm}$
(correspondingly $m_0<m<1.0005\, m_0$) everything happens as if the bosons were
noninteracting. To our knowledge, there is no strong constraint on a repulsive
self-interaction (see, however, the Bullet Cluster constraint mentioned in
Sec. \ref{sec_tf}) so that large positive values of $a_s$ ($\gg
0.001\,
a'_*=4.95\times 10^{-65}\, {\rm fm}$), corresponding to large values of the
boson masse $m$ ($\gg 1.0005\, m_0=2.25\times 10^{-22}\, {\rm
eV}/c^2$), are possible in
principle.

\section{Conclusion}
\label{sec_con}

In this paper, we have analytically derived the core mass -- halo mass relation
of
fermionic and bosonic DM halos from an effective thermodynamical approach. We
have modeled the DM halos by a quantum core of mass $M_c$ surrounded by an
isothermal atmosphere of uniform density. We first determined
an analytical
expression of the free energy $F(M_c)$ and entropy $S(M_c)$ of the DM halos.
The equilibrium core
mass $M_c$ is then
obtained by extremizing the free energy $F(M_c)$ at fixed mass or
by extremizing the entropy $S(M_c)$ at fixed mass and energy. By representing
the quantum core by a polytrope of index $n$
we have developed a unified description for fermions ($n=3/2$), noninteracting
bosons ($n=2$) and self-interacting bosons in the TF approximation ($n=1$). This
allowed us to 
treat fermionic and bosonic DM halos with the same formalism.
In the generic
case, the extremization problem determines three solutions
corresponding to a gaseous phase (G), a core-halo phase (CH) and a condensed
phase phase (C). The most important solution is the core-halo phase. 
We showed
that this phase  is always
unstable in the canonical ensemble (maximum of free energy at
fixed mass) while
it is stable in
the microcanonical ensemble (maximum of entropy at fixed mass
and energy) when $M_h<(M_h)_{\rm MCP}$ (the gaseous and
the condensed phases are stable in all statistical ensembles). When
$M_h>(M_h)_{\rm MCP}$ the core-halo phase
is unstable in all statistical ensembles. In that case, the quantum core may
be replaced by a supermassive black hole \cite{modeldm} resulting from a
gravothermal
catastrophe \cite{lbw} followed by a dynamical instability of general
relativistic origin \cite{balberg}.

Our thermodynamical approach leads to the velocity dispersion tracing
relation (\ref{ana31b}) put forward heuristically in
\cite{mocz,bbbs,modeldm}. Therefore, this
relation can be
{\it justified} by an effective
thermodynamical approach (maximum entropy principle). For
noninteracting bosons, we obtain the
mass-radius relation (\ref{sat6}) which is consistent to the one found by Schive
{\it et
al.} \cite{ch3} (see also \cite{veltmaat}). For fermions, we obtain the
mass-radius relation (\ref{sat4})
which is
consistent to the one
found by Ruffini {\it et al.} \cite{rar}. For bosons with an attractive
self-interaction in the
TF limit, we predict the mass-radius relation (\ref{sat8}) which still has to be
confirmed numerically. Combining the velocity dispersion tracing
relation \cite{mocz,bbbs,modeldm} with the core mass -- core radius relation
derived in \cite{prd1} we have obtained an approximate general core mass --
halo mass relation $M_c(M_v)$ [see Eq. (\ref{mrr12})] that is valid for bosons
with arbitrary repulsive or
attractive self-interaction. For an attractive self-interaction, corresponding
to axions \cite{marshrevue}, we have determined
the maximum halo mass $(M_v)_{\rm max}$ [see Eq. (\ref{mrr20})] that can harbor
a stable quantum core (dilute axion star).

The mass $(M_h)_{\rm min}$ of the minimum halo (ground state)
determines the parameters of the DM particle (depending on the strength of the
self-interaction). Observations reveal that $(M_h)_{\rm min}\sim 10^8\,
M_{\odot}$. For fermions, we find $m=165 \, {\rm
eV}/c^2$ [see Eq. (\ref{mf})]. For noninteracting bosons
we find $m=1.44\times 10^{-22}
\, {\rm
eV}/c^2$ [see Eq. (\ref{mbni})]. For self-interacting bosons in the TF limit we
find $a_s/m^3=1.76\times
10^3\, {\rm fm}/({\rm eV}/c^2)^3$ [see Eq. (\ref{mbft})]. We can then use the 
core mass -- halo mass relation to determine the characteristics of the quantum
core residing in a given DM halo. Let us consider a DM halo of mass
$M_h=10^{12}\,
M_{\odot}$ similar to the one that surrounds our Galaxy. For
fermions, we
find $M_c=4.74\times 10^9\, M_{\odot}$ and $R_c=301\, {\rm pc}$ [see Eqs.
(\ref{fdm3}) and (\ref{sat3})]. For noninteracting bosons
we find $M_c=1.84\times 10^9\, M_{\odot}$ and $R_c=118\, {\rm pc}$ [see Eqs.
(\ref{t6}) and (\ref{sat5})]. For self-interacting bosons in the TF limit we
find $M_c=1.16\times 10^{10}\, M_{\odot}$ and $R_c=733\, {\rm pc}$ [see Eqs.
(\ref{tf3}) and (\ref{sat7})]. In our model, the quantum core represents a bulge
or a nucleus. It cannot mimic a black hole as it has been sometimes suggested.

Finally, we have argued that the mass scale
of noninteracting DM bosons is determined in terms of fundamental constants by
$m_{\Lambda}=\hbar\sqrt{\Lambda}/c^2=2.08\times 10^{-33}\,
{\rm eV/c^2}$ while  the mass
scale of fermions is determined by
$m_{\Lambda}^*=({\Lambda\hbar^3}/{Gc^3})^{1/4}=\sqrt{m_{\Lambda}M_P}
=5.04\times 10^{-3}\, {\rm eV/c^2}$. We
found that the prefactor between the actual DM particle mass and these
fundamental mass
scales can be very large ($11$ orders of magnitude for bosons and $4$ orders of
magnitude for fermions). However, these fundamental mass scales can explain the
intrinsic difference of mass between bosonic and fermionic DM
particles. Their ratio is
$m_{\Lambda}^*/m_{\Lambda}=(c^5/G\hbar\Lambda)^{1/4}=2.42\times 10^{30}$,
corresponding to a difference of $30$ orders of magnitude.
Finally, in the
case of self-interacting bosons, we have found
that the fundamental scale of the ratio $a_s/m^3$ is 
${r_{\Lambda}}/{m_{\Lambda}^3}={2Gc^2}/{
\Lambda\hbar^2}=6.11\times
10^{17}\, {\rm fm\, (eV/c^2)^{-3}}$.

In the present paper, we have developed an analytical model in which the
isothermal atmosphere has a uniform density. This approximation is sufficient
to obtain the correct scaling of the  core mass -- halo
mass relation. However, in order to develop more accurate models of fermionic
and bosonic DM halos, and in particular to be able to determine their density
and circular velocity profiles, we need to solve a generalized Emden equation
numerically. The case of self-gravitating BECs with a repulsive self-interaction
has been treated in detail in \cite{modeldm}. The case of noninteracting bosons
and fermions can be treated with the same method. These models are being
presently investigated \cite{inpreparation}.

It will be important in future works to determine if DM is made
of fermions or bosons (either noninteracting, with a repulsive self-interaction,
or with
an attractive self-interaction). All these models are very interesting from a
physical point of view, with fascinating properties, but it is possible that
some of them will be ruled out by observations. Alternatively, all these models
could be of interest if DM is made of
several types of particles (bosons and fermions) as suggested in
\cite{modeldm}.

It will also be important in future works to go beyond certain
approximations made in this paper. For example, we have ignored the influence of
baryons. They may have several significant effects on DM halos and, in
particular, they can alleviate the cusp problem \cite{romano}. It would be
interesting to check the
validity of our results in the presence of baryons. We have also used a
nonrelativistic approach. As we have shown, this nonrelativistic approach is
expected to be sufficient in most applications of DM halos. However, in the case
of bosons with an attractive self-interaction, if the mass of the quantum core
$M_c$ becomes greater than the maximum mass $(M_c)_{\rm max}$ \cite{prd1}, the
core collapses. In
that case, relativistic effects can be important and the GPP equations must be
replaced by the Klein-Gordon-Einstein (KGE) equations
\cite{tkachevprl,helfer,phi6,visinelli,moss}. The KGE equations
must also be used  \cite{mabc} in order to study boson stars and scalar field
configurations in highly relativistic
environments, such as scalar field dark matter laying in the vicinity of compact
objects like black holes or neutron stars.

\appendix

\section{Polytropic spheres}
\label{sec_gl}

In this Appendix, we recall general results pertaining to
self-gravitating polytropic spheres \cite{emden,chandrabook}. We apply
them to the quantum models of DM halos at $T=0$ (ground states) discussed in
Sec. \ref{sec_qm}.

For classical self-gravitating systems, or for quantum self-gravitating
systems in the TF approximation, the condition of hydrostatic equilibrium
\begin{eqnarray}
\label{gl1}
\nabla P+\rho\nabla\Phi={\bf 0}
\end{eqnarray}
combined with the Poisson equation
\begin{equation}
\label{gl2}
\Delta \Phi=4\pi G\rho
\end{equation}
leads to the fundamental differential equation
\begin{equation}
\label{gl3}
\nabla\cdot \left (\frac{\nabla P}{\rho}\right )=-4\pi G\rho.
\end{equation}

For a polytropic equation of state of the form
\begin{equation}
\label{gl4}
P=K\rho^{\gamma},
\end{equation}
where $K$ is the polytropic constant and $\gamma=1+1/n$ is the polytropic
index, the differential equation (\ref{gl3}) becomes
\begin{equation}
\label{gl5}
K(n+1)\Delta\rho^{1/n}=-4\pi G\rho.
\end{equation}
In the following, we restrict ourselves to spherically symmetric distributions.
We also assume $K>0$ and $6/5<\gamma<+\infty$ (i.e. $0\le n<5$) for reasons
explained below.
With the substitution
\begin{equation}
\label{gl6}
\rho=\rho_0\theta^n,\qquad \xi=\frac{r}{r_0},
\end{equation}
where $\rho_0$ is the central density and
\begin{equation}
\label{gl6b}
r_0=\left\lbrack \frac{K(n+1)}{4\pi
G\rho_0^{1-1/n}}\right\rbrack^{1/2}
\end{equation}
is the polytropic radius, we obtain the Lane-Emden equation
\begin{equation}
\label{gl7}
\frac{1}{\xi^{2}}\frac{d}{d\xi}\left (\xi^{2}\frac{d\theta}{d\xi}\right
)=-\theta^n
\end{equation}
with the boundary conditions 
\begin{equation}
\label{gl8}
\theta(0)=1,\qquad \theta'(0)=0.
\end{equation}

According to the general results of Refs.
\cite{emden,chandrabook}, a polytrope of index $n$ has
a compact support  provided that
$0\le n<5$. In that case, the density falls off to zero at a finite
radius. This characterizes a complete polytrope.\footnote{Polytropes with
$n>5$ have an infinite mass. The polytrope $n=5$ is unbounded but has
a finite mass. For this index, the
Lane-Emden equation has  a simple analytical expression discovered by Schuster
\cite{schuster}. It was used by
Plummer \cite{plummer} to fit the density profile of globular clusters
(Plummer's
model).}
 It is customary to denote by $\xi_1$
the
normalized radius at which the density vanishes: $\theta_1=0$. The radius
and the mass of a complete polytrope then are 
\begin{equation}
\label{rpol}
R=\xi_1\left\lbrack \frac{K(n+1)}{4\pi
G}\right\rbrack^{1/2}\frac{1}{\rho_0^{(n-1)/2n}},
\end{equation}
\begin{equation}
\label{mpol}
M=-4\pi \frac{\theta'_1}{\xi_1} \rho_0 R^3.
\end{equation}
Eliminating $\rho_0$ between these two relations, we find that the mass-radius
relation of a complete polytrope of index $n$
is
\begin{equation}
\label{gl9}
M^{(n-1)/n}R^{(3-n)/n}=\frac{K(1+n)}{G(4\pi)^{1/n}}\omega_n^
{(n-1)/n},
\end{equation}
where  $\omega_n=-\xi_1^{(n+1)/(n-1)}\theta'_1$ is a constant determined by the
Lane-Emden equation (\ref{gl7}). A complete
polytrope of index $n$ is dynamically stable with respect to
the Euler-Poisson equations if $n<3$ and linearly unstable if $n>3$.

For the polytrope $n=3/2$ (fermion stars), using Eq. (\ref{fdm2}), we find
\begin{equation}
\xi_1=3.65375,\qquad \theta'_1=-0.203302,
\label{p8}
\end{equation}
\begin{equation}
\label{rpolf}
R=0.35885\, \frac{h}{m^{4/3}G^{1/2}\rho_0^{1/6}},
\end{equation}
\begin{equation}
\label{mpolf}
M=0.699218\, \rho_0 R^3,
\end{equation}
\begin{equation}
\label{mrpolf}
MR^3=0.0014931\, \frac{h^6}{G^3 m^8}.
\end{equation}

For the polytrope $n=2$ (noninteracting boson stars), using Eq.
(\ref{t4}), we find
\begin{equation}
\xi_1=4.353,\qquad \theta'_1=-0.1272,
\label{p10}
\end{equation}
\begin{equation}
\label{rpolfni}
R=1.94415\, \frac{\hbar^{1/2}}{m^{1/2}G^{1/4}\rho_0^{1/4}},
\end{equation}
\begin{equation}
\label{mpolfni}
M=0.367205\, \rho_0 R^3,
\end{equation}
\begin{equation}
\label{mrpolfni}
MR=5.24594\, \frac{\hbar^2}{G m^2}.
\end{equation}

For the polytrope $n=1$ (self-interacting boson stars), using
$\theta(\xi)=\sin(\xi)/\xi$ and Eq.
(\ref{tf2}), we find
\begin{equation}
\xi_1=\pi,\qquad \theta'_1=-1/\pi,
\label{p12}
\end{equation}
\begin{equation}
\label{rpolfsi}
R=\pi\, \left (\frac{a_s\hbar^2}{Gm^3}\right )^{1/2},
\end{equation}
\begin{equation}
\label{mpolfsi}
M=\frac{4}{\pi}\, \rho_0 R^3.
\end{equation}

\section{Total energy, eigenenergy and virial theorem of a self-gravitating BEC}
\label{sec_vt}

We consider a self-gravitating BEC described by the GPP
equations with a self-interaction corresponding to a power-law potential
\cite{ggp}. In
the hydrodynamic representation of the GPP equations, a power-law potential of
interaction gives rise to a polytropic equation of state. In that
case, the total energy of the BEC is given by
\begin{equation}
\label{vt1}
E_{\rm tot}=\Theta_Q+U+W.
\end{equation}
This is the sum of the quantum kinetic energy
\begin{equation}
\label{qk}
\Theta_Q=\frac{\hbar^2}{8m^2}\int \frac{(\nabla\rho)^2}{\rho}\, d{\bf
r}=\frac{\hbar^2}{2m^2}\int (\nabla\sqrt{\rho})^2\, d{\bf
r},
\end{equation}
the internal energy
\begin{equation}
\label{vt3}
U=\frac{1}{\gamma-1}\int P\, d{\bf r}=\frac{K}{\gamma-1}\int \rho^{\gamma}\,
d{\bf r},
\end{equation}
and the gravitational energy
\begin{equation}
\label{vt4}
W=\frac{1}{2}\int\rho\Phi\, d{\bf r}.
\end{equation}
The eigenenergy $E$ satisfies the relation
\begin{equation}
\label{vt5}
NE=\Theta_Q+\gamma U+2W.
\end{equation}
On the other hand, the equilibrium scalar virial theorem writes
\begin{equation}
\label{vt6}
2\Theta_Q+3(\gamma-1)U+W=0.
\end{equation}

For classical self-gravitating systems, or for  self-gravitating  BECs in the
TF approximation where the quantum potential can be neglected, the foregoing
equations reduce to
\begin{equation}
\label{vt7}
E_{\rm tot}=U+W,
\end{equation}
\begin{equation}
\label{vt8}
NE=\gamma U+2W,
\end{equation}
\begin{equation}
\label{vt9}
3(\gamma-1)U+W=0.
\end{equation}
From these equations, we obtain the relations
\begin{equation}
\label{vt10}
E_{\rm tot}=-\frac{3-n}{n}U=\frac{3-n}{3}W=\frac{3-n}{5-n}NE.
\end{equation}

\section{Betti-Ritter formula}
\label{sec_br}

For classical self-gravitating systems, or for
self-gravitating BECs in the TF approximation, the condition of hydrostatic
equilibrium is given by Eq. (\ref{gl1}). For a polytropic equation of
state (\ref{gl4}) we have
\begin{eqnarray}
\label{br1}
\frac{\nabla P}{\rho}=(n+1)\nabla\left (\frac{P}{\rho}\right
).
\end{eqnarray}
As a result, the condition of hydrostatic equilibrium (\ref{gl1}) can be
integrated into
\begin{eqnarray}
\label{br2}
(n+1)\frac{P}{\rho}+\Phi=\frac{E}{m},
\end{eqnarray}
where $E$ is a constant of integration. For a self-gravitating BEC
it represents the eigenenergy \cite{ggp}. Multiplying Eq. (\ref{br2}) by
$\rho$ and
integrating over the
whole
configuration, we obtain Eq. (\ref{vt8}). Assuming
$6/5<\gamma<+\infty$ (i.e.
$0\le n<5$) so that $P/\rho=0$ on the boundary of the system
$r=R$ where
the density vanishes, we find from Eq. (\ref{br2}) that
\begin{eqnarray}
\label{br3}
\frac{E}{m}=\Phi(R)=-\frac{GM}{R}.
\end{eqnarray}
This equation determines the
eigenenergy $E$.
As a result, Eq. (\ref{vt8}) can be rewritten as 
\begin{eqnarray}
\label{br4}
-\frac{GM^2}{R}=\gamma U+2W.
\end{eqnarray}
Combining this relation with the equilibrium scalar virial theorem
(\ref{vt9}) we obtain the Betti-Ritter formula
\begin{eqnarray}
\label{br5}
W=-\frac{3}{5-n}\frac{GM^2}{R}
\end{eqnarray}
determining the gravitational energy $W$ of a polytropic sphere
\cite{chandrabook}.
From Eqs. (\ref{br5}) and (\ref{vt10}), we get
\begin{eqnarray}
\label{br6}
U=\frac{n}{5-n}\frac{GM^2}{R}
\end{eqnarray}
and 
\begin{eqnarray}
\label{br7}
E_{\rm tot}=-\frac{3-n}{5-n}\frac{GM^2}{R}.
\end{eqnarray}
From the last relation, we can directly conclude that complete polytropes with
index $n<3$, i.e. $\gamma>4/3$,  are stable (because $E_{\rm tot}<0$) while
complete polytropes with
index
$n>3$, i.e. $\gamma<4/3$, are unstable (because $E_{\rm tot}<0$)
\cite{chandrabook}.

\section{Ledoux formula}
\label{sec_ledoux}

The complex pulsation of a polytrope of index
$0<n<5$ is approximately given by the
Ledoux
formula \cite{ledouxpekeris}:\footnote{This formula can also be obtained from a
variational principle
based on a Gaussian ansatz \cite{ggp}.}
\begin{equation}
\omega^2=(4-3\gamma)\frac{W}{I},
\label{ledoux1}
\end{equation}
where $I=\int \rho r^2\, d{\bf r}$ is the moment of inertia of the system.
Using the results of Appendix \ref{sec_gl} it can be written as
\begin{equation}
I=\kappa_{n} MR^2\quad {\rm with}\quad
\kappa_{n}=\frac{\int_0^{\xi_1}\theta^n\xi^{4}\,
d\xi}{\xi_1^2\int_0^{\xi_1}\theta^n\xi^{2}\, d\xi}.
\label{ledoux2}
\end{equation}
Combining this equation with the Betti-Ritter formula (\ref{br5}), we can
rewrite Eq.
(\ref{ledoux1}) as 
\begin{equation}
\omega^2=-\frac{3(n-3)}{(5-n)n\kappa_n}\frac{GM}{
R^{3}}.
\label{ledoux3}
\end{equation}
For $n=3/2$, we find $\kappa_{3/2}=0.306899$. For $n=2$, we
find $\kappa_{2}=0.232332$. For $n=1$, we find $\kappa_{1}=1-6/\pi^2=0.392073$.

\section{Density profile of a noninteracting self-gravitating BEC with a
compact support: polytrope $n=2$}
\label{sec_comp}

The fundamental differential equation of quantum hydrostatic equilibrium
determining the density profile of a  noninteracting self-gravitating BEC is
given by Eq. (\ref{ni1}). On the other hand, the fundamental differential
equation of classical hydrostatic equilibrium determining the density profile of
a polytrope of index $n$ is given by Eq. (\ref{gl5}). For $n=2$ it becomes
\begin{equation}
\label{t2}
3K\Delta\sqrt{\rho}=-4\pi G\rho.
\end{equation}
Dividing Eq. (\ref{t2}) by $\sqrt{\rho}$, applying the Laplacian operator, and
using
Eq. (\ref{t2}) again, we obtain
\begin{equation}
\label{t3}
\Delta\left (\frac{\Delta\sqrt{\rho}}{\sqrt{\rho}}\right )=\left (\frac{4\pi
G}{3K}\right )^2\rho.
\end{equation}
Remarkably, this equation coincides with Eq. (\ref{ni1}) provided that
we make the identification
\begin{equation}
\label{t4}
K=\left (\frac{2\pi
G\hbar^2}{9m^2}\right )^{1/2}.
\end{equation}
As a result, the density profile of a  polytrope of index $n=2$ and polytropic
constant given by Eq. (\ref{t4}) is a particular solution of Eq.
(\ref{ni1}).\footnote{We note that Eq. (\ref{t2}) implies Eq. (\ref{t3}) but the
converse is wrong. As a result Eqs. (\ref{ni1}) and (\ref{t2}) are not
equivalent.} Using the variables defined in Appendix \ref{sec_gl}, it can be
written as
\begin{equation}
\label{t5}
\rho(r)=\rho_0\theta(\xi)^2,\quad \xi=r/r_0,\quad r_0=\left
(\frac{\hbar^2}{8\pi Gm^2\rho_0}\right
)^{1/4},
\end{equation}
where $\theta(\xi)$ is the solution of the Lane-Emden equation (\ref{gl7}) of
index $n=2$.

A polytrope of index $n=2$ has a compact support and is stable.
Furthermore, according to Eq. (\ref{gl9}), the mass-radius relation is
\begin{equation}
\label{t6}
MR=\frac{1}{2}\omega_2\frac{\hbar^2}{Gm^2}=5.25\, \frac{\hbar^2}{Gm^2},
\end{equation}
where we have used $\omega_2=10.4950...$ Eq. (\ref{t6}) displays the same
scaling as the
mass-radius relation from Eq. (\ref{ni2}) but the prefactor is different. This
is because the
density profile given by Eq. (\ref{t5}) is different from the density profile
of the soliton that has been reported previously in the
literature \cite{rb,membrado,gul0,gul,prd2,ch2,ch3,pop,hui}
(see Sec. \ref{sec_ni}). Indeed, the authors of Refs. 
\cite{rb,membrado,gul0,gul,prd2,ch2,ch3,pop,hui} have looked
for a solution of Eq. (\ref{ni1})
corresponding to a density profile that goes to zero at infinity. Actually,
there exists another solution of Eq. (\ref{ni1}), given by Eq.
(\ref{t5}), corresponding to a density profile with a  compact
support that vanishes at a finite radius $R$.\footnote{In Ref. \cite{prd2} we
found a solution of Eq. (\ref{ni1}) for which ``the program breaks down because
the density achieves too small values ($<10^{-11}$).'' This solution corresponds
to the density profile (\ref{t5}) with a  compact support.} The two
profiles are plotted in
Fig. \ref{rhobosonNI}.
Apparently, they are both stable. It
may
therefore be useful to compare their
respective energy in
order to determine which profile has the lowest energy (ground
state).

The quantum kinetic energy of a BEC is given by Eq. (\ref{qk}). Integrating the
second
expression by parts, we obtain
\begin{equation}
\label{t7}
\Theta_Q=\frac{\hbar^2}{4m^2}\oint \nabla\rho\cdot d{\bf
S}-\frac{\hbar^2}{2m^2}\int \sqrt{\rho}\Delta\sqrt{\rho}\, d{\bf r}.
\end{equation}
Since $\rho'(R)\propto \theta_1\theta'_1$ for a polytrope of index $n=2$ [see
Eq. (\ref{t5})] and
since $\theta_1=0$ and 
$|\theta'_1|<+\infty$ (see Appendix \ref{sec_gl}), we have $\rho'(R)=0$. As a
result, the surface
term vanishes and Eq. (\ref{t7}) reduces to
\begin{equation}
\label{t8}
\Theta_Q=-\frac{\hbar^2}{2m^2}\int \sqrt{\rho}\Delta\sqrt{\rho}\, d{\bf r}.
\end{equation}
Using Eq. (\ref{t2}) we then find that
\begin{equation}
\label{t9}
\Theta_Q=3K\int\rho^{3/2}\, d{\bf r}.
\end{equation}
This result can be compared to the internal energy (\ref{vt3}) of a polytrope of
index $n=2$ which is
\begin{equation}
\label{t10}
U=2K\int\rho^{3/2}\, d{\bf r}.
\end{equation}
We have the relation
\begin{equation}
\label{t11}
\Theta_Q=\frac{3}{2}U.
\end{equation}
Therefore, the energy $E_{\rm tot}=\Theta_Q+W$ of a self-gravitating
noninteracting BEC with the density profile (\ref{t5}) is
different from the energy  $E_{\rm tot}=U+W$ of the corresponding $n=2$
polytrope.

For a  self-gravitating
noninteracting BEC, we have (see Sec. \ref{sec_ni})
\begin{equation}
\label{t12}
E_{\rm tot}=-\Theta_Q=\frac{1}{2}W.
\end{equation}
On the other hand, the gravitational energy of  a polytrope $n=2$ is (see
Appendix \ref{sec_br})
\begin{eqnarray}
\label{t13}
W=-\frac{GM^2}{R}.
\end{eqnarray}
Combining Eqs. (\ref{t12}) and (\ref{t13}) and using the mass-radius
relation from Eq. (\ref{t6}), we find that the
energy of a self-gravitating
noninteracting BEC with the density profile (\ref{t5}) is
\begin{equation}
\label{t14}
E_{\rm
tot}=-\frac{GM^2}{2R}=-\frac{1}{\omega_2}\frac{G^2M^3m^2}{\hbar^2}=-0.0953
\frac{G^2M^3m^2}{\hbar^2}.
\end{equation}
This energy is lower than the one given by Eq. (\ref{ni6}).
Therefore, the
solution of Eq. (\ref{ni1}) that has a compact support [see Eq. (\ref{t5})] has
a lower energy than the solution of Eq. (\ref{ni1}) that extends to infinity
(see
Sec. \ref{sec_ni}). The ground state of the self-gravitating
noninteracting BEC corresponds therefore to the solution considered in this
Appendix, not to the solution that has been
considered in Refs.
\cite{rb,membrado,gul0,gul,prd2,ch2,ch3,pop,hui}
(see Sec. \ref{sec_ni}). However, in the case of systems with long-range
interactions, we know that a metastable state (i.e. a local but not a global
energy minimum) may have a very long lifetime and can be fully relevant. This
may be the case of the solution considered in Refs.
\cite{rb,membrado,gul0,gul,prd2,ch2,ch3,pop,hui}
which seems to be selected in direct numerical simulations \cite{ch2,ch3}.

{\it Remark:} the energy of a polytrope
of index $n=2$ can be obtained from the relations (see Appendices \ref{sec_vt}
and \ref{sec_br}):
\begin{eqnarray}
\label{t14b}
E_{\rm tot}=U+W,
\end{eqnarray}
\begin{eqnarray}
\label{t15}
NE=\frac{3}{2}U+2W,
\end{eqnarray}
\begin{eqnarray}
\label{t16}
\frac{3}{2}U+W=0,
\end{eqnarray}
implying
\begin{eqnarray}
\label{t17}
E_{\rm tot}=-\frac{1}{2}U=\frac{1}{3}W=-\frac{GM^2}{3R}.
\end{eqnarray}
It differs from Eq. (\ref{t14}) by a factor $2/3$.

\section{Gravitational energy}
\label{sec_ge}

The gravitational energy of a spherically symmetric system is given by
(see, e.g., Appendix G of \cite{ggp})
\begin{eqnarray}
\label{ge1}
W=-\int_{0}^{+\infty} \rho(r)\frac{GM(r)}{r}\, 4\pi r^2\, dr,
\end{eqnarray}
where
\begin{eqnarray}
\label{ge2}
M(r)=\int_{0}^{r} \rho(r)\, 4\pi r^2\, dr
\end{eqnarray}
is the mass contained within the sphere of radius $r$. In our model,
the system is made of a core of mass $M_c$ and radius $R_c$ and a
uniform atmosphere of density $\rho_a$ and mass $M_a=M-M_c$ contained within the
spheres of radius
$R_c$ and $R$.  Therefore, we can write
\begin{eqnarray}
\label{ge3}
W=W_c-4\pi G \rho_a\int_{R_c}^{R} M(r)\, r\, dr,
\end{eqnarray}
where the first term $W_c$  is the gravitational energy of the core and the
second term  $W_a$ is the gravitational energy of the atmosphere in the presence
of
the core. For $r\ge R_c$, the cumulated mass is 
\begin{eqnarray}
\label{ge4}
M(r)=M_c+\frac{4}{3}\pi \rho_a
(r^3-R_c^3).
\end{eqnarray}
Substituting Eq. (\ref{ge4}) into
Eq. (\ref{ge3}) and evaluating the integral we get
\begin{equation}
\label{ge5}
W=W_c-4\pi G\rho_a\left\lbrack M_c
\frac{r^2}{2}+\frac{4}{15}\pi\rho_a
r^5-\frac{2}{3}\pi\rho_a R_c^3r^2\right\rbrack_{R_c}^{R}
\end{equation}
with
\begin{eqnarray}
\label{ge6}
\rho_a=\frac{3(M-M_c)}{4\pi (R^3-R_c^3)}.
\end{eqnarray}
These expressions are exact. For our problem, it is a good approximation to
assume that $R\gg R_c$. As a
result, the foregoing equation simplifies into
\begin{eqnarray}
\label{ge7}
W=W_c-4\pi G\rho_a\left (M_c \frac{R^2}{2}+\frac{4}{15}\pi\rho_a
R^5\right )
\end{eqnarray}
with
\begin{eqnarray}
\label{ge8}
\rho_a=\frac{3(M-M_c)}{4\pi R^3}.
\end{eqnarray}
We finally obtain 
\begin{eqnarray}
\label{ge9}
W=W_c-\frac{3GM_c(M-M_c)}{2R}-\frac{3G(M-M_c)^2}{5R}.
\end{eqnarray}
The first term is the potential energy of the core, the third term is the
potential energy of the atmosphere and the second term is the interaction
energy. This is as if we had a point mass $M_c$ at the center of a distribution
of mass $M-M_c$ \cite{inpreparation}.

\section{The minimum halo radius, the minimum halo mass and the maximum halo
central density}
\label{sec_p}

In this Appendix, we determine the radius, the mass and the central
density of the ``minimum halo'' assuming that it corresponds to the ground
state ($T=0$) of a self-gravitating quantum gas. We have seen in Sec.
\ref{sec_qm} that it can be represented by a polytropic sphere. For the sake of
generality we treat the case of an arbitrary polytropic index $n$, then consider
particular cases corresponding to fermions, noninteracting bosons and
self-interacting bosons in the TF limit.

\subsection{General results}

The halo radius $r_h$ is defined as the distance at which the
central density  $\rho_0$  is divided by $4$. Using Eqs. (\ref{gl6}) and
(\ref{gl6b}), it is given by
\begin{eqnarray}
\label{p1}
r_h=\xi_h\left\lbrack \frac{K(n+1)}{4\pi
G}\right\rbrack^{1/2}\frac{1}{\rho_0^{(n-1)/2n}},
\end{eqnarray} 
where
$\xi_h$ is determined by the equation
\begin{eqnarray}
\label{p2}
\theta(\xi_h)^{n}=\frac{1}{4}.
\end{eqnarray} 
The value of $\xi_h$ can be obtained by solving the Lane-Emden equation
(\ref{gl7}). The  halo mass,
which is the
mass contained within the sphere of radius $r_h$, is given by
\begin{eqnarray}
\label{p3}
M_h=-4\pi\frac{\theta'(\xi_h)}{{\xi_h}}\rho_0 r_h^3.
\end{eqnarray} 
Eliminating the central density between Eqs. (\ref{p1}) and (\ref{p3}), we
obtain the minimum halo mass-radius relation 
\begin{eqnarray}
\label{p4}
M_h
r_h^{(3-n)/(n-1)}=-4\pi\theta'(\xi_h)\xi_h^{(n+1)/(n-1)}\nonumber\\
\times\left\lbrack
\frac{K(n+1)}{4\pi
G}\right\rbrack^{n/(n-1)},
\end{eqnarray} 
which is analogous to the mass-radius relation (\ref{gl9}). On the
other hand, using Eqs. (\ref{p1}) and (\ref{p3}) and introducing the universal
surface
density of DM halos (\ref{p5}) we find that the minimum halo radius,
the minimum halo mass, and the maximum
halo central density are given by
\begin{equation}
\label{p6}
(r_h)_{\rm min}=\xi_h^{2n/(n+1)}\left\lbrack
\frac{K(n+1)}{4\pi
G}\right\rbrack^{n/(n+1)}\frac{1}{\Sigma_0^{(n-1)/(n+1)}}
\end{equation} 
and
\begin{eqnarray}
\label{p7}
(M_h)_{\rm min}=&-&4\pi \theta'(\xi_h) \xi_h^{(3n-1)/(n+1)}\nonumber\\
&\times&\left\lbrack
\frac{K(n+1)}{4\pi
G}\right\rbrack^{2n/(n+1)}\Sigma_0^{(3-n)/(n+1)}.\nonumber\\
\end{eqnarray} 
\begin{equation}
\label{sad2}
(\rho_0)_{\rm max}=\frac{1}{\xi_h^{2n/(n+1)}}\left\lbrack
\frac{4\pi
G}{K(n+1)}\right\rbrack^{n/(n+1)}\Sigma_0^{2n/(n+1)}.
\end{equation}

\subsection{Fermions}

For the polytrope $n=3/2$ (fermion stars) using Eq. (\ref{fdm2}), we have
\begin{equation}
\xi_h=2.27, \qquad \theta'_h=-0.360,
\label{p9}
\end{equation}
\begin{equation}
\label{frpolf}
r_h=  0.223   \, \frac{h}{m^{4/3}G^{1/2}\rho_0^{1/6}},
\end{equation}
\begin{equation}
\label{fmpolf}
M_h=1.99\, \rho_0 r_h^3,
\end{equation}
\begin{equation}
\label{fmrpolf}
M_hr_h^3=2.45\times 10^{-4}\, \frac{h^6}{G^3 m^8}.
\end{equation}
Using Eq. (\ref{p5}), we obtain Eqs.
(\ref{fpv2})-(\ref{fpv3b}). Inversely, assuming that the mass
$(M_h)_{\rm min}$ of the minimum halo is known, we obtain the fermion
mass
\begin{equation}
\label{mf}
m=1.60\, \frac{\hbar^{3/4}\Sigma_0^{3/16}}{G^{3/8}(M_h)_{\rm min}^{5/16}}.
\end{equation}
If we take $(M_h)_{\rm min}=10^8\, M_{\odot}$ we obtain $m=165 \, {\rm
eV}/c^2$.

\subsection{Noninteracting bosons}

For the polytrope $n=2$ (noninteracting boson stars) we have
\begin{equation}
\xi_h=2.092, \qquad \theta'_h=-0.3216,
\label{p11}
\end{equation}
\begin{equation}
\label{brpolf}
r_h=  0.934333     \, \frac{\hbar^{1/2}}{m^{1/2}G^{1/4}\rho_0^{1/4}},
\end{equation}
\begin{equation}
\label{bmpolf}
M_h=  1.93181    \, \rho_0 r_h^3,
\end{equation}
\begin{equation}
\label{bmrpolf}
M_h r_h=1.47221\, \frac{\hbar^2}{G m^2}.
\end{equation}
Using Eq. (\ref{p5}), we obtain Eqs.
(\ref{nipv2})-(\ref{nipv3b}). Inversely, assuming that the mass
$(M_h)_{\rm min}$ of the minimum halo is known, we obtain the
noninteracting boson mass
\begin{equation}
\label{mbni}
m=1.43\, \frac{\hbar\Sigma_0^{1/4}}{G^{1/2}(M_h)_{\rm min}^{3/4}}.
\end{equation}
If we take $(M_h)_{\rm min}=10^8\, M_{\odot}$ we obtain $m=1.44\times 10^{-22}
\, {\rm
eV}/c^2$.

\subsection{Self-interacting bosons in the TF limit}

For the polytrope $n=1$ (self-interacting boson stars) we have
\begin{equation}
\xi_h=2.4746, \qquad \theta'_h=-0.41853,
\label{p13}
\end{equation}
\begin{equation}
\label{tfrpolf}
r_h=2.4746\, \left (\frac{a_s\hbar^2}{Gm^3}\right )^{1/2},
\end{equation}
\begin{equation}
\label{tfmpolf}
M_h=2.12535\, \rho_0 r_h^3.
\end{equation}
Using Eq. (\ref{p5}), we obtain Eqs.
(\ref{sipv2})-(\ref{sad1}). Inversely, assuming
that the mass
$(M_h)_{\rm min}$ of the minimum halo is known, we obtain the
self-interacting boson parameter (in the TF limit)
\begin{equation}
\label{mbft}
\frac{a_s}{m^3}=0.0769\, \frac{G(M_h)_{\rm
min}}{\hbar^2\Sigma_0}.
\end{equation}
If we take $(M_h)_{\rm min}=10^8\, M_{\odot}$ we obtain $a_s/m^3=1.76\times
10^3\, {\rm fm}/({\rm eV}/c^2)^3$.

\section{Results of the quantum Jeans instability theory}
\label{sec_qj}

In this Appendix, we recapitulate the main results of the quantum Jeans
instability theory developed in Refs. \cite{prd1,abriljeans}, restricting
ourselves to
the
nonrelativistic regime. We refer to these papers for details about their
derivation and for some generalizations.

\subsection{Fermions}
\label{sec_jf}

If DM  is made of fermions, the Jeans length and the
Jeans mass are given by
\begin{equation}
\label{jf1}
\lambda_J=\frac{1}{2}\left (\frac{\pi}{3}\right )^{1/6} \,
\frac{h}{G^{1/2}m^{4/3}\rho^{1/6}},
\end{equation}
\begin{equation}
\label{jf2}
M_J=\frac{4}{3}\pi\rho \left (\frac{\lambda_J}{2}\right
)^{3}=\frac{1}{16}\left (\frac{\pi}{3}\right )^{3/2} \,
\frac{h^3\rho^{1/2}}{G^{3/2}m^{4}}.
\end{equation}
Eliminating the density between these expressions, we get the Jeans mass-radius
relation
\begin{equation}
\label{jf3}
M_J \lambda_J^3=\frac{\pi^2}{1152} \frac{h^6}{G^3 m^8}.
\end{equation}
We can also define a Jeans surface density 
\begin{equation}
\label{jf4}
\Sigma_J=\rho \frac{\lambda_J}{2}=\frac{1}{4}\left (\frac{\pi}{3}\right )^{1/6}
\,
\frac{h\rho^{5/6}}{G^{1/2}m^{4/3}}.
\end{equation}
These equations display the same scalings as Eqs. (\ref{frpolf})-(\ref{fmrpolf})
and (\ref{fpv3b}) for DM halos.

\subsection{Noninteracting bosons}
\label{sec_jb}

If DM is made of noninteracting bosons, the Jeans length and the Jeans mass are
given by
\begin{equation}
\label{jb1}
\lambda_J=2\pi\left (\frac{\hbar^2}{16\pi G\rho m^2}\right )^{1/4},
\end{equation}
\begin{equation}
\label{jb2}
M_J=\frac{4}{3}\pi\rho \left (\frac{\lambda_J}{2}\right
)^{3}=\frac{\pi}{6}\left (\frac{\pi^3\hbar^2\rho^{1/3}}{Gm^2}\right )^{3/4}.
\end{equation}
Eliminating the density between these expressions, we get the Jeans mass-radius
relation
\begin{equation}
\label{jb3}
M_J \lambda_J=\frac{\pi^4}{6} \frac{\hbar^2}{G m^2}.
\end{equation}
The Jeans surface density is 
\begin{equation}
\label{jb4}
\Sigma_J=\rho \frac{\lambda_J}{2}=\pi\left (\frac{\hbar^2\rho^3}{16\pi G
m^2}\right )^{1/4}.
\end{equation}
These equations display the same scalings as Eqs. (\ref{brpolf})-(\ref{bmrpolf})
and (\ref{nipv3b}) for DM halos.

\subsection{Self-interacting bosons in the TF limit}
\label{sec_sib}

If DM is made of self-interacting bosons in the TF limit, the Jeans length and
the Jeans mass
are
given by
\begin{equation}
\label{sib1}
\lambda_J=2\pi\left (\frac{a_s\hbar^2}{Gm^3}\right )^{1/2},
\end{equation}
\begin{equation}
\label{sib2}
M_J=\frac{4}{3}\pi\rho \left (\frac{\lambda_J}{2}\right
)^{3}=\frac{\pi}{6}\rho\left (\frac{4\pi^2a_s\hbar^2}{Gm^3}\right
)^{3/2}.
\end{equation}
The Jeans surface density is
\begin{equation}
\label{sib3}
\Sigma_J=\rho \frac{\lambda_J}{2}=\pi\rho \left (\frac{a_s\hbar^2}{G
m^3}\right )^{1/2}.
\end{equation}
These equations display the same scalings as Eqs.
(\ref{tfrpolf}), (\ref{tfmpolf})
and (\ref{sad1}) for DM halos.

\section{The mass of the DM particle}
\label{sec_mdm}

In this Appendix, we relate the mass $m$ of the DM particle to the
cosmological constant $\Lambda$ and to the other fundamental constants of
physics.

\subsection{Fermions}
\label{sec_msf}

We have seen in Sec. \ref{sec_fdm} that the minimum mass of DM halos made of
fermions is given by
\begin{equation}
\label{msf1}
(M_{h})_{\rm min}=4.47\, \left (\frac{\hbar^{12}\Sigma_0^3}{G^6m^{16}}\right
)^{1/5}.
\end{equation}
It is obtained by requiring that the smallest DM halo in the Universe
corresponds to the ground state of the self-gravitating Fermi gas.

On the other hand, the minimum mass of DM
halos may be obtained from a
quantum Jeans instability theory (see, e.g., Refs. \cite{prd1,abriljeans})
leading to
the Jeans mass (\ref{jf2}). Let us compute the Jeans mass at the present epoch
where
$\rho_{\rm m,0}=2.66\times 10^{-24}\, {\rm g \, m^{-3}}$. For a  fermion
mass $m=170\, {\rm eV/c^2}$, the Jeans mass is $M_J=1.10\times 10^5\,
M_{\odot}$  \cite{abriljeans}. It is $3$ orders of
magnitude smaller than the
minimum mass $(M_h)_{\rm min} \sim
10^8\, M_{\odot}$ of observed DM halos. Actually, we cannot
expect to have a perfect agreement between the Jeans mass computed at the
present epoch and the observed
minimum mass of DM halos because the linear Jeans  instability took place at
an earlier epoch (see Appendix \ref{sec_com}) and the present DM halos result
from a nonlinear evolution. Therefore, we write $(M_h)_{\rm
min}=\chi_{\rm F}
M_J$, where $\chi_{\rm F}$ is a dimensionless factor that is difficult to
predict theoretically (for fermions the previous estimate gives $\chi_{\rm
F}\sim 10^3$). Using Eq. (\ref{jf2}), we obtain
\begin{equation}
\label{msf2}
(M_{h})_{\rm min}=16.6\,\chi_{\rm F} \,
\frac{\hbar^3\rho_{\rm m,0}^{1/2}}{G^{3/2}m^{4}}.
\end{equation}

Comparing (\ref{msf1}) and (\ref{msf2}) we get
\begin{equation}
\label{msf3}
m=5.16\, \chi_{\rm F}^{5/4}\, \frac{\hbar^{3/4}\rho_{\rm
m,0}^{5/8}}{\Sigma_0^{3/4}G^{3/8}}.
\end{equation}
This relation gives the surface density $\Sigma_0$ of the smallest DM halo if
we know the
fermion mass $m$. Inversely, since $\Sigma_0$ appears to have a universal value
(see Eq. (\ref{p5})),  we can use Eq. (\ref{msf3})
to obtain the fermion
mass $m$. More precisely, since  $\rho_{\rm m,0}$ and $\Sigma_0$ can be
expressed in terms of the cosmological constant $\Lambda$ by Eqs. (\ref{sc1b})
and (\ref{sc12}), we find that the mass of
the fermionic particle is given by
\begin{equation}
m=7.62\, \chi_{\rm F}^{5/4}\,\left (\frac{\Lambda\hbar^3}{Gc^3}\right )^{1/4}.
\label{msf4}
\end{equation}
It is equal to the mass
scale
$m_{\Lambda}^*=5.04\times 10^{-3}\, {\rm eV/c^2}$ given by Eq. (\ref{sc9})
multiplied by a large numerical factor of
order $4\times 10^{4}$ (for $\chi_{\rm F}\sim 10^3$). This gives $m\sim 200\,
{\rm eV/c^2}$ which is the correct order
of magnitude of the fermion mass usually advocated in DM models (see
Appendix D of \cite{suarezchavanis3}). We note
that, up to the dimensionless factor $\chi_{\rm F}$, this mass scale has been
predicted in terms of the fundamental constants of physics independently from
the
observations.

{\it Remark:} We can obtain these results (without the prefactor) directly from
the Jeans scales of Appendix \ref{sec_jf}. From Eq. (\ref{jf4}), we have
\begin{equation}
\label{jf5}
\Sigma\sim \frac{\hbar \rho^{5/6}}{m^{4/3}G^{1/2}}\quad {\rm
i.e.}\quad  m\sim \frac{\hbar^{3/4}\rho^{5/8}}{\Sigma^{3/4}G^{3/8}}.
\end{equation}
If we take $\rho\sim \rho_{\rm m,0}\sim \Lambda/G$ and $\Sigma\sim
c\sqrt{\Lambda}/G$ (see  Eqs. (\ref{sc1b})
and (\ref{sc12})), we get
\begin{equation}
\label{jf6}
m\sim \left (\frac{\Lambda\hbar^3}{Gc^3}\right )^{1/4}\sim m_{\Lambda}^*.
\end{equation}
Inversely, if we assume that $\rho\sim \rho_{\rm m,0}\sim \Lambda/G$
and $m\sim m_{\Lambda}^*$ we find that $\Sigma\sim c\sqrt{\Lambda}/G$. We also
note that $\lambda\sim \Sigma/\rho\sim c/\sqrt{\Lambda}$ and $M\sim
\rho\lambda^3\sim c^3/G\sqrt{\Lambda}$ are the cosmological scales corresponding
to the size and to the mass of the Universe \cite{ouf}.

\subsection{Noninteracting bosons}
\label{sec_msbni}

We have seen in Sec. \ref{sec_ni} that the minimum mass of DM halos made of
noninteracting bosons is
given by
\begin{equation}
\label{msbni1}
(M_{h})_{\rm min}=1.61\, \left (\frac{\hbar^4\Sigma_0}{G^2m^4}\right
)^{1/3}.
\end{equation}
On the other hand, the
quantum Jeans instability theory \cite{prd1,abriljeans} leads to
the Jeans mass (\ref{jb2}). For a boson
mass $m=2.92\times 10^{-22}\, {\rm eV}/c^2$, the Jeans mass computed 
at the present epoch where
$\rho_{\rm m,0}=2.66\times 10^{-24}\, {\rm g \, m^{-3}}$ is $M_J=3.07\times
10^6\,
M_{\odot}$ \cite{abriljeans}. It is $1$ order
of
magnitude smaller than the
minimum mass $(M_h)_{\rm min} \sim
10^8\, M_{\odot}$ of observed DM halos. Writing  $(M_h)_{\rm min}=\chi_{\rm B}
M_J$ with $\chi_{\rm B}\sim 10$, we obtain
\begin{equation}
\label{msbni2}
(M_{h})_{\rm min}=6.88\,\chi_{\rm B} \,
\frac{\hbar^{3/2}\rho_{\rm m,0}^{1/4}}{G^{3/4}m^{3/2}}.
\end{equation}
Comparing Eqs. (\ref{msbni1}) and (\ref{msbni2}) we get
\begin{equation}
\label{msbni3}
m=6089\, \chi_{\rm B}^{6}\, \frac{\hbar\rho_{\rm
m,0}^{3/2}}{\Sigma_0^{2}G^{1/2}}.
\end{equation}
Using Eqs. (\ref{sc1b})
and (\ref{sc12}), we find that the mass of
the bosonic particle (in the noninteracting case) is given by
\begin{equation}
m=33748\, \chi_{\rm B}^{6}\, \frac{\hbar\sqrt{\Lambda}}{c^2}.
\label{msbni4}
\end{equation}
It is equal to the mass scale
$m_{\Lambda}=2.08\times 10^{-33}\, {\rm eV/c^2}$ given by Eq. (\ref{sc15})
multiplied by a huge numerical factor of
order $3\times 10^{10}$ (for $\chi_{\rm B}\sim 10$). This is because $\chi_B$
is raised in Eq.
(\ref{msbni4}) to the power $6$. This gives $m\sim 
10^{-22}\, {\rm eV/c^2}$ which is the correct
order
of magnitude of the mass of ultralight axions usually advocated in DM models
(see
Appendix D of \cite{suarezchavanis3}).

{\it Remark:} We can obtain these results (without the prefactor) directly from
the Jeans scales of Appendix \ref{sec_jb}. From Eq. (\ref{jb4}), we have
\begin{equation}
\label{jb5}
\Sigma\sim \left (\frac{\hbar^2 \rho^{3}}{G m^2}\right )^{1/4}\quad {\rm
i.e.}\quad  m \sim \left (\frac{\hbar^2 \rho^{3}}{G \Sigma^4}\right )^{1/2}.
\end{equation}
If we take $\rho\sim \rho_{\rm m,0}\sim \Lambda/G$ and $\Sigma\sim
c\sqrt{\Lambda}/G$ (see  Eqs. (\ref{sc1b})
and (\ref{sc12})), we get
\begin{equation}
\label{jb6}
m\sim \frac{\hbar\sqrt{\Lambda}}{c^2}\sim m_{\Lambda}.
\end{equation}
Inversely, if we assume that $\rho\sim \rho_{\rm m,0}\sim \Lambda/G$
and $m\sim m_{\Lambda}$ we find that $\Sigma\sim c\sqrt{\Lambda}/G$.

\subsection{Self-interacting bosons in the TF limit}
\label{sec_msbsi}

We have seen in Sec. \ref{sec_tf} that the minimum mass of DM halos made of
self-interacting bosons in the TF limit is
given by
\begin{equation}
\label{msbsi1}
(M_{h})_{\rm min}=13.0\, \frac{a_s\hbar^2\Sigma_0}{G m^3}.
\end{equation}
On the other hand, the
quantum Jeans instability theory \cite{prd1,abriljeans} leads to
the Jeans mass (\ref{sib2}). For a
ratio $a_s/m^3=3.28\times 10^3\, {\rm fm\, (eV/c^2)^{-3}}$, the Jeans mass
computed at the present epoch where $\rho_{\rm m,0}=2.66\times 10^{-24}\, {\rm g
\, m^{-3}}$ is $M_J=165 M_{\odot}$ \cite{abriljeans}. It is $6$ order
of
magnitude smaller than the
minimum mass $(M_h)_{\rm min} \sim
10^8\, M_{\odot}$ of observed DM halos. Writing $(M_h)_{\rm min}=\chi_{\rm TF}
M_J$ with $\chi_{\rm TF}\sim 10^6$, we obtain 
\begin{equation}
\label{msbsi2}
(M_{h})_{\rm min}=130\,\chi_{\rm TF}\,\rho_{\rm m,0}\left
(\frac{a_s\hbar^2}{Gm^3}\right
)^{3/2}.
\end{equation}
Comparing (\ref{msbsi1}) and (\ref{msbsi2}) we get
\begin{equation}
\label{msbsi3}
\frac{a_s}{m^3}=\frac{0.01}{\chi_{\rm TF}^2}\frac{G\Sigma_0^2}{\hbar^2\rho_{\rm
m,0}^2}.
\end{equation}
Using Eqs. (\ref{sc1b})
and (\ref{sc12}), we find that the ratio $a_s/m^3$  is
given by
\begin{equation}
\frac{a_s}{m^3}=\frac{0.0135}{\chi_{\rm TF}^2}\frac{Gc^2}{\Lambda\hbar^2}.
\label{msbsi4}
\end{equation}
It is equal to the
scale ${r_{\Lambda}}/{m_{\Lambda}^3}={2Gc^2}/{\Lambda\hbar^2}=6.11\times
10^{17}\, {\rm fm\, (eV/c^2)^{-3}}$ given by Eq. (\ref{rlml3})
multiplied by a very small numerical factor of
order $10^{-14}$ (for $\chi_{\rm TF}\sim 10^6$). This gives 
$a_s/m^3\sim 10^3\, {\rm fm\, (eV/c^2)^{-3}}$ which is the correct
order
of magnitude of the parameter $a_s/m^3$ of self-interacting bosons usually
advocated in DM models (see
Appendix D of \cite{suarezchavanis3}).

{\it Remark:} We can obtain these results (without the prefactor) directly from
the Jeans scales of Appendix \ref{sec_sib}. From Eq. (\ref{sib3}), we have
\begin{equation}
\label{sib4}
\Sigma\sim \rho \left (\frac{a_s\hbar^2}{G
m^3}\right )^{1/2}\quad {\rm
i.e.}\quad  \frac{a_s}{m^3} \sim \frac{G\Sigma^2}{\hbar^2\rho^2}.
\end{equation}
If we take $\rho\sim \rho_{\rm m,0}\sim \Lambda/G$ and $\Sigma\sim
c\sqrt{\Lambda}/G$ (see  Eqs. (\ref{sc1b})
and (\ref{sc12})), we get
\begin{equation}
\label{sib5}
\frac{a_s}{m^3}\sim \frac{Gc^2}{\Lambda\hbar^2}\sim
\frac{r_\Lambda}{m_\Lambda^3}.
\end{equation}
Inversely, if we assume that $\rho\sim \rho_{\rm m,0}\sim \Lambda/G$
and $a_s/m^3\sim r_\Lambda/m_{\Lambda}^3$ we find that $\Sigma\sim
c\sqrt{\Lambda}/G$.

\subsection{The epoch where we can apply the Jeans instability theory}
\label{sec_com}

Let us consider how the minimum halo mass $(M_h)_{\rm min}$ depends on the
density $\rho_{\rm m}$ and on the dimensionless constant $\chi$ in Eqs.
(\ref{msf2}), (\ref{msbni2}) and (\ref{msbsi2}). For fermions, they appear in
the combination $\chi_{\rm F}
\rho_{\rm m,0}^{1/2}=(\chi_{\rm F}^2 \rho_{\rm
m,0})^{1/2}=(10^6 \rho_{\rm
m,0})^{1/2}$. For noninteracting bosons, we have $\chi_{\rm B} \rho_{\rm
m,0}^{1/4}=(\chi_{\rm B}^4 \rho_{\rm
m,0})^{1/4}=(10^4 \rho_{\rm
m,0})^{1/2}$. For self-interacting bosons in the TF limit, we have
$\chi_{\rm TF} \rho_{\rm
m,0}=10^6 \rho_{\rm
m,0}$. Remarkably, the last terms in these equalities involve the {\it same}
typical number $\sim 10^5\, \rho_{\rm m,0}$. These relations suggest that
$(M_h)_{\rm min}$ is not equal to the
present Jeans length but rather to the Jeans length at the epoch where the
density of the Universe is $\sim 10^5 \rho_{\rm m,0}$. Indeed, if we calculate
the Jeans mass at the epoch where the density of the Universe is  $10^5
\rho_{\rm m,0}$, we find in the three cases considered above that $M_J\sim
(M_h)_{\rm min}$. Using $\rho_{\rm
m}=\rho_{\rm m,0}/a^3$ and $z+1=1/a$ (redshift), this epoch corresponds to
\begin{equation}
\rho_{\rm m}=2.66\times 10^{-19}\, {\rm
g \, m^{-3}} , \quad a=0.0215, \quad z=45.5.
\label{mdm5}
\end{equation}
It is intermediate between the epoch of radiation-matter equality ($\rho_{\rm
eq}=8.77\times 10^{-14}\, {\rm
g \, m^{-3}}, a_{\rm eq}=2.95\times 10^{-4}, z_{\rm eq}=3389$) and the
present epoch ($\rho_{m,0}=2.66\times 10^{-24}\, {\rm
g \, m^{-3}}, a_0=1, z_{0}=0$). It may correspond to the epoch where we can
apply the linear Jeans instability theory to predict the typical mass of DM
halos.

\end{document}